 \documentclass[iop]{emulateapj}
\usepackage{url}
\usepackage{graphicx}
\usepackage{url}
\usepackage[T1]{fontenc}
\usepackage{hyperref}

\usepackage{rotating}
\def\nborg{93106}
\def\borgstars{1308}
\def\borgarea{209.9}
\def\borgnorth{134.5}
\def\borgsouth{72.1}

\def\Mdwarfs{274}
\def\Ldwarfs{30}
\def\Tdwarfs{1}
\def\MLTdwarfs{29}

\def\nM{{274}} % number of M-dwarfs identified in the BoRG fields
\def\nL{{30}} % number of L-dwarfs identified in the BoRG fields
\def\nT{{1}} % number of T-dwarfs identified in the BoRG fields

\def\Nfields{68}
\def\f125w{{\em F125W}}
\def\f160w{{\em F160W}}
\def\f098m{{\em F098M}}
\def\f606w{{\em F606W}}
\def\f600lp{{\em F600LP}}

\slugcomment{ApJ {\em accepted} + ERRATUM}

\shorttitle{Milky Way M-Dwarfs}
\shortauthors{Holwerda et al.}

\begin{document}

%% LaTeX will automatically break titles if they run longer than
%% one line. However, you may use \\ to force a line break if
%% you desire.

\title{Milky Way Red Dwarfs in the BoRG Survey;\\
Galactic scale-height and the distribution of dwarfs stars in WFC3 imaging}

%%%%%%%%%%%%%%%%%%%%%%%%%%%%%%%%%%%%%%%%%%%%%%%%%%
% v0 - first nearly completed draft
% v1 -  found out about f125w zeropoint mismatch, fix
% v2 - new sample; BoRG 2.0 59 fields total.

% v5 - commentss Michele, remove field throught the bulge.
% v6 - comments from Nor
%%%%%%%%%%%%%%%%%%%%%%%%%%%%%%%%%%%%%%%%%%%%%%%%%%

\author{B. W.Holwerda\altaffilmark{1,2}, 
M. Trenti\altaffilmark{3}, 
W. Clarkson\altaffilmark{4}, 
K. Sahu\altaffilmark{5},  
L. Bradley\altaffilmark{5}, 
M. Stiavelli\altaffilmark{5},
N. Pirzkal\altaffilmark{5}, 
\\
G. De Marchi\altaffilmark{2},
M. Andersen\altaffilmark{6},
R. Bouwens\altaffilmark{1}, 
R. Ryan\altaffilmark{5}
I. van Vledder\altaffilmark{1},
D. van der Vlugt\altaffilmark{1}
}

% \author{W. Clarkson\altaffilmark{4}, M. Andersen\altaffilmark{1}, G. de Marchi\altaffilmark{1} }

%% Notice that each of these authors has alternate affiliations, which
%% are identified by the \altaffilmark after each name.  Specify alternate
%% affiliation information with \altaffiltext, with one command per each
%% affiliation.

\altaffiltext{1}{Leiden Observatory, Leiden University, P.O. Box 9513, 2300 RA Leiden, The Netherlands\email{benne.holwerda@rssd.esa.int}, twitter: \@ benneholwerda }
\altaffiltext{2}{Research Fellow, European Space Agency, ESA-ESTEC, Keplerlaan 1, 2200 AG Noordwijk, the Netherlands}
\altaffiltext{3}{University of Melbourne?}
\altaffiltext{4}{Department of Natural Sciences College of Arts, Sciences and Letters, University of Michigan-Dearborn 4901 Evergreen Road, Dearborn, MI 48128, USA}
% \altaffiltext{3}{Center for Astrophysics and Space Astronomy, University of Colorado, 389 UCB, Boulder, Colorado 80309-0389, USA}
\altaffiltext{5}{Space Telescope Science Institute, Baltimore, MD 21218, USA
\altaffiltext{6}{UJF-Grenoble 1/CNRS-INSU, Institut de Plan\'etologie et d'Astrophysique de Grenoble (IPAG) UMR 5274, Grenoble, F-38041, France}
}

\begin{abstract}
We present a tally of Milky Way late-type dwarf stars in \Nfields~ WFC3 pure-parallel fields (227 arcmin$^2$) from the Brightest of Reionizing Galaxies (BoRG) survey for high-redshift galaxies.
Using spectroscopically identified M-dwarfs in two public surveys, the CANDELS and the ERS mosaics, we identify a morphological selection criterion using the half-light radius ($r\_{50}$), a near-infrared J-H, G-J color region where M-dwarfs are found, and a V-J relation with M-dwarf subtype. 

We apply this morphological selection of stellar objects, color-color selection of M-dwarfs and optical-near-infrared color subtyping to compile a catalog of \nM\ M-dwarfs belonging to the disk of the Milky Way with a limiting magnitude of $m_{F125W} < 24 (AB)$. Based on the M-dwarfs statistics, we conclude that 
 (a) the previously identified North/South discrepancy in M-dwarf numbers persists in our sample; there are more M-dwarfs in the Northern fields on average than in Southern ones, 
  (b) the Milky Way's single disk scale-height for M-dwarfs is 0.3-4 kpc, depending on sub-type,
 (c) the scale-height depends on M-dwarf subtype with early type (M0-4) high scale-height ($z_0 =3-4$ kpc) and later types M5 and above in the thin disk ($z_0 = 0.3-0.5$ kpc). 
 (d) a second component is visible in the vertical distribution, with a different, much higher scale-height in the Southern Fields compared to the Northern ones.
 We report the M-dwarf component of the Sagittarius stream in one of our fields with 11 confirmed M-dwarfs, 7 of which are at the stream's distance
In addition to the M-dwarf catalog, we report the discovery of \nT\ T-dwarf and \nL\ L-dwarfs from their near-infrared colors. 
The dwarf scale-height and the relative low incidence in our fields of L- and T-dwarfs in these fields makes it unlikely that these stars will be interlopers in great numbers in color-selected samples go high-redshift galaxies.
The relative ubiquity of M-dwarfs however will make them ideal tracers of Galactic Halo substructure with EUCLID and reference stars for JWST observations. 
\end{abstract}


\section{Introduction}

Counting stars to infer the shape and size of our Milky Way Galaxy is a classic experiment in Astronomy. However, it is also among the most prone to insufficient data as well as conceptual shortfalls \citep[e.g., Herschel 1785;][]{Kapteyn22}. Over time, it was established that the Milky Way's scale height is inversely proportional to the masses of stars sampled \citep{Gilmore83,Gilmore84,Siegel02}.  Initially, many of the Galactic models were focused on relatively luminous giant stars and rarely addressed (sub)stellar objects due to completeness issues \citep[see for reviews of star counts and Galactic structure;][]{Bahcall86,Gilmore89, MWasGal, Majewski93}. 

Interest has now shifted to the spatial distribution of (sub)stellar objects, in part because they are possibly the most significant interlopers in studies of extremely high redshift objects \citep[see e.g.,][]{Caballero08}, but also because these are physically some of the longest lived objects. Consequently, any deep high-redshift survey, especially in the near-infrared, undertaken with the Hubble Space Telescope automatically becomes a census of the smallest Milky Way stars.
More importantly, these objects constitute the most numerous and oldest population of stars in the Galactic disk.
For instance, \cite{Pirzkal05} determined the scale-height of different types of dwarfs from the Hubble Ultra Deep Field \citep[HUDF,][]{Beckwith06}.
\cite{Ryan05} found L and T dwarfs in a small set of ACS parallel observations. \cite{Stanway08} and \cite{Pirzkal09} determined the Galactic scale-height of 
M-dwarfs from the Great Observatories Origins Deep Survey fields \citep[GOODS,][]{goods}. \cite{Ryan11} added a search of early WFC3 pure-parallel fields for L and T dwarfs. These studies gradually improved statistics on L, T and M-dwarfs to several dozens of objects. 

The identification of these Galactic dwarfs in images is difficult as one needs both high-quality data to distinguish from extragalactic sources and a good understanding of the initial mass function or the local density of these objects as a class. 
Two of the above studies \citep{Pirzkal05,Pirzkal09} benefited from spectroscopic identification of the dwarf subtype in the 
HST/ACS grism spectra of the HUDF and GOODS fields (the GRAPES\footnote{Grism ACS Program for Extragalactic Science, see \cite{Pirzkal04}} and
PEARS\footnote{Probing Evolution And Reionization Spectroscopically, see \cite{Pirzkal09}.} projects respectively). Combined with an index of stellarity in the direct image, 
this made identification of the dwarfs near certain. Alternatively, \cite{Kilic05} used high-precision proper motion to identify the Galactic objects.
Once an object has been identified as a candidate dwarf, its luminosity distance must be estimated, which requires accurate (sub)typing. Without direct spectra, we will show that this can be readily achieved with sufficient color coverage from the optical to the near-infrared. 
A dwarf census including subtype and distance, can answer several outstanding issues regarding the shape of the Milky Way traced by its most numerous stellar. members.

\cite{Pirzkal09} found, like \cite{Stanway08}, an odd discrepancy in the numbers of M-dwarfs: there are 24\% more M-dwarfs in the Northern than in the  Southern GOODS field. Since our vantage point in the Milky Way is above the plane of the disk, one would expect the Southern GOODS fields to have more M-dwarfs. However, as present, this discrepancy has only been determined from two sight-lines out of the Galaxy, so it is still possible this may be source count field-to-field variance.  By studying the older stellar population content of our Galaxy along many sight-lines down to the faintest possible magnitude, we can directly observe and measure the shape of the thin Galactic disk as traced by the old stellar population. 

In order to improve our understanding of the contribution of faint dwarfs to Galactic structure, we will need much improved statistics to compare against 
Milky Way structure models. We use the recent {\em HST/WFC3} Pure Parallel searches for bright high-redshift galaxies to find Milky Way dwarfs and determine the parameters of their distribution in the Milky Way disk \citep{Trenti11,Yan10b, Bradley12}. 
Thus far, based on {\em HST} imaging, the disk scale height for dwarfs has been found to be: 300 pc (Table \ref{t:z0}). \citep{Zheng01}, 400 $\pm$ 100 pc \citep{Pirzkal05}, 350 $\pm$ 50 pc, \citep[L, T dwarfs][]{Ryan05}, 370 $\pm$ 65 pc (for M4--M9 dwarfs) and 300 $\pm$ 70 pc \citep[for M0--M9 dwarfs,][]{Pirzkal09} and 300 $\pm$ 25 ($\pm$ 31 systematic) pc. \citep[$>$M8, ][]{Ryan11}.

The goal of this paper is to explore the reality of the North-South difference in M-dwarf counts and to accurately determine the scale-height of our Milky Way as traced by M-dwarfs.
Previously, \cite{Ryan11} investigated the L and T dwarf content, here we focus our analysis on the M dwarfs. 
% If the reader is suddenly overwhelmed by a strong sense of deja-vu, \cite{Ryan11} tackles a good number of our fields in a very similar manner but they look for L and T dwarfs, ignoring the bounty of M-dwarfs in the sample. We focus on the latter.

The paper is organized as follows, 
\S \ref{s:data} reports the properties of the {\em HST/WFC3} observations we use, 
\S \ref{s:cat} discusses how we generate the relevant {\sc sextractor} catalogs, 
\S \ref{s:cat:candels} presents our observational calibration of morphological identification of stars, M-dwarfs specifically, and subsequent sub-typing of the M-dwarfs. 
\S \ref{s:MWstars} presents our catalog of dwarfs in the BoRG fields and briefly discusses the validity of our morphological selection.
In \S \ref{s:disc} we discuss the North/South difference, the Galactic scale-height as a function of dwarf subtype. 
\S \ref{s:concl} lists our conclusions and outlines future work.

% We propose to use the multi-filter WFC3 data from HST Proposal 11700, a pure parallel SNAP proposal, to put a limit the oblateness of the M-dwarf MW disk. 
% We have a reasonable idea already of the thin disk height from HUDF and GOODS-N/S data (GRAPES and PEARS grism projects respectively).

\begin{table}[htdp]
\caption{\label{t:z0}The scale-heights for the thin disk of dwarf stars from the literature.}
\begin{center}
\begin{tabular}{l l l}
Milky Way Scale  	& Reference	& type stars \\ 
Height ($z_0$) (pc.)							& & \\
\hline
300  							& \cite{Zheng01} 	& \\
400 $\pm$ 100  				& \cite{Pirzkal05} 	& \\
350 $\pm$ 50 					& \cite{Ryan05} 	& L, T dwarfs\\
370 $\pm$ 65 		 			& \cite{Pirzkal09} 	& M4--M9 dwarfs\\
300 $\pm$ 70  					& \cite{Pirzkal09}	& M0--M9 dwarfs \\
300 $\pm$ 25 $\pm$ 31 			& \cite{Ryan11} & $>$M8 \\
\hline
\end{tabular}
\end{center}
\end{table}%

\section{WFC3 Observations}
\label{s:data}

We use three WFC3 data-sets for this project: the CANDELS public v0.5 data-release, and the Early Release Science mosaic, both in the GOODS-South field for calibration and characterization of known M-dwarfs,
and the BoRG pure-parallel observations for the detection of new Galactic M-dwarfs.

\begin{table}[htdp]
\caption{\label{t:surveys} Summary of the WFC3 observations of BoRG, CANDELS (v0.5) and ERS on GOODS-S.}
\begin{center}
\begin{tabular}{l l l l}

			& BoRG							& CANDELS 		& ERS\\
\hline
Filters		& {\em F125W, } 					& {\em F125W} 		& {\em F125W} \\
			& {\em F160W, }					& {\em F160W}		& {\em F160W} \\
			& {\em F098M,} 					& 				& {\em F098M} \\
			& {\em F606W	(F600L)}				& 				& \\
pixelscale		& 0\farcs08						& 0\farcs06 		& 0\farcs08 \\
			&								&(resampled to 0\farcs08) & \\
PSF			& 0\farcs12						& 0\farcs12		& 0\farcs12 \\
Exposure times	& 								& 				& \\
(seconds)		&								&				& \\
\ \ {\em F125W}	& 700-5100						& 1000 			& 5017.61\\
\ \ {\em F160W}	& 500-3900						& 1050 			& 5017.61\\		
\hline
\end{tabular}
\end{center}
\end{table}%

\begin{figure}[htbp]
\begin{center}
	\includegraphics[width=0.5\textwidth]{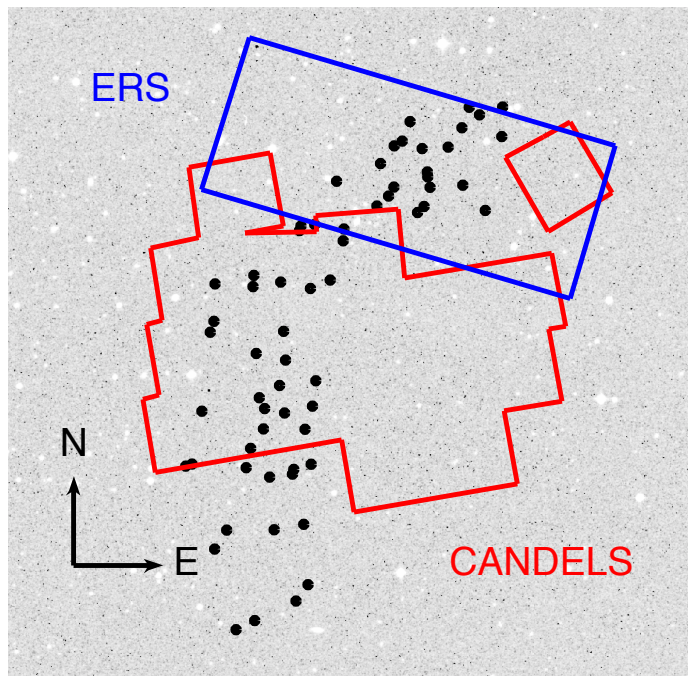}
\caption{The GOODS South field with the The CANDELS v0.5 data-release of the deep part of the survey (red) and the Early Release Science mosaics (blue) overlaid on a digital sky survey image. The black circles mark the PEARS-identified stars. }
\label{f:candels}
\end{center}
\end{figure}

\subsection{CANDELS GOODS-S observations}
\label{s:candels}

The Cosmic Assembly Near-IR Deep Extragalactic Legacy Survey \citep[CANDELS,][]{Grogin11, Koekemoer11} is designed to explore galaxy evolution, from redshift $z=1.5$ to $\sim8$. The survey is designed to cover approximately 800 arcmin$^2$ and is divided into two parts; CANDELS/Deep survey on both GOODS-N and GOODS-S and CANDELS/Wide covering GOODS as well as the EGS, COSMOS, and UDS fields.  Data from the survey are non-proprietary and are published in a series of data-releases as soon as reduced to scientific quality (see \url{http://candels.ucolick.org}). In this paper we use the v0.5 data-release of the GOODS-South field, the first epoch of the CANDELS/Deep survey. 
Exposure times and sensitivity are very similar for this initial mosaic to the BoRG pure-parallel observations (Table \ref{t:surveys}) in the two near-infrared channels ({\em F125W} and {\em F160W}).
We resampled the CANDELS {\em F125W} and {\em F160W} mosaics to a pixel scale of 0\farcs08 using {\sc drizzle} \citep{drizzle} to conform to the BoRG standard. For subsequent analysis, we only use a sub-section of this field (Figure \ref{f:candels}), where PEARS and CANDELS coverage overlap.

% We considered the Early Release Science mosaic \citep{Windhorst11} as a calibration data-set as it also overlaps with the PEARS-South observations, and supplies {\em F098M} filter information as well. However, the mosaic is drizzled to a different pixel scale, the WCS information is still off, making cross-correlation difficult and still not available to the general public.

\subsection{Early Release Science WFC3 mosaic in Goods-South}
\label{s:ers}
% As a reference for the PEARS-S identifications, we use the Early Science Release \citep{Windhorst10,Windhorst11} WFC3 data 

The WFC3 imaging of the GOODS-S field is from the Early Science Release \citep[ERS][]{Windhorst11} (see the footprint information at \url{http://candels.ucolick.org/survey/tile\_maps/GOODS-S.html} and Figure \ref{f:candels}). The 10 WFC3 fields in the ERS program were also imaged with the {\em F125W, F160W} and {\em F098M} filters, the reason it was included in the \cite{Ryan11} sample. We retrieve the data from the Hubble Legacy Archive (\url{www.hla.stsci.edu}), and combined these into a mosaic using {\sc swarp} (\url{http://www.astromatic.net/software/swarp}) to the same pixel scale as the BoRG fields. Figure \ref{f:candels} shows the overlap of this mosaic with the PEARS-S survey \citep{Pirzkal09}.

\begin{figure}[htbp]
\begin{center}
	\includegraphics[width=0.5\textwidth]{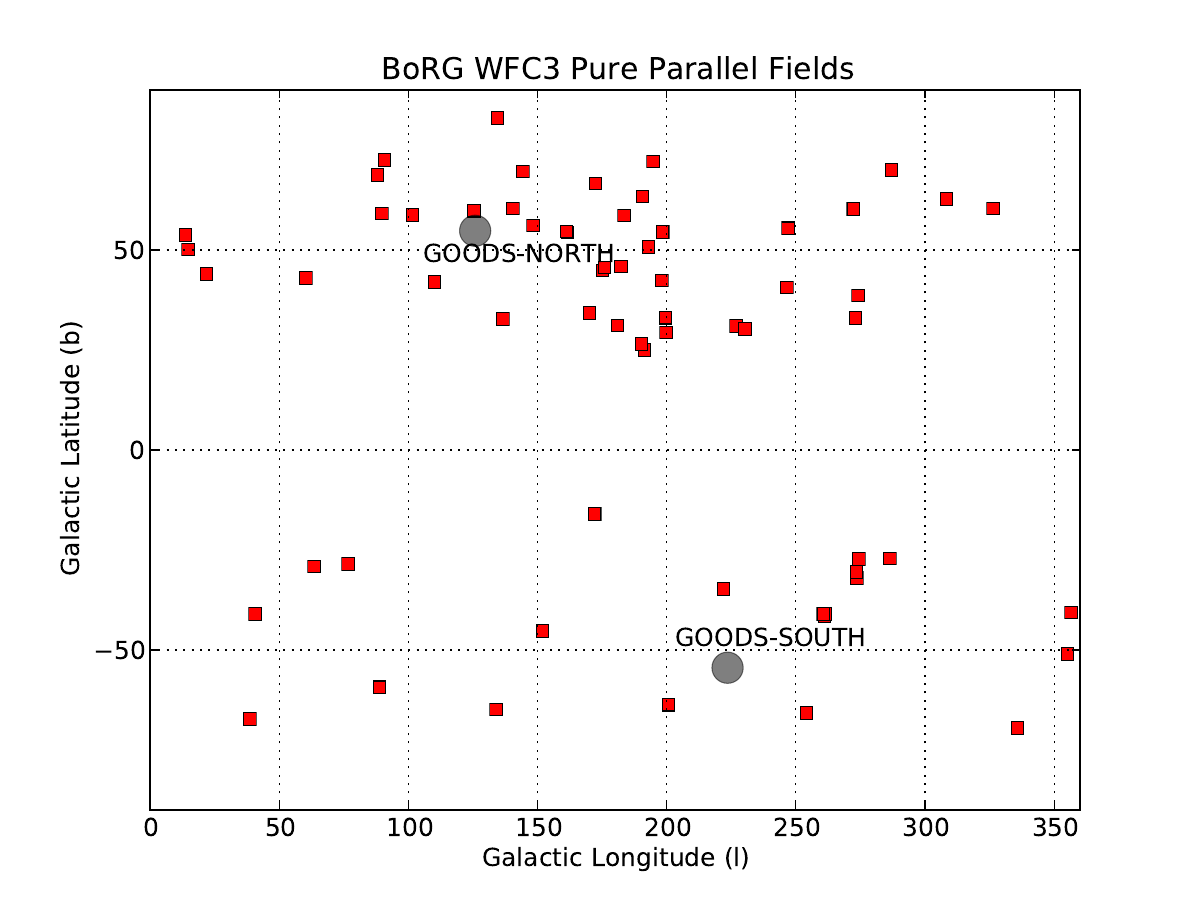}
\caption{The position of the \Nfields~ unique WFC3 Pure Parallel Fields in Galactic coordinates (red squares). The sampling of the Milky Way disk structure of substellar objects is much better than the two lines-of-sight by the GOODS/PEARS fields (black circles). We discard one field (borg\_1815-3244) for its low latitudes and line-of-sight through the plane of the disk and close of the center of the bulge.}
	\label{f:fields}
\end{center}
\end{figure}

\begin{figure}[htbp]
\begin{center}
	\includegraphics[width=0.5\textwidth]{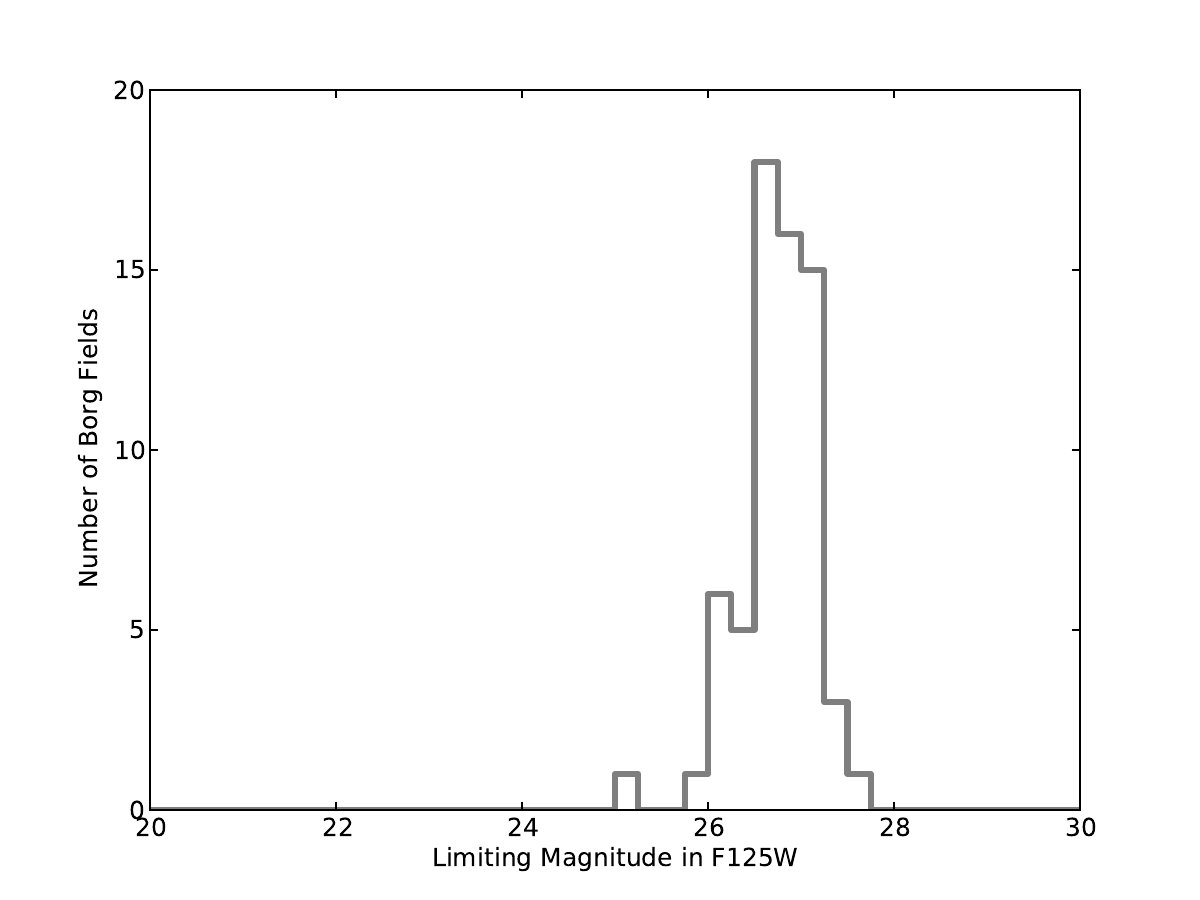}
\caption{The distribution of limiting magnitude in the {\em F125W} filter from \cite{Bradley12}. Photometric detections are reliable down to 26 magnitude.}
	\label{f:limmag}
\end{center}
\end{figure}

\subsection{BoRG}
\label{s:borg}

Our principal data-set is the WFC3 data from the BoRG \citep[Brightest of Reionizing Galaxies, HST GO/PAR-11700,][]{Trenti11, Bradley12} survey  to identify Milky Way dwarf stars from their morphology and color. The BoRG observations are undithered HST/WFC3 conducted in pure-parallel with the telescope pointing to a primary spectroscopic target with the Cosmic Origin Spectrograph (typically a high-z QSO at high Galactic latitude). The limitations for such observations are primarily that no dithering strategy can be used (final images are at WFC3 native pixel scale) and total exposure times are dictated by the primary program.

The program's initial aims were to obtain 176 arcmin$^2$ of sky in 38 pointings, using four WFC3 filters ({\em F606W, F098M, F125W}, and {\em F160W}). An extension of the BoRG program in Cycle-19 brought more parallel observations and the full reproduced set of \Nfields~ images is presented in \cite{Bradley12}.
Because lines of sight are independent and well separated on the sky mostly at high Galactic latitudes ($|b| > 20^\circ$, see Figure \ref{f:fields}), the BoRG survey samples the Milky Way disk away from the plane better than single sight-lines (e.g., the GOODS or H(U)DF fields). The BoRG data-set used here is the second data-release of \Nfields~ WFC3 fields (Table \ref{t:fields}) for a total of approximately \borgarea\ arcmin$^2$ as described in detail by \cite{Trenti11} and \cite{Bradley12}, including some reprocessed fields from another pure-parallel program with similar science goals \citep[Hubble Infrared Pure Parallel Imaging Extragalactic Survey, HIPPIES, HST GO/PAR-11702,][]{Yan10b}.
We use the WFC3 data-products generated by the BoRG team \citep[see for details][]{Bradley12}. This is a standard multi-drizzle reduction of these undithered WFC3 data with Laplacian edge detection \citep{van-Dokkum01} to the individual FLT files to mitigate detector hot pixels and cosmic rays. Stellar objects are not affected by this filtering.

The BoRG survey is designed to identify relatively bright ($m_{F125W} \leq 27$) high-redshift galaxies from their broad-band colors using the Lyman-Break technique \citep{Steidel96}. The primary aim of the survey is to select redshift $z\sim7.5$ galaxies as {\em F098M} dropouts. Two near-infrared filters ({\em F125W} and {\em F160W}) are used for source detection and characterization down to 26 mag (Figure \ref{f:limmag}). One optical filter ({\em F606W} or {\em F600LP} in the case of the HIPPIES survey) is used to control contamination from lower redshift  sources; $z\sim1.5$ compact galaxies, AGNs, and cool Milky Way stars. Our interest here now goes to the latter of these interlopers.
Given the near-random pointing nature of the pure-parallel HST program (higher Galactic Latitude objects are preferred for COS targets), the BoRG fields are minimally affected by field-to-field (cosmic) variance \citep{Trenti08}. Therefore, this catalog is uniquely positioned to set constraints on the number density of unresolved sources, either Milky Way stars or $z\sim8$ galaxies.

% BoRG 2013 data
A recent BoRG observing campaign (BoRG13) is presented by Schmidt et al. ({\em in preparation}), which includes approximately 50 arcmin$^2$ of new data (13 fields) and deeper observations of two previous BoRG pointings
This brings the total J-band area of the BoRG survey to $\sim350$ arcmin$^2$ makes BoRG the largest existing near-infrared survey with HST, e.g., compare with CANDELS wide (260 arcmin$^2$) or deep (120 arcmin$^2$). We include nine BoRG13 fields as additional lines-of-sight in our sample (see Table \ref{t:fields}).

\section{Catalog Generation}
\label{s:cat}

To construct the catalogs for the BoRG Fields, we ran {\sc sextractor} \citep{se,seman} with similar settings as \cite{Trenti11} but set to include {\sc mu\_max }, $r\_{50}$ and other morphological information. 
% NB DO I NEED TO LIST THE SEXTRACOR SETTINGS?
% Tables \ref{t:se:in} and \ref{t:se:out} lists the typical input parameters and the output.
% detection
Source detection was done in the {\em F125W} image, with the other filters run in dual-image mode. The multi-drizzle weight files were used as RMS maps  \citep[once normalized, following the prescription in][{\sc WEIGHT\_TYPE=MAP\_RMS}]{Casertano00}, but scales appropriately to reflect the noise correlation introduced.  
% A requirement of S/N $> 8$ in {\em F125W} and S/N $> 3$ in {\em F160W} was used for the final catalog. 
The photometric zero-points are from \cite{Dressel10, Windhorst11}: {\em F606W}: 26.08, {\em F600LP}: 25.85, {\em F098W}: 25.68, {\em F125W}: 26.25, and {\em F160W}: 25.96 respectively with $A\_V$ corrections derived for each individual field (Table \ref{t:fields}). These settings resulted in \nborg\ objects in the BoRG survey.

\section{CANDELS and ERS Calibration Catalogs}
\label{s:cat:candels}

We generated {\sc sextractor} v2.8 catalogs with the same settings as above for the section of the resampled CANDELS mosaic, which includes the PEARS-S coverage (Figure \ref{f:candels}) to obtain {\em F125W} and {\em F160W} luminosities and morphological information for the spectroscopically identified M-dwarfs. Optical magnitudes for these are from ACS photometry presented in \cite{Pirzkal09}. The overlap between the PEARS-S survey and CANDELS mosaic is a sample of 24 M-dwarfs (Table \ref{t:candels:Mdwarfs}). We will use their morphology to select bona-fide stars in the {\em F125W} images and their colors to select and sub-type the M-dwarfs.

\begin{figure*}
\begin{center}
	\includegraphics[width=0.49\textwidth]{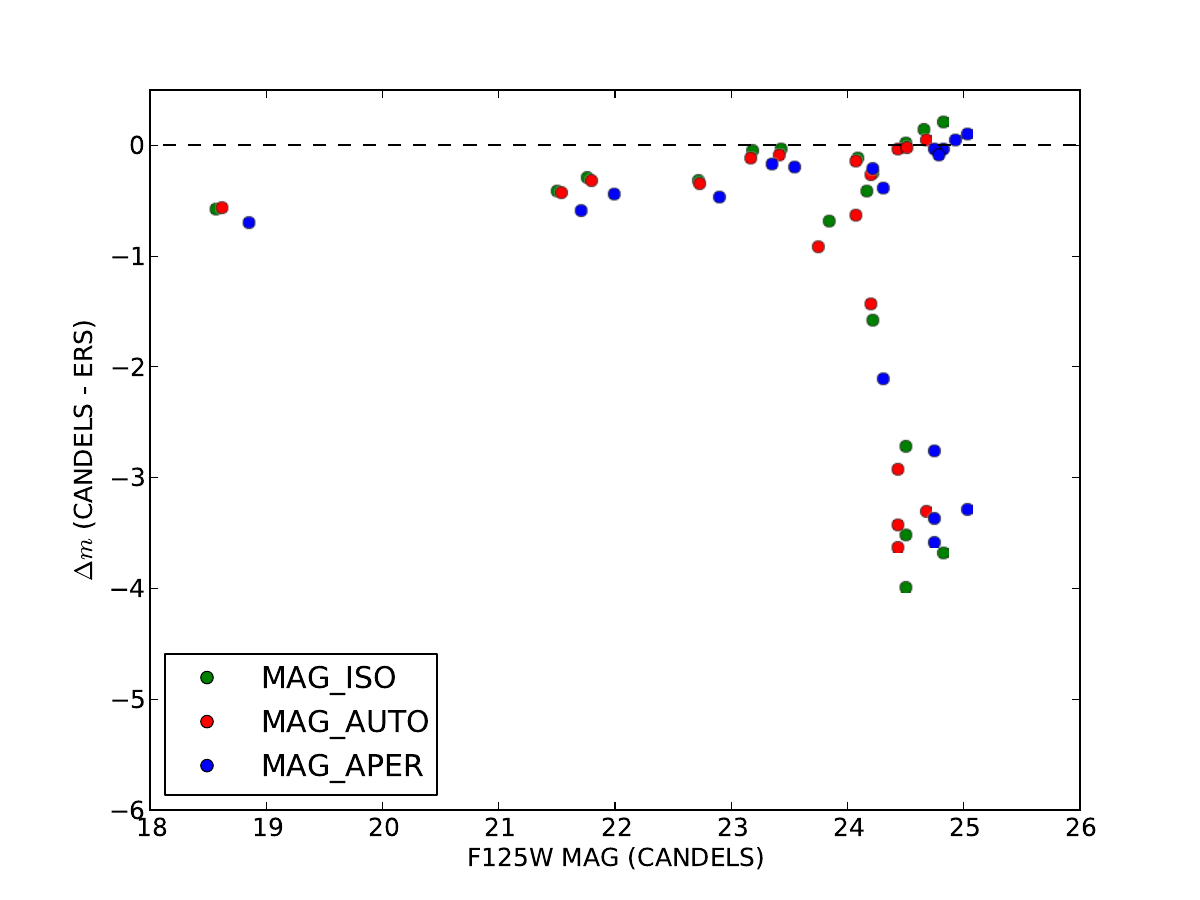}
	\includegraphics[width=0.49\textwidth]{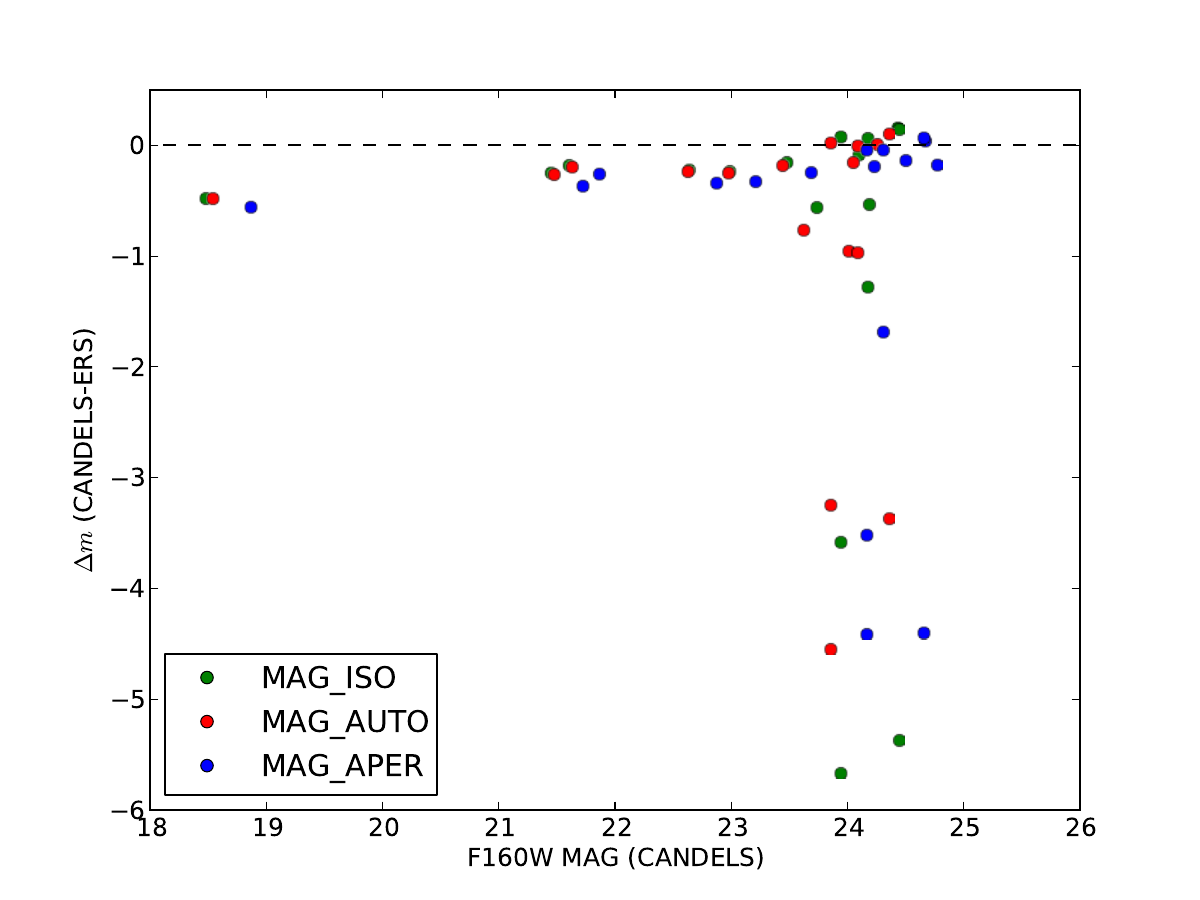}
\caption{The difference in magnitude between the CANDELS and ERS mosaics. The CANDELS {\sc sextractor} catalog was generated from the 0\farcs06 pixel scale, full mosaic. Cross-correlation with the ERS mosaic yielded 13 stellar objects. There is an offset in both filters between the mosaics.}
\label{f:zpcheck}
\end{center}
\end{figure*}

% ERS
A second calibration catalog of 22 bona-fide PEARS-identified M-dwarfs is constructed based on the ERS mosaics (Table \ref{t:ers:Mdwarfs}). The ERS mosaic pixel scale was set to be identical to the BoRG fields and all the {\sc sextractor} parameters are kept identical to the BoRG ones. Similar to the CANDELS and BoRG catalog, detection was done in {\em F125W} with the catalogs in {\em F160W}, {\em F098M} and {\em F606W} or {\em F600LP} in dual mode for the respective photometry. 
Cross-correlation with the PEARS identified sources used the published right ascension and declination from \cite{Pirzkal09} converted to pixels but due to some lingering issues with the wcs coordinates in the ERS data, we are less certain about the cross-identification than in the CANDELS data ($\sim$0\farcs1 uncertainty in position). Because of these issues and the uncertainty in the flux calibration between CANDELS and the ERS mosaics (Figure \ref{f:zpcheck}), we adopt the CANDELS mosaic as our primary calibrator for morphological selection and sub-typing and the ERS to calibrate the color-color selection and as a check of the CANDELS data.

\begin{figure}
\begin{center}
	\includegraphics[width=0.49\textwidth]{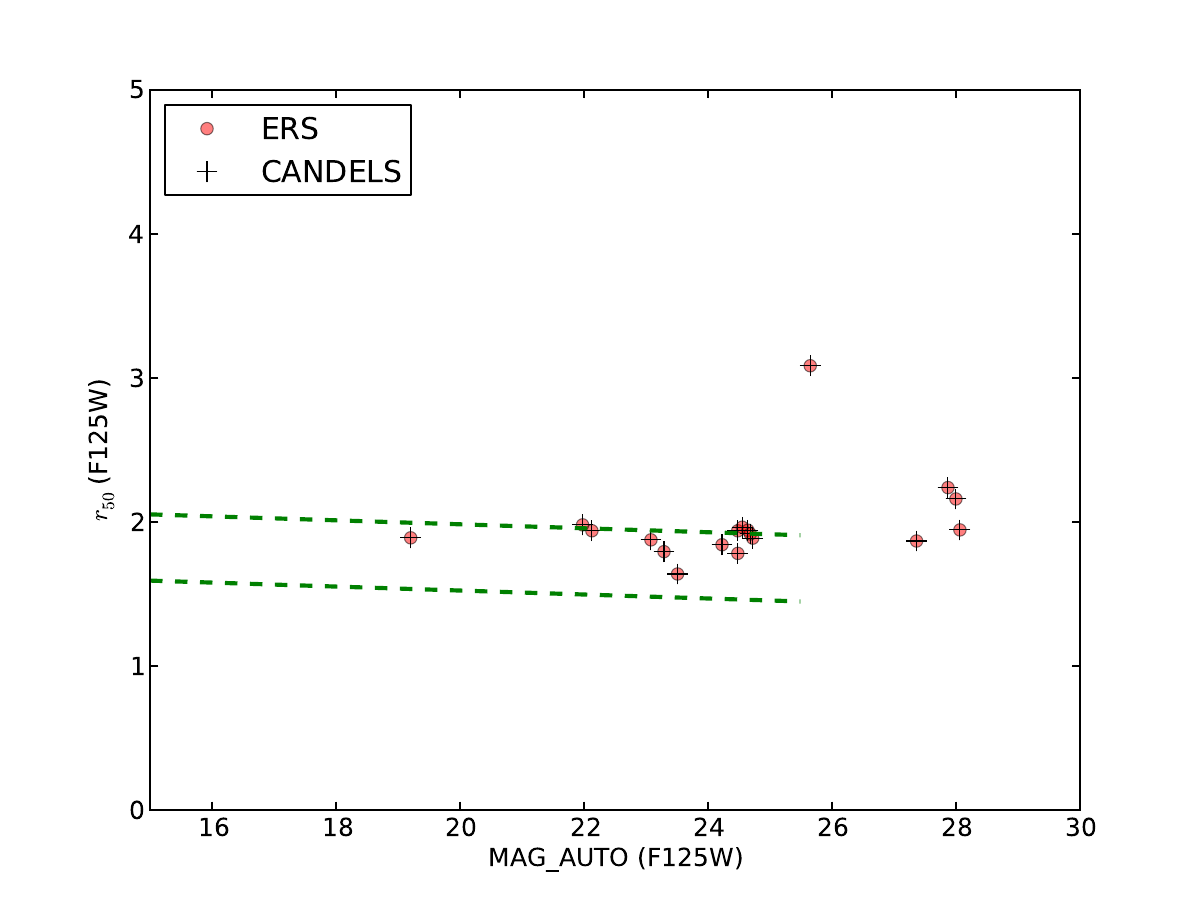}
\caption{The relation between MAG\_AUTO and half-light radius ($r\_{50}$) for the common objects between the CANDELS v0.5 field and the ERS (see Figure \ref{f:candels}). The green lines are the stellarity criterion for the half-light radius. Beyond MAG\_AUTO$\sim$24, the stellarity of these objects is doubtful.}
\label{f:zpcheck:r50}
\end{center}
\end{figure}

\subsection{CANDELS and ERS photometry}
\label{s:candelsphot}

We used the same zeropoints for all the filters as those used for the BoRG survey \citep{Trenti11, Bradley12} and ERS \citep{Windhorst11}. However, the ERS data was taken when the WFC3 instrument was still cooling down and hence one could expect slightly different WFC3 performance between the CANDELS and ERS photometry catalog. To check, we identified stars (using the $r\_{50}$ criterion below) in the full ERS mosaic and cross-correlated these with the full CANDELS catalog (derived from the 0\farcs06 mosaic). The overlap is 13 point-sources. Figure \ref{f:zpcheck} shows the difference {\sc sextractor} magnitudes for these stars between the CANDELS and ERS mosaics in the {\em F125W} and {\em F160W} filters.

Discarding the faintest objects as their stellarity becomes uncertain (Figure \ref{f:zpcheck:r50}), the mean difference between the overlap stars for different apertures ($m_{F125W} < 24$, $m_{F160W} < 23.5$) are listed in Table \ref{t:zpcheck}. We take the differences in the MAG\_APER as the most reliable as these were computed for identical-size apertures. Because we used the ASSOC option in {\sc sextractor}, the placement of these apertures could conceivably be slightly different (typically less than a pixel difference) but this is not enough to explain the offset in photometry.

Thus, for the following, we correct the ERS photometry by subtracting 0.43 and 0.39 for the {\em F125W} and {\em F160W} respectively. 
However, we lack a similar information for the {\em F098M} filter and adopt a correction of 0.4 magnitude, the mean of the correction for the {\em F125W} and {\em F160W} correction.

\begin{table}[htdp]
\caption{The difference and rms between the CANDELS and ERS magnitudes for the F125W and F160W filters. The aperture for MAG\_APER is 0\farcs48 (8 and 6 pixels respectively).}
\begin{center}
\begin{tabular}{l l l l}
Filter		& MAG\_ISO			& MAG\_AUTO		& MAG\_APER \\
\hline
\hline
F125W	& -0.34 (0.23)			&  -0.40 ( 0.26)		& -0.43 (0.19) \\
F160W	& -0.26 (0.11)			& -0.28  (0.10)		& -0.39 (0.10)\\
\hline
\end{tabular}
\end{center}
\label{t:zpcheck}
\end{table}%

\begin{figure}[htbp]
\begin{center}
	\includegraphics[width=0.5\textwidth]{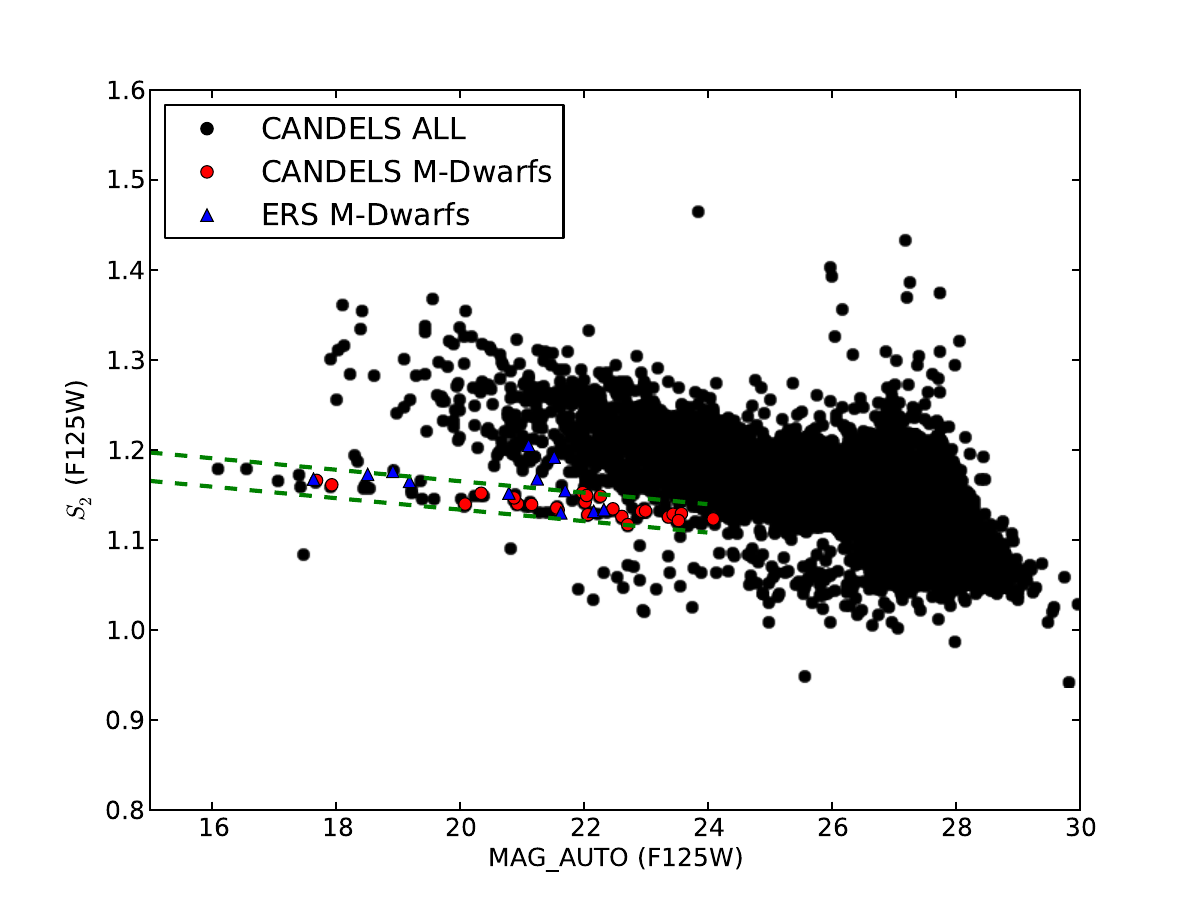}
\caption{The $S\_2$ parameter for the CANDELS subfield of Figure \ref{f:candels}, (crosses) and the bona-fide M-dwarfs identified in this field by PEARS (red points). The green dashed lines indicate our $S\_2$ selection criterion.}
\label{f:s2}
\end{center}
\end{figure}

\begin{figure}[htbp]
\begin{center}
	\includegraphics[width=0.5\textwidth]{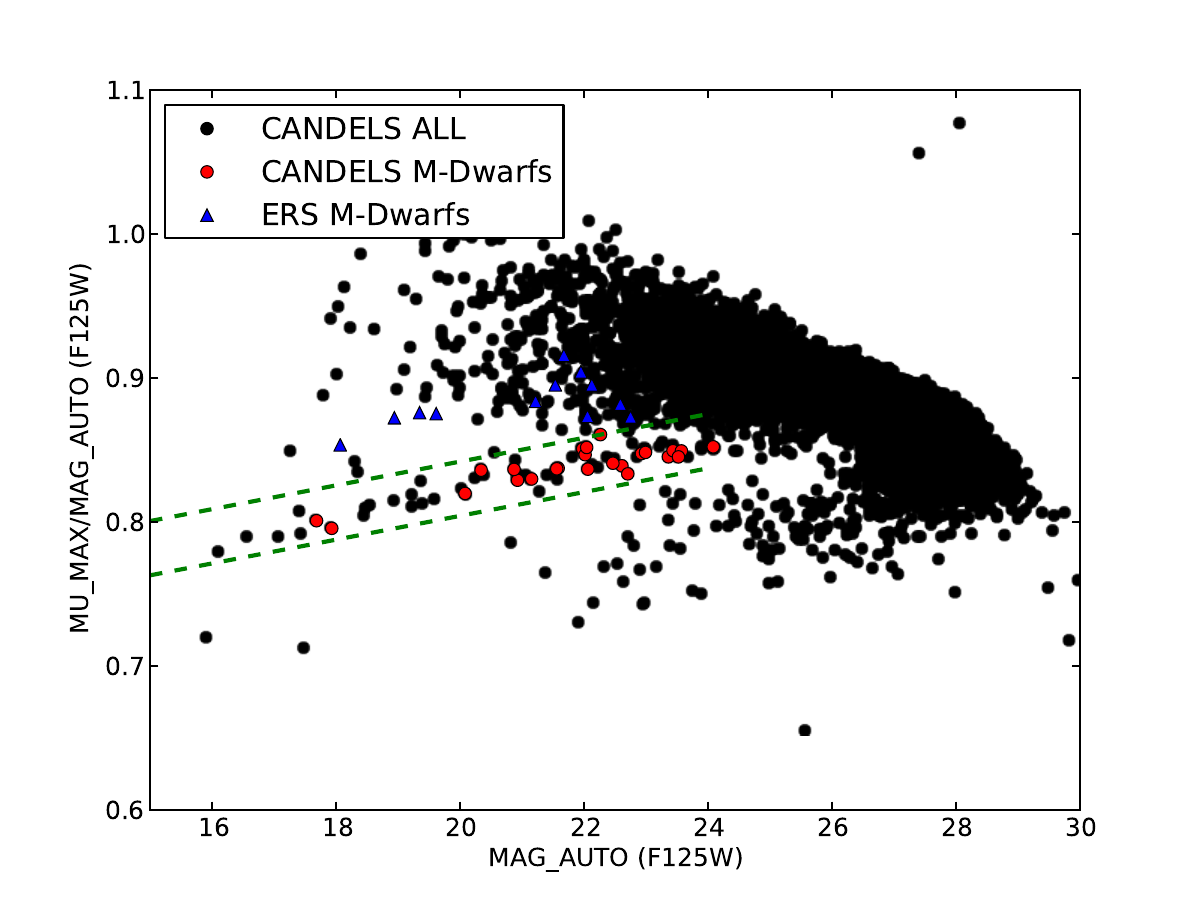}
\caption{The {\sc mu\_max/mag\_auto} parameter for the CANDELS subfield of Figure \ref{f:candels}, (crosses) and the bona-fide M-dwarfs identified in this field by PEARS (red circles). The green dashed lines indicate our {\sc mu\_max/mag\_auto} selection criterion.}
\label{f:murat}
\end{center}
\end{figure}

\begin{figure}[htbp]
\begin{center}
	\includegraphics[width=0.5\textwidth]{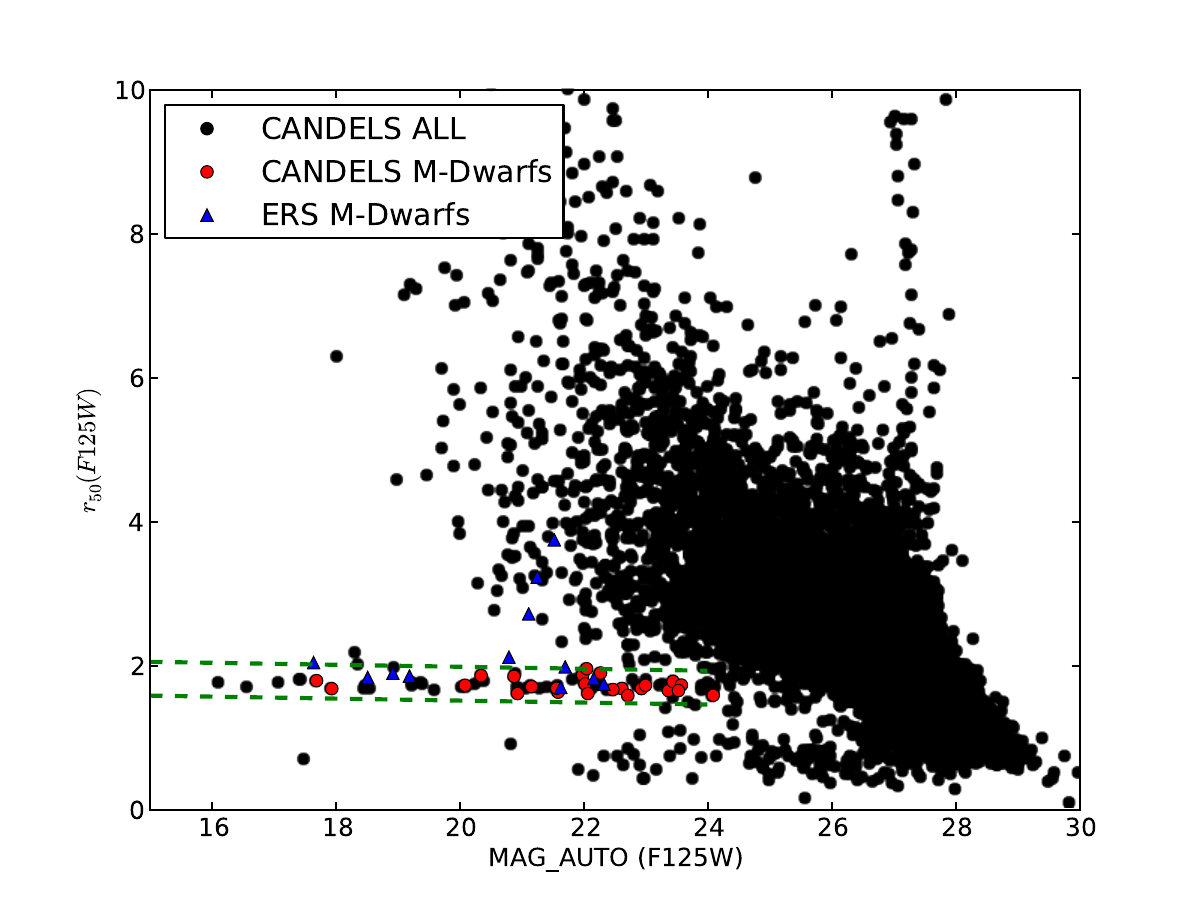}
\caption{The $r\_{50}$ parameter (half-light radius) for the CANDELS subfield of Figure \ref{f:candels}, (crosses) and the bona-fide M-dwarfs identified in this field by PEARS (circles). The green dashed lines indicate our $r\_{50}$ selection criterion.}
\label{f:r50}
\end{center}
\end{figure}

\begin{figure}[htbp]
\begin{center}
	\includegraphics[width=0.5\textwidth]{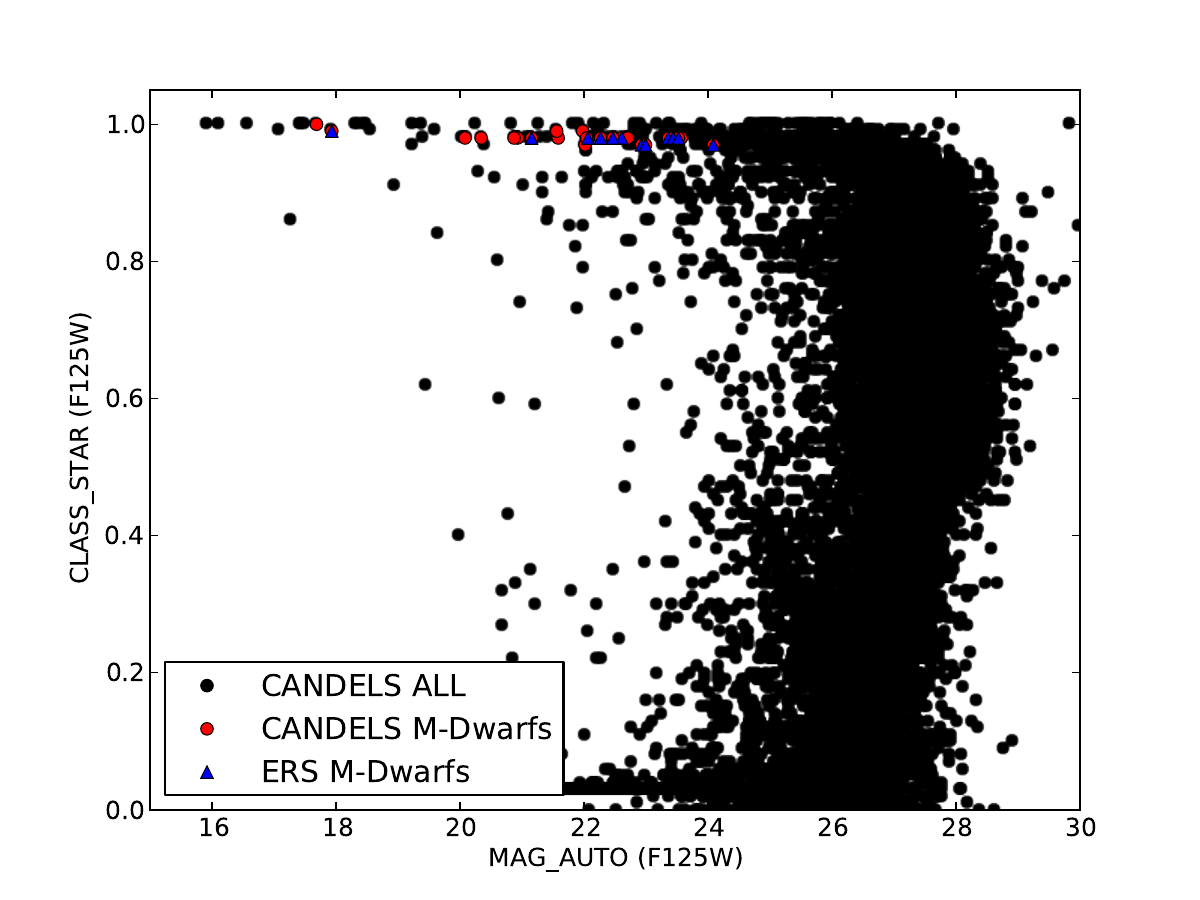}
\caption{The {\sc class\_star} parameter from {\sc sextractor} for the CANDELS subfield of Figure \ref{f:candels}, (crosses) and the bona-fide M-dwarfs identified in this field by PEARS (circles). }
\label{f:classstar}
\end{center}
\end{figure}

\subsection{Morphological Identification of Stars}
\label{s:morph}

To identify the low-mass dwarfs in our catalogs, we opt for a strict morphological selection to allow the color information to be used for identification of the stellar type and subtype. 
There are several {\sc sextractor} parameters cuts in use to identify unresolved sources, all of them a variation on a concentration index; the {\sc class\_star}, native to {\sc sextractor}, the half-light or effective radius ($r\_{50}$) used by \cite{Ryan11}, the flux ratio between to pre-defined apertures, e.g., the stellarity index $\rm S\_2 =  MAG\_{aper(1 pix)}/ MAG\_{ISO} (F125W)$) in \cite{Pirzkal09}, and the relation between the brightest pixel surface brightness and total source luminosity \citep[{\sc mu\_max/mag\_auto},][who used it to discard stars.]{seman,Holwerda05b, Leauthaud07}.

Using the M-dwarf catalog for CANDELS, we define three morphological criteria: one based on the stellarity index ($S\_2$), one on the {\sc mu\_max/mag\_auto} ratio, and one for the half-light radius ($r\_{50}$) for {\em F125W} catalog. If we fit the relation between $S\_2$, {\sc mu\_max/mag\_auto} and $r\_{50}$ with {\sc mag\_auto} for these M-dwarf we get:
\begin{equation}
\label{eq:s2}
| S\_2+0.0065 \times MAG\_AUTO - 1.18 | ~ < ~ 0.016 ~ (0.018),
\end{equation}
the relation between {\sc mu\_max} and the luminosity ({\sc mag\_auto}),
\begin{equation}
\label{eq:mu}
|{ MU\_MAX \over MAG\_AUTO} - 0.0085 \times MAG\_AUTO -0.65 | < 0.019 
\end{equation}
\noindent and the half-light radius ($r\_{50}$),
\begin{equation}
\label{eq:r50}
| r\_{50} -0.014 \times MAG\_AUTO-2.03 | < 0.23.
\end{equation} 
\noindent with the variance in the relation determined from the scatter of the M-dwarfs around the fit. 
Thus, by definition these criteria will typically select all 24 M-dwarfs in the CANDELS field. 
Figure \ref{f:s2}, \ref{f:murat}, and \ref{f:r50} show these three criteria for objects in the CANDELS sub-field 
catalog with the PEARS M-dwarfs highlighted. For comparison, the ERS M-dwarfs are also marked.
Table \ref{t:crit} shows how many stars they select in the CANDELS field. % We opt for the $r\_{50}$ criterion, similar to \cite{Ryan11}

Our goal is to define criteria that would successfully select the PEARS stars. We note that our calibrators are the 
confirmed M-dwarfs and that some of those objects selected as stars in the CANDELS field are still Galactic stars, just not M-dwarfs. 
All three criteria work remarkably better than {\sc class\_star}, selecting stars with little contamination for objects 
brighter than $m_{F125W} < 24$ (Figures \ref{f:s2}--\ref{f:classstar}). However, there is a marked difference in 
efficiency between the $S\_2$ and {\sc mu\_max/mag\_auto} criteria and the $r\_{50}$ criterion in the 24-25.5 
mag range. 
The half-light radius criterion is much stricter than the other two (Table \ref{t:crit}), i.e., it includes most of the PEARS identified M-dwarfs but includes much fewer objects from CANDELS than the other two criteria. 

The half-light parameter was used by \cite{Ryan11} in some of these fields to select faint dwarf stars ($1.2 \leq r\_{50}  \leq 1.8$ pixels with their pixelscale set to 0\farcs09). Our calibration here shows that this selection criterion is even a little lenient (the selection range can even be narrower), allowing for the different pixel scales, but works very well for the M-dwarfs in these fields.

\subsubsection{ERS catalog check}

The 22 M-dwarfs in the ERS catalog are also plotted in Figures \ref{f:s2}--\ref{f:classstar} to serve as a check for the morphological selection. Their positions in these figures reveal a sensitivity in both the $S\_2$ and the {\sc mu\_max-mag\_auto} relations for unresolved objects to the pixel scale.
 The $r\_{50}$ criterion is less sensitive to the scale of the pixels. Most likely, this is because it is computed from the growth curve of an objects, i.e., the ordered list of pixels. The other two criteria both depend on an aperture of a single pixel and hence are more sensitive to the pixelscale. 
 Several of the identified objects in the ERS catalog are well away ($r\_{50}>2$ pixels) from those values expected for stars, i.e., unresolved objects. These are likely either mis-identifications in the ERS field or blends with other objects or artifacts.

\begin{table}
\caption{\label{t:crit}The number of selected objects in the CANDELS/GOODS-S mosaic section (Figure \ref{f:candels}) for the three different criteria separately and in combinations. The mosaic contains 24 bona-fide M-dwarfs and we note their selected number as well. Stricter limits resulted in some of those not being selected.}
\begin{center}
\begin{tabular}{l l l l}

criterion        		& CANDELS		&  PEARS M-dwarfs \\
				& $<25.5$	 		& \\
\hline
\hline
$S\_2$			&  216			& 24 \\
{\sc mu\_max/mag\_auto} & 331         	& 23 \\
$r\_{50}$              	& 120  			& 23 \\
%All              			& 71  	& 131	& 24 \\
\hline
\end{tabular}
\end{center}
\end{table}%

The comparison shows that both the $r\_{50}$ and the {\sc mu\_max/mag\_auto} work well for stellar selection brighter than 24 mag and the $r\_{50}$ includes much less interlopers down to 25.5 mag. 
The practical limit of 24 mag for PEARS identified M-dwarfs in {\em F124W} would limit identifications to distances from 3.2 to 36.3 kpc depending on M-dwarf subtype (M9-M0 respectively), or some 10-100 thin disk scale-heights.  We treat our sample as uncontaminated, photometrically accurate, and calibrated with the CANDELS M-dwarfs up to this conservative limit, which is still two magnitudes above the BoRG photometric detection limit (Figure \ref{f:limmag}).

\begin{table*}
\caption{\label{t:candels:Mdwarfs} The M-dwarfs identified by PEARS in the CANDELS field. The optical magnitudes are from \cite{Pirzkal09} and the {\em F125W} and {\em F160W} magnitudes from our {\sc sextractor} catalog of the CANDELS field. }
\begin{center}
\begin{tabular}{l l l l l l l l l l }
ID & Type & ra & dec & $m_{F435W}$ & $m_{F606W}$ & $m_{F775W}$ & $m_{F850W}$ & $m_{F125W}$ & $m_{F160W}$ \\ 
\hline
\hline
% \input{../CANDELS\_Mdwarfs.tex}
% ID & Type & ra & dec & $m_{F435W}$ & $m_{F606W}$ & $m_{F775W}$ & $m_{F850W}$ & $m_{F125W}$ & $m_{F160W}$ \\ 
48173 & 4.0 	 & 53.220073 & -27.854046 & 25.22 & 24.96 & 23.73 & 23.15 & 22.61 & 22.63 \\ 
53237 & 4.0 	 & 53.194890 & -27.848121 & 25.19 & 25.45 & 24.14 & 23.49 & 22.93 & 22.88 \\ 
58796 & 1.5 	 & 53.173782 & -27.841699 & 25.70 & 25.04 & 24.17 & 23.85 & 23.36 & 23.29 \\ 
58826 & 4.3 	 & 53.189870 & -27.841487 & 24.76 & 24.34 & 23.04 & 22.39 & 21.98 & 21.82 \\ 
63079 & 1.0 	 & 53.181657 & -27.836074 & 25.49 & 24.44 & 23.76 & 23.47 & 22.99 & 23.00 \\ 
63028 & 3.7 	 & 53.213644 & -27.835281 & 24.29 & 23.97 & 22.78 & 22.15 & 21.58 & 21.59 \\ 
63752 & 3.9 	 & 53.189345 & -27.834509 & 23.80 & 23.43 & 22.37 & 21.74 & 21.16 & 21.12 \\ 
63993 & 2.0 	 & 53.170843 & -27.833788 & 23.81 & 23.13 & 22.27 & 21.91 & 21.56 & 21.46 \\ 
66572 & 8.4 	 & 53.191395 & -27.830774 & 24.60 & 26.71 & 24.80 & 23.58 & 22.47 & 22.53 \\ 
69522 & 6.0 	 & 53.183514 & -27.826627 & 24.89 & 24.60 & 23.41 & 22.64 & 22.02 & 21.91 \\ 
70032 & 3.9 	 & 53.169502 & -27.825087 & 22.88 & 22.52 & 21.40 & 20.77 & 20.34 & 20.25 \\ 
74670 & 1.0 	 & 53.181116 & -27.817942 & 25.79 & 24.96 & 24.20 & 23.97 & 23.44 & 23.31 \\ 
74928 & 4.0 	 & 53.192508 & -27.815594 & 20.55 & 20.06 & 19.10 & 18.51 & 17.93 & 17.92 \\ 
79699 & 4.0 	 & 53.210190 & -27.808231 & 22.75 & 22.38 & 21.29 & 20.66 & 20.08 & 20.02 \\ 
80618 & 2.0 	 & 53.181804 & -27.808089 & 26.31 & 25.72 & 24.88 & 24.48 & 24.08 & 24.06 \\ 
82885 & 0.2 	 & 53.208736 & -27.804499 & 25.77 & 24.76 & 24.13 & 23.83 & 23.57 & 23.45 \\ 
91263 & 0.9 	 & 53.171308 & -27.793465 & 24.35 & 23.43 & 22.76 & 22.49 & 22.27 & 22.09 \\ 
92395 & 2.1 	 & 53.193512 & -27.792703 & 24.44 & 23.83 & 22.93 & 22.54 & 22.06 & 22.08 \\ 
93532 & 2.0 	 & 53.208197 & -27.791671 & 26.19 & 25.55 & 24.52 & 24.07 & 23.52 & 23.50 \\ 
93841 & 4.9 	 & 53.182786 & -27.791200 & 25.10 & 25.14 & 23.97 & 23.34 & 22.71 & 22.74 \\ 
94206 & 4.3 	 & 53.163637 & -27.790605 & 24.71 & 24.21 & 23.00 & 22.37 & 22.04 & 21.75 \\ 
95752 & 3.8 	 & 53.193173 & -27.788834 & 23.42 & 23.24 & 22.12 & 21.53 & 20.93 & 20.93 \\ 
104030 & 0.0 	 & 53.158515 & -27.777252 & 22.95 & 21.97 & 21.34 & 21.16 & 20.87 & 20.69 \\ 
104673 & 0.0 	 & 53.158184 & -27.773143 & 19.76 & 18.90 & 18.48 & 18.04 & 17.69 & 17.48 \\ 
\hline
\end{tabular}
\end{center}
\end{table*}%

\begin{table*}
\caption{\label{t:ers:Mdwarfs} The M-dwarfs identified by PEARS in the ERS mosaic. The optical magnitudes are from \cite{Pirzkal09} and the {\em F125W} and {\em F160W} magnitudes from our {\sc sextractor} catalog of the ERS mosaic. }
\begin{center}
\begin{tabular}{l l l l l l l l l l }
ID & Type & ra & dec & $m_{F435W}$ & $m_{F606W}$ & $m_{F775W}$ & $m_{F850W}$ & $m_{F125W}$ & $m_{F160W}$ \\ 
\hline
\hline
% \input{../ERS\_Mdwarfs.tex}
%ID & Type & ra & dec & $m_{F435W}$ & $m_{F606W}$ & $m_{F775W}$ & $m_{F850W}$ & $m_{F125W}$ & $m_{F160W}$ \\ 
108497 & 1.7 	 & 53.093027 & -27.735838 & 21.39 & 20.58 & 19.78 & 19.42 & 19.19 & 19.07 \\ 
108014 & 2.0 	 & 53.057966 & -27.735053 & 19.76 & 18.81 & 18.37 & 17.80 & 17.64 & 17.47 \\ 
110839 & 2.0 	 & 53.089621 & -27.733082 & 23.02 & 22.36 & 21.48 & 21.06 & 21.52 & 21.30 \\ 
111269 & 3.0 	 & 53.113985 & -27.732740 & 24.68 & 23.98 & 23.05 & 22.58 & 21.99 & 22.05 \\ 
111982 & 6.0 	 & 53.107958 & -27.728253 & 21.63 & 21.19 & 19.94 & 19.11 & 18.51 & 18.41 \\ 
114688 & 0.0 	 & 53.086462 & -27.724381 & 24.06 & 23.11 & 22.49 & 22.22 & 22.73 & 22.41 \\ 
115223 & 4.0 	 & 53.105121 & -27.724124 & 25.11 & 24.64 & 23.51 & 22.90 & 22.16 & 22.18 \\ 
114563 & 6.0 	 & 53.069511 & -27.723391 & 22.27 & 21.88 & 20.69 & 19.90 & 19.37 & 19.12 \\ 
116612 & 4.0 	 & 53.134575 & -27.721317 & 24.55 & 23.85 & 22.70 & 22.24 & 21.63 & 21.53 \\ 
117391 & 1.3 	 & 53.087638 & -27.719656 & 23.01 & 22.13 & 21.42 & 21.04 & 20.66 & 20.67 \\ 
119050 & 2.7 	 & 53.087816 & -27.717329 & 26.03 & 24.77 & 23.84 & 23.42 & 23.02 & 22.81 \\ 
120814 & 4.2 	 & 53.111938 & -27.713369 & 24.19 & 23.99 & 22.87 & 22.19 & 21.70 & 21.61 \\ 
123686 & 5.3 	 & 53.090399 & -27.706629 & 23.60 & 23.44 & 22.25 & 21.52 & 20.79 & 20.72 \\ 
124539 & 8.7 	 & 53.076943 & -27.706018 & 25.28 & 25.75 & 24.27 & 23.22 & 22.64 & 22.52 \\ 
124624 & 1.0 	 & 53.104778 & -27.705292 & 23.78 & 23.41 & 22.76 & 22.42 & 21.25 & 21.08 \\ 
125478 & 9.0 	 & 53.100622 & -27.703149 & 24.03 & 24.37 & 22.53 & 21.29 & 20.38 & 20.34 \\ 
126754 & 1.0 	 & 53.049225 & -27.701327 & 24.91 & 23.94 & 23.19 & 22.88 & 22.32 & 22.21 \\ 
128173 & 3.3 	 & 53.069739 & -27.697205 & 23.69 & 23.12 & 22.11 & 21.57 & 21.11 & 21.09 \\ 
128247 & 2.0 	 & 53.096541 & -27.694200 & 21.16 & 20.43 & 19.62 & 19.24 & 18.92 & 18.78 \\ 
130804 & 1.0 	 & 53.060606 & -27.691367 & 24.23 & 23.33 & 22.66 & 22.35 & 21.82 & 21.71 \\ 
132690 & 3.0 	 & 53.066145 & -27.687565 & 24.21 & 23.86 & 22.91 & 22.44 & 27.13 & 26.77 \\ 
132690 & 3.0 	 & 53.065896 & -27.687864 & 24.21 & 23.86 & 22.91 & 22.44 & 22.15 & 22.29 \\ 
\hline
\end{tabular}
\end{center}
\end{table*}%

\begin{figure}[htbp]
\begin{center}
	\includegraphics[width=0.5\textwidth]{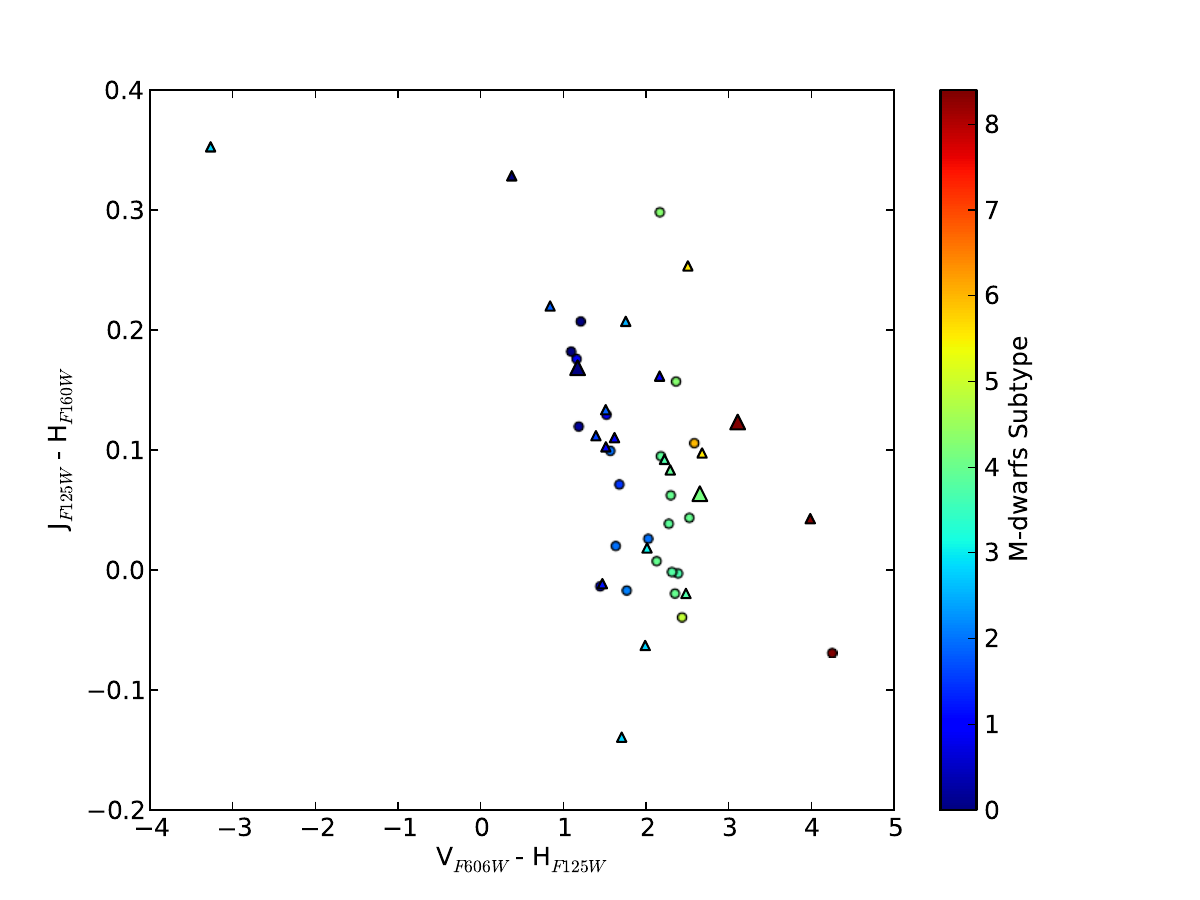}
\caption{The relation between the {\em F606W-F125W} and {\em F125W-F160W} colors for the M-dwarf sub-type (0-8) 
determined from the GOODS, CANDELS  and ERS data for the PEARS spectroscopically identified M-dwarfs. }
\label{f:vjh}
\end{center}
\end{figure}

\begin{figure}[htbp]
\begin{center}
	\includegraphics[width=0.5\textwidth]{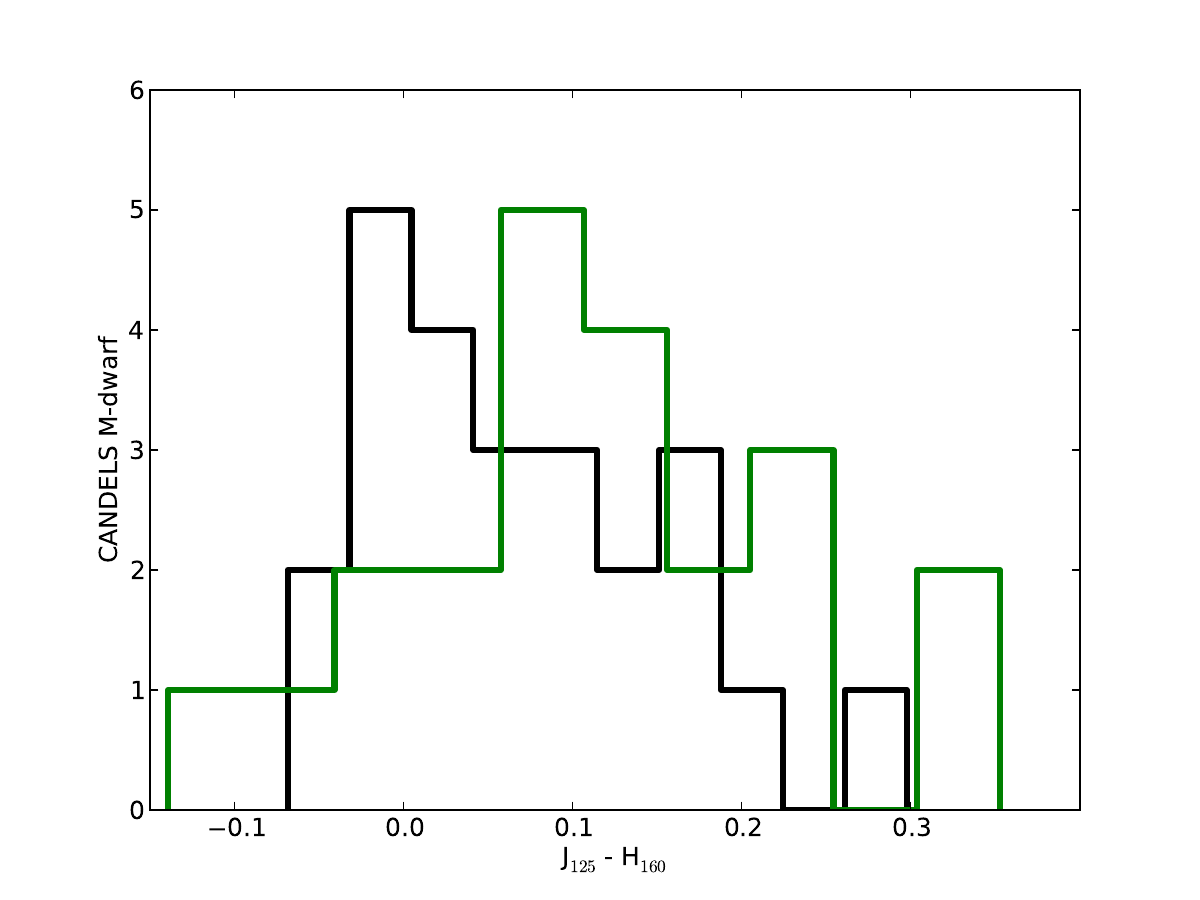}
\caption{The histogram of the {\em F125W-F160W} colors from CANDELS and ERS of the PEARS spectroscopically identified M-dwarfs.}
\label{f:jh:hist}
\end{center}
\end{figure}

\begin{figure}[htbp]
\begin{center}
	\includegraphics[width=0.5\textwidth]{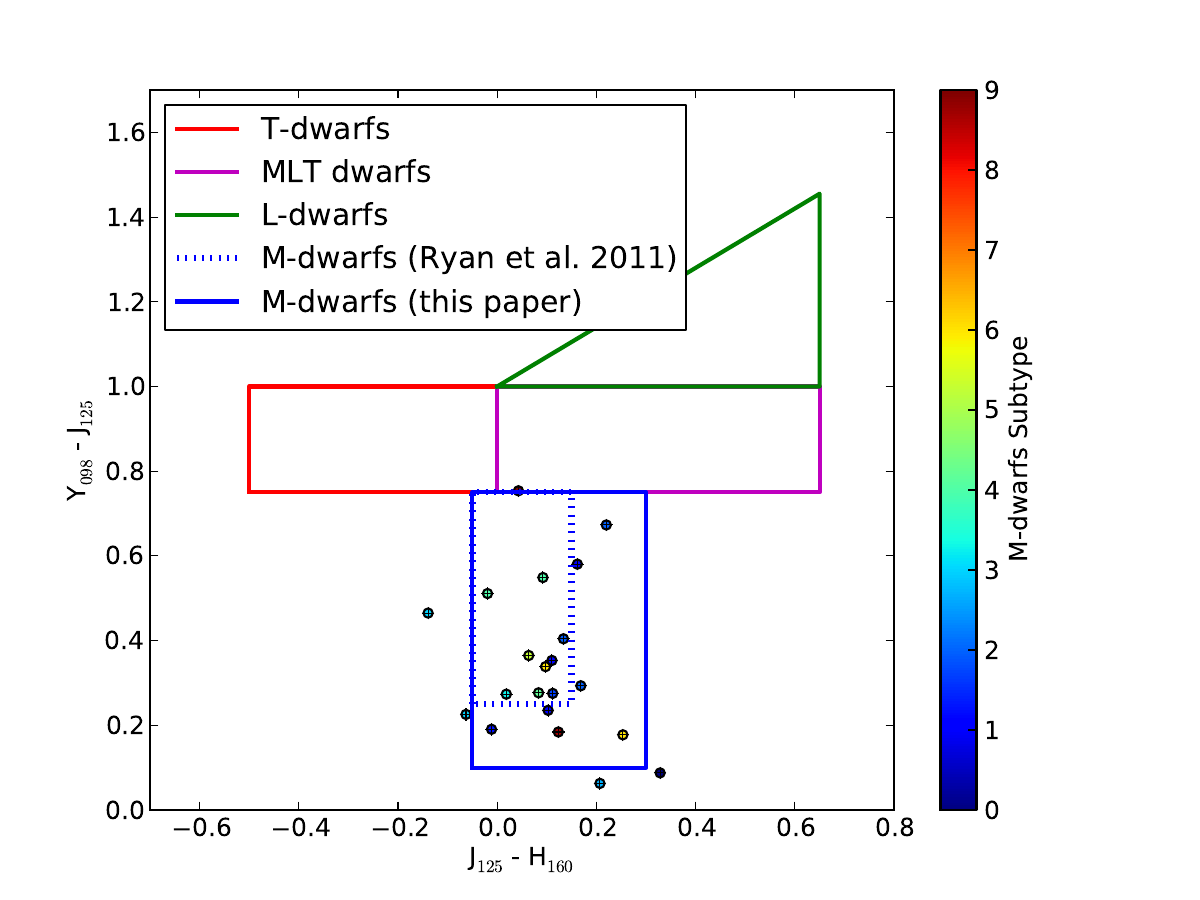}
\caption{A plot of the {\em F125W-F160W} and {\em F098M-F125W} colors from the ERS of the 22 PEARS spectroscopically identified M-dwarfs.}
\label{f:jh:gj:ers}
\end{center}
\end{figure}

\subsection{M-Dwarf NIR color-color selection}
\label{s:dialM}

The second part of the calibration is an empirical color-color selection for the M-dwarfs in the GOODS-S field. 
We base the M-dwarf selection in near-infrared colors because of the wide range in optical-near-infrared colors
M-dwarf subtypes can display (Figure \ref{f:vjh}).
% 
% How do we select M-dwarfs
To identify M-dwarfs, we require their $J_{F125W}-H_{F160W}$ and $Y_{F098M} - J_{F125W}$ color to be similar to the distribution of the PEARS-identified M-dwarfs. 
To find the J-H color range, we can use both the CANDELS as the ERS catalog but for the Y-J color, we must rely on the ERS catalog alone.

% J-H
Figure \ref{f:jh:hist} shows the distribution of $J_{F125W}-H_{F160W}$ color for the PEARS M-dwarfs in the CANDELS and ERS fields. These are not in the range 
one would infer from \cite{Ryan11}, in their Figures 3 and 4 but wider by a few tenths of magnitudes. The ERS and CANDELS J-H colors are still different by 
$\sim0.1$ magnitude (Figure \ref{f:jh:hist}).
We suspect that the PEARS M-dwarf {\em F125W-F160W} color distribution in CANDELS in Figure \ref{f:jh:hist} is different than the distribution in \cite{Ryan11} because their colors were derived from spectra convolved with appropriate filter response functions. There is however an under-reported drop in WFC3 detector sensitivity on the red side of the {\em F160W} filter (Andersen et al. {\em in preparation, private communication}). The {\em F098M} and {\em F125W} filters however, do not suffer from the detector sensitivity degradation at the red side. This could in part account for the small difference in J-H color of the CANDELS M-dwarfs, whose photometry we trust, with the position of M-dwarfs in Figure 3 in \cite{Ryan11} or the difference in  {\em F125W-F160W} color with the ERS as the detector was still cooling down (Figure \ref{f:jh:hist})

% Y-J
The ERS mosaics is in the same NIR filters as the BoRG fields ({\em F125W, F160W} and {\em F098M}) and Figure \ref{f:jh:gj:ers} shows the position of the 22 objects matched to PEARS M-dwarfs in a J-H, Y-J color-color plot. We show the same color criteria as Figures 3 and 4 in \cite{Ryan11}. The width and position of the Y-J color criterion is close to what one would infer from Figure 3 in \cite{Ryan11}.

Based on both Figure \ref{f:jh:hist} and \ref{f:jh:gj:ers}, we define a near-infrared color-color criterion to select M-dwarfs in the BoRG survey; $-0.05 < J-H < 0.3 $ and $0.1 < G-J < 0.75$ (the solid blue boxes in Figure \ref{f:jh:gj:ers} and \ref{f:jh:gj:borg} in addition to the L, T and MLT boxes from \cite{Ryan11}.

\begin{figure}[htbp]
\begin{center}
	\includegraphics[width=0.5\textwidth]{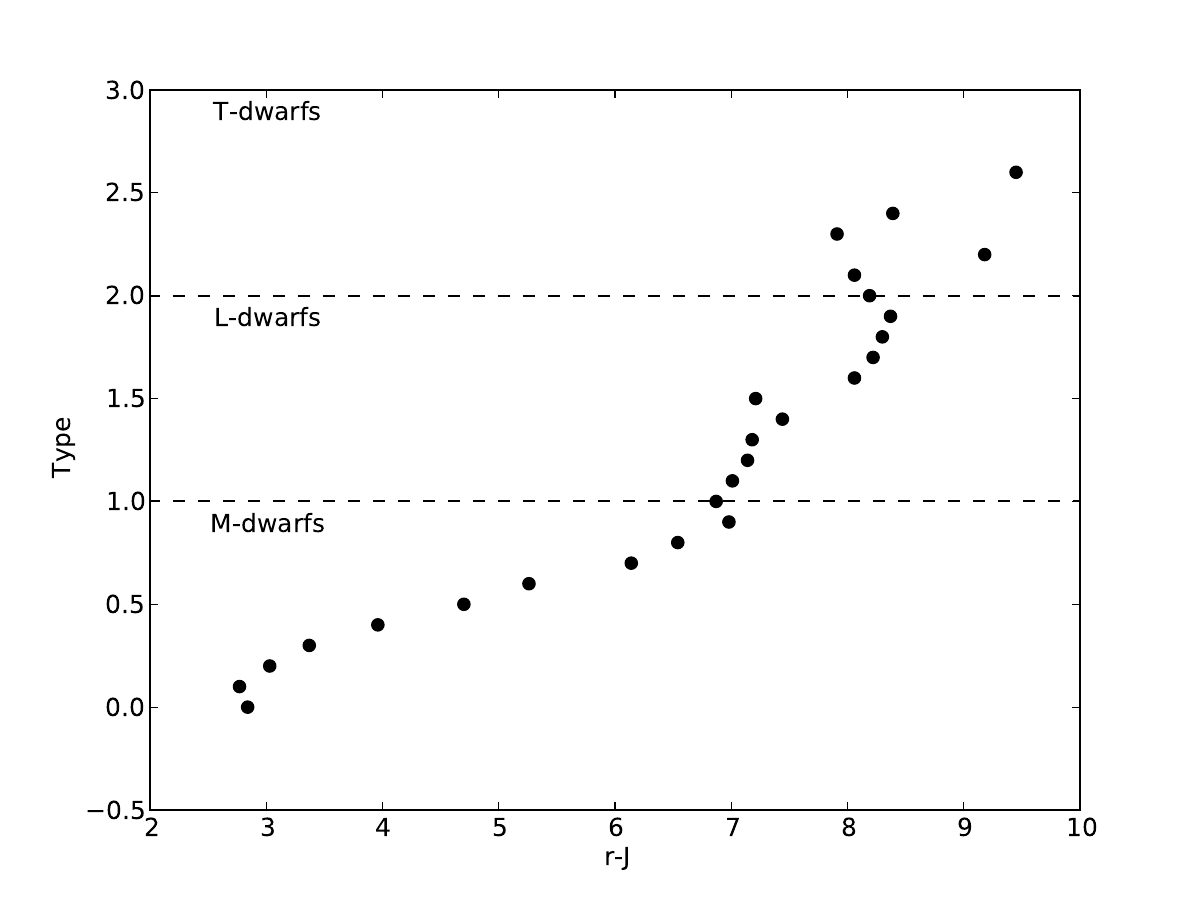}
\caption{The relation between {\em sdss-r}-J color and spectral type from \cite{Hawley02}. M-dwarfs are 0-1 with subtype as the decimal, L-dwarfs are 1-2 with the subtypes as the decimal and T dwarfs are type 3 with the decimal the assigned subtype as molecular lines start to dominate the spectrum. The later type dwarfs do not show a good linear relation between a wide optical-near-infrared color and sub-type. However, they do show a redder color than all the M-dwarf subtypes and hence we can exclude all most stellar objects based on color. }
\label{f:dtypes}
\end{center}
\end{figure}

\begin{figure}[htbp]
\begin{center}
	\includegraphics[width=0.5\textwidth]{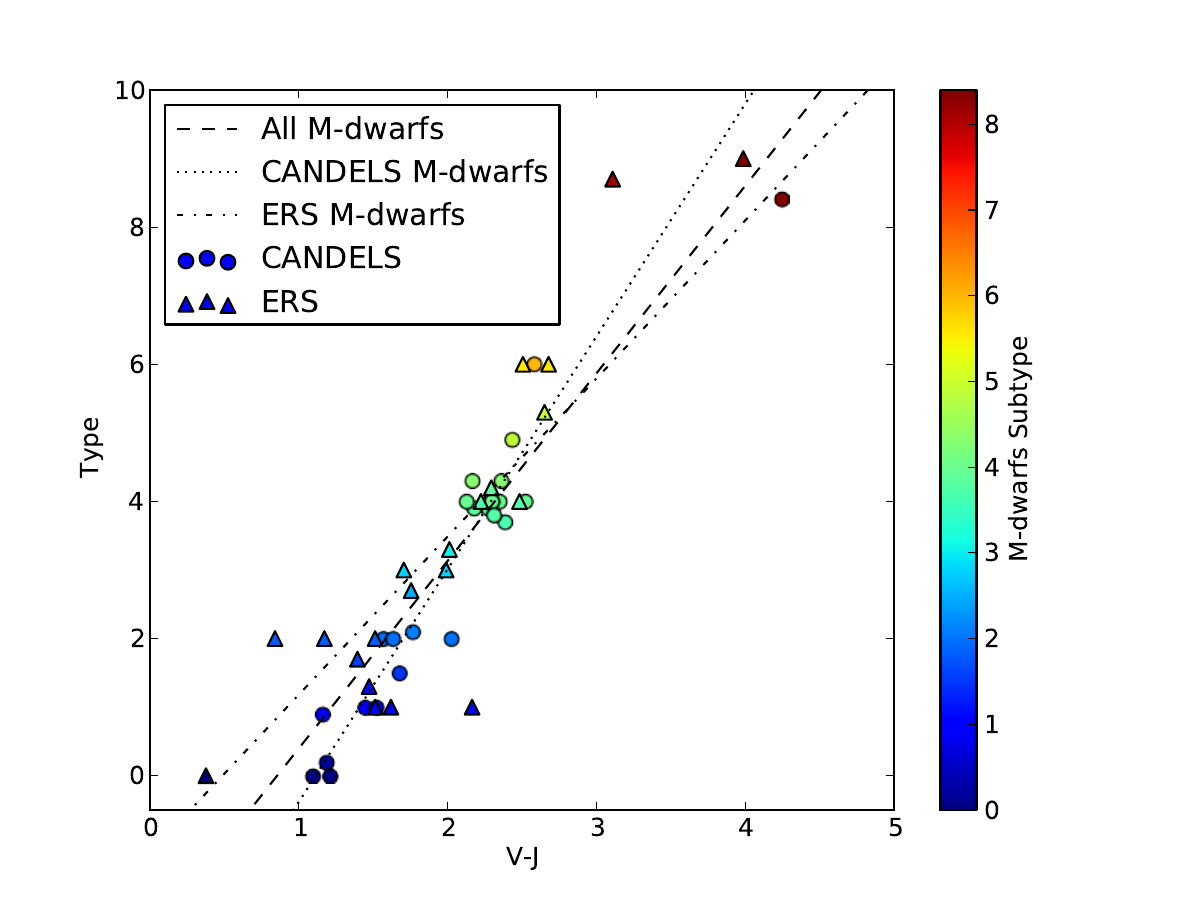}
\caption{The relation between M-dwarf sub-type (0-8) and the {\em F606W-F125W} color relation as determined from the GOODS, CANDELS and ERS data for the PEARS spectroscopically identified M-dwarfs. There is a linear relation between this color and the subtype which we use to type the dwarfs in the BoRG WFC3 parallel fields. We show three fits; one based on all M-dwarfs (t<6), only the CANDELS M-dwarfs (t$<$6) and only the ERS M-dwarfs. We adopt the CANDELS relation as this sample is the cleanest with accurate photometry and similar exposure times as the BoRG fields.}
\label{f:typing}
\end{center}
\end{figure}

\begin{figure}[htbp]
\begin{center}
	\includegraphics[width=0.5\textwidth]{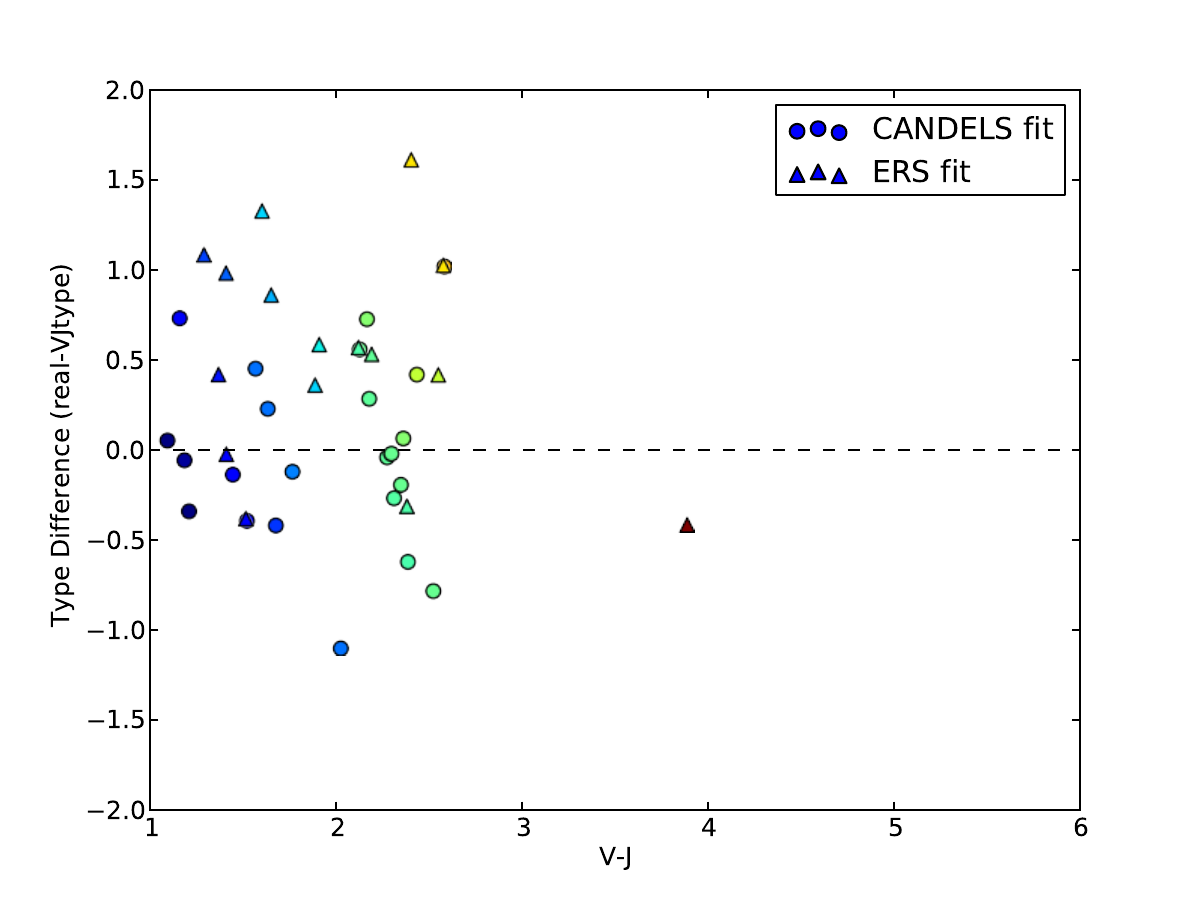}
\caption{The residual between V-J inferred subtype and the spectroscopic subtype from PEARS for the M-dwarfs in the ERS and CANDELS data. The ERS {\em F125W} photometry may be shifted slightly still with respect to the CANDELS's (Figure \ref{f:jh:hist}) and this may account for the higher scatter for this subset. }
\label{f:deltatyping}
\end{center}
\end{figure}

\subsection{Sub-typing M-dwarfs}
\label{s:subtype}

Our last calibration involving the PEARS identified M-dwarfs in the CANDELS and ERS mosaics of the GOODS-S field is to derive a color-subtype relation.
% SUBTYPING
Figure \ref{f:dtypes} shows the relation between {\em r-J} color and dwarf type from \cite{Hawley02}. M-dwarfs show a linear relation with a color, provided the baseline is long enough, i.e. optical to infrared. The later L and T dwarfs do not have such a clear relation. 

Figure \ref{f:typing} shows the relation between the $V_{F606W} - J_{F125W} $ color and the spectroscopic sub-type as determined by \cite{Pirzkal09} for the 24 PEARS M-dwarfs in CANDELS and the 22 PEARS M-dwarfs in the ERS. 
They obtain a type from template fits to each ACS grism spectrum obtained for these objects. Because of design of the PEARS observations, there are ofttimes several spectra taken at different roll-angles. Cross-comparison between the spectral fits show that the uncertainty in the PEARS type classification is less than an M-dwarf subtype. We therefore take these spectral types as our gold standard for the classification of the stellar objects in the WFC3 parallel fields.

Figure \ref{f:typing} shows the relation between the V--J color and spectroscopic type from PEARS. We fit a linear relation to the CANDELS sample because it is the cleanest selection and photometry.
The linear relation in the PEARS-identified CANDELS M-dwarfs can be expressed as:
\begin{equation}
\label{eq:type}
M\_{type} = 3.39 \times [V_{F606W}-J_{F125W}] - 3.78
% 3.39340752208*dw\_VJ -3.77607265231
\end{equation}

% ERS
The M-dwarfs identified by PEARS in the ERS mosaic display more scatter that the CANDELS M-dwarfs but after correction of the {\em F125W} magnitudes (see above), the V-J colors and their PEARS subtype generally agree well with the relation derived from the CANDELS M-dwarfs, although some difference may remain (Figure \ref{f:jh:hist}).

% Accuracy
Figure \ref{f:deltatyping} shows the residual between the V-J color type and PEARS spectroscopic type. 
The PEARS M-dwarfs cover mostly subtype 0-5 and only a few M ($t>6$) are in this sample. 
Consequently, the color-subtype relation is most reliably for the earlier subtypes. % (Figure \ref{f:jh:gj:borgerr}).
Based in Figure \ref{f:deltatyping}, we take that our subtype based on V-J color is accurate within a subtype.  
We adopt this relation to type M-dwarfs in the BoRG fields and compute type uncertainties as a result of the photometric uncertainty in the {\em F606W} and {\em F125W} magnitudes. 
We use the absolute magnitudes from \cite{Hawley02} to compute distance moduli (Table \ref{t:M}) and infer distances.

\begin{table}[htdp]
\caption{\label{t:M} Absolute 2MASS J-band (F125W) magnitudes of M-dwarfs from \cite{Hawley02}.}
\begin{center}
\begin{tabular}{l l }
Subtype & $M\_J$ \\
\hline
0 & 6.45 \\
1 & 6.72 \\
2 & 6.98 \\
3 & 7.24 \\
4 & 8.34 \\
5 & 9.44 \\
6 & 10.18 \\
7 & 10.92 \\
8 & 11.14 \\
9 & 11.43 \\
10 & 11.43 \\
\hline
\end{tabular}
\end{center}
\end{table}%

\begin{figure}[htbp]
\begin{center}
	\includegraphics[width=0.5\textwidth]{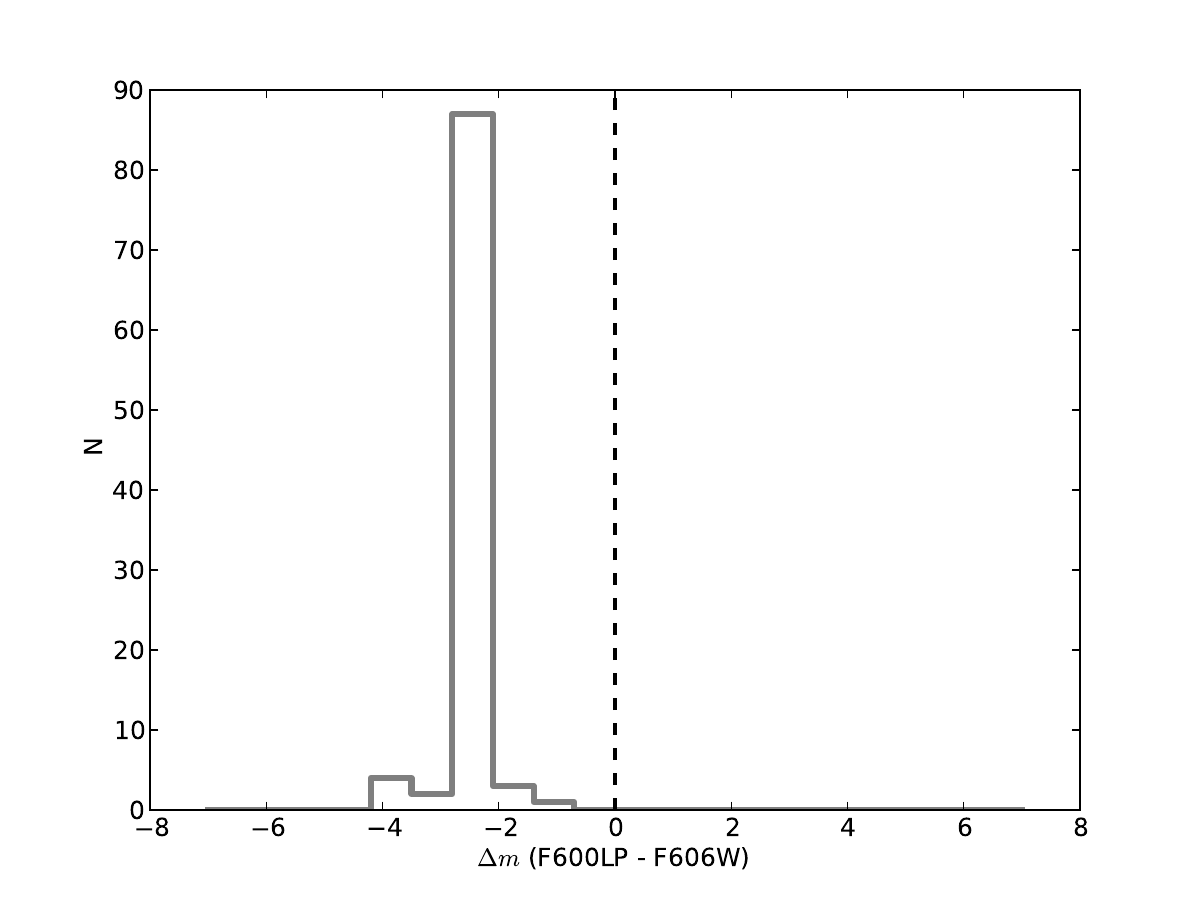}
\caption{A histogram of the difference in F600W and F600LP MAGAUTO values for those stars in the M-dwarf near-infrared criteria for the three fields in the BoRG survey that have data in both filters.  }
\label{f:f600lp}
\end{center}
\end{figure}

\subsection{Substituting F600LP for F606W photometry}

In the case of 24 fields in the BoRG survey, F606W photometry is not available. These fields come from the HIPPIES or COS-GTO parallel observations which opted for F600LP rather than F606W.  The F600LP filter has a slightly different width compared to the F606W and a different central wavelength (closer to Johnson-I). Fortunately, there are three fields for which both F606W and F600LP data is available. We compared the F606W-F600LP colors for those objects already pre-selected by their near-infrared color and morphology as likely M-dwarf to assess the impact of the change of filters. Figure \ref{f:f600lp} shows the histogram of the color difference between the two bandpasses for M-dwarfs. The difference is substantial: F600LP - F606W =  -2.65 mag with a spread of 0.88 mag. 
Therefore we correct our F600LP photometry with this difference and increase the uncertainty in the photometry accordingly as well (which should be reflected in the subsequent automated typing).

\begin{figure}[htbp]
\begin{center}
	\includegraphics[width=0.5\textwidth]{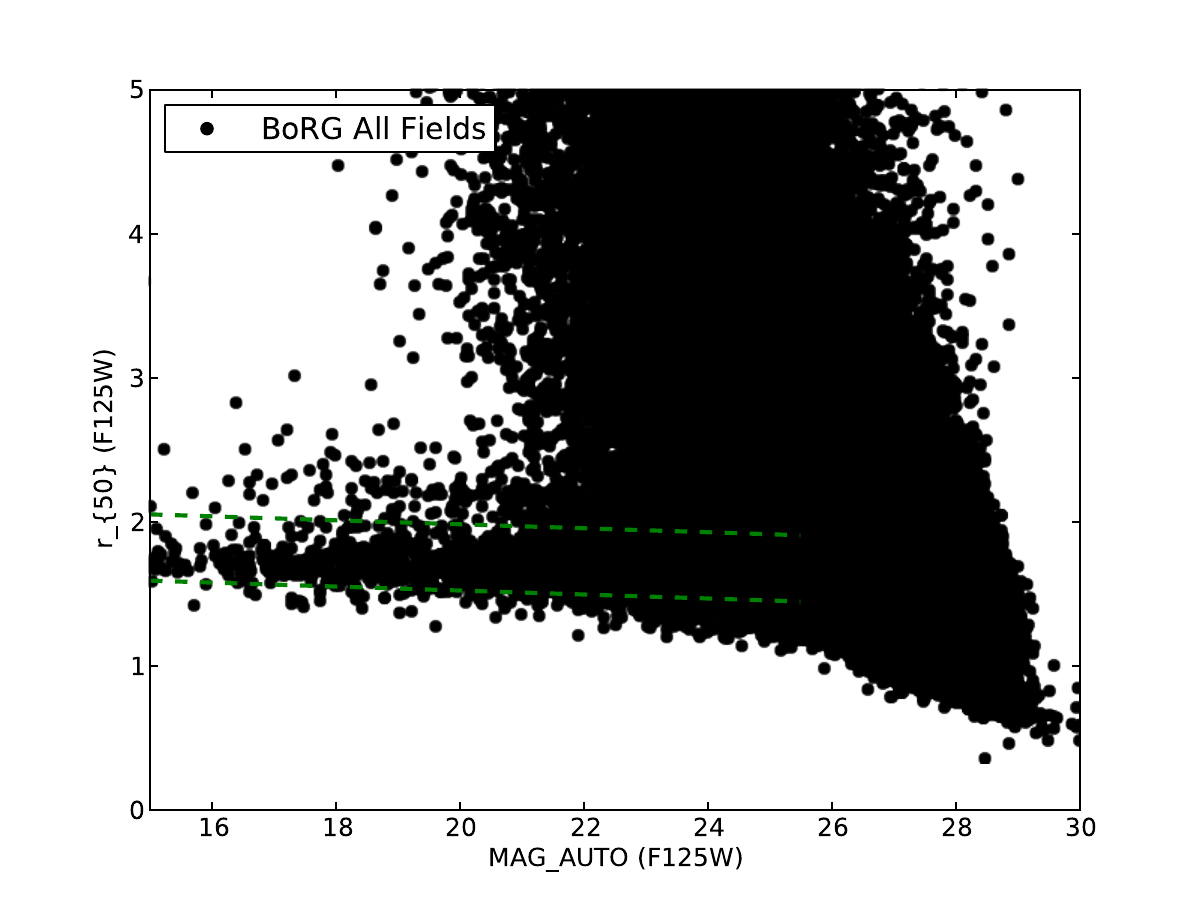}
\caption{The {\em F125W} half-light radius as a function of magnitude for all the BoRG fields. The green dashed lines are the selection criterion from equation \ref{eq:r50} to identify stars in these fields. Objects below the bottom green line are more likely to be defects (e.g. remaining hot pixels etc.).  }
\label{f:borg:r50}
\end{center}
\end{figure}

\begin{figure*}
\begin{center}
	\includegraphics[width=0.49\textwidth]{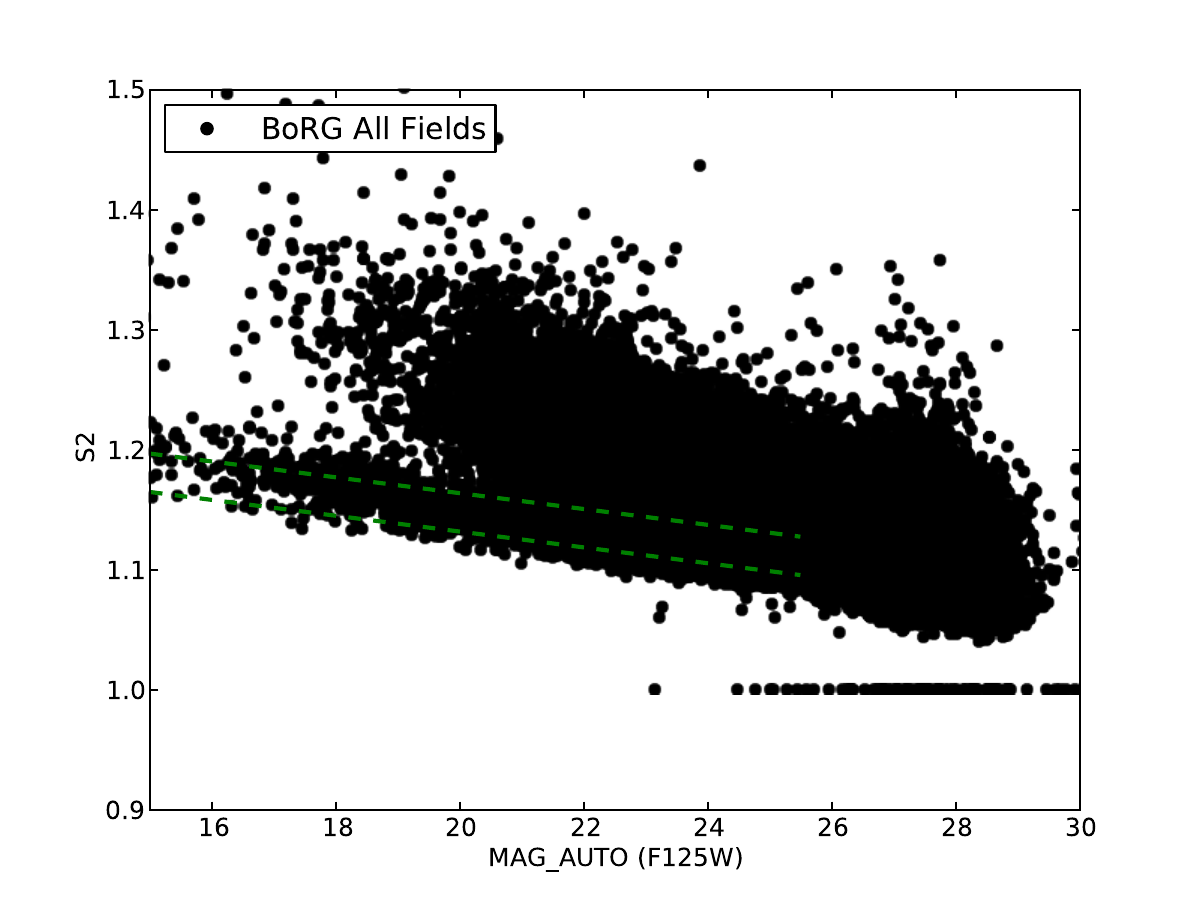}
	\includegraphics[width=0.49\textwidth]{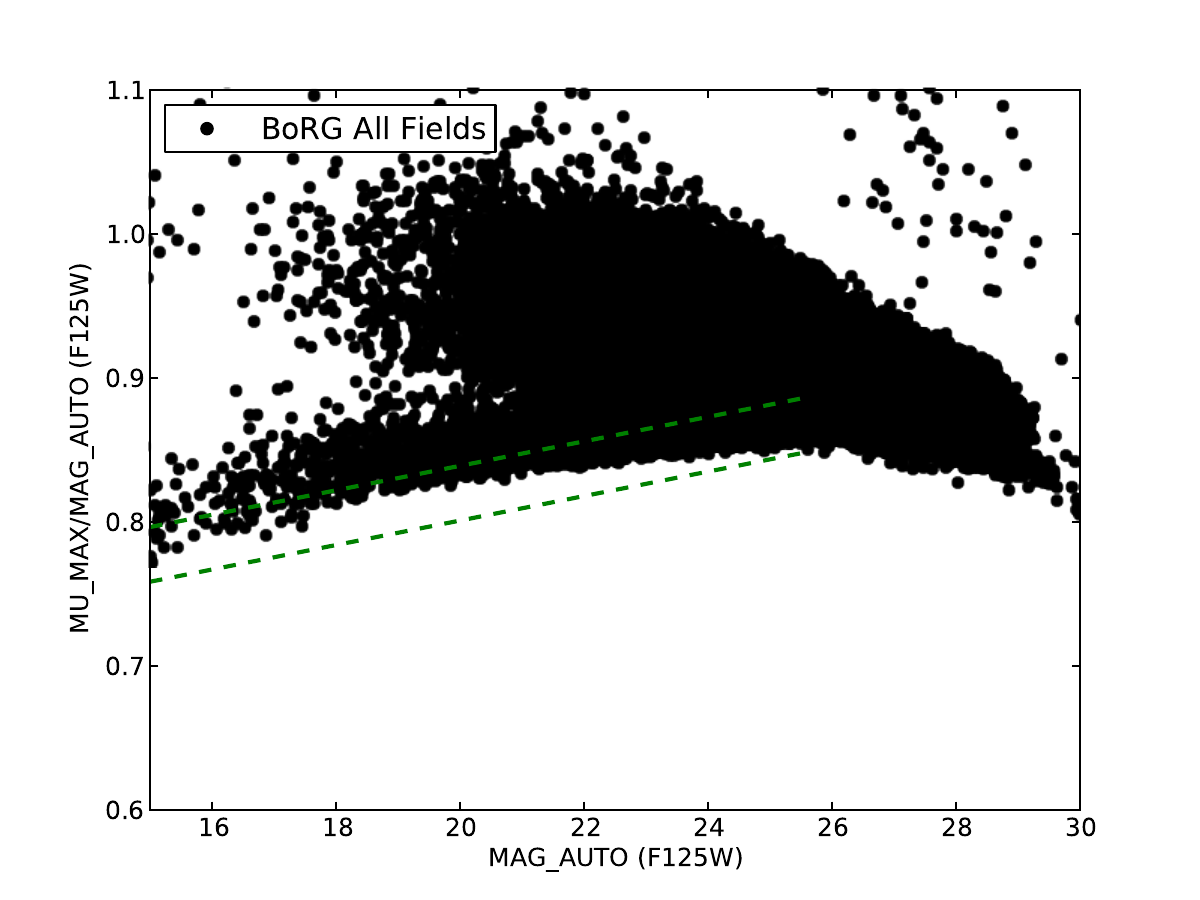}
\caption{The two alternative morphological selections for star, for the BoRG data: the $S\_2$ and the mu\_max/mag\_auto ratio.}
\label{f:borg:morph}
\end{center}
\end{figure*}

\begin{figure}[htbp]
\begin{center}
	\includegraphics[width=0.5\textwidth]{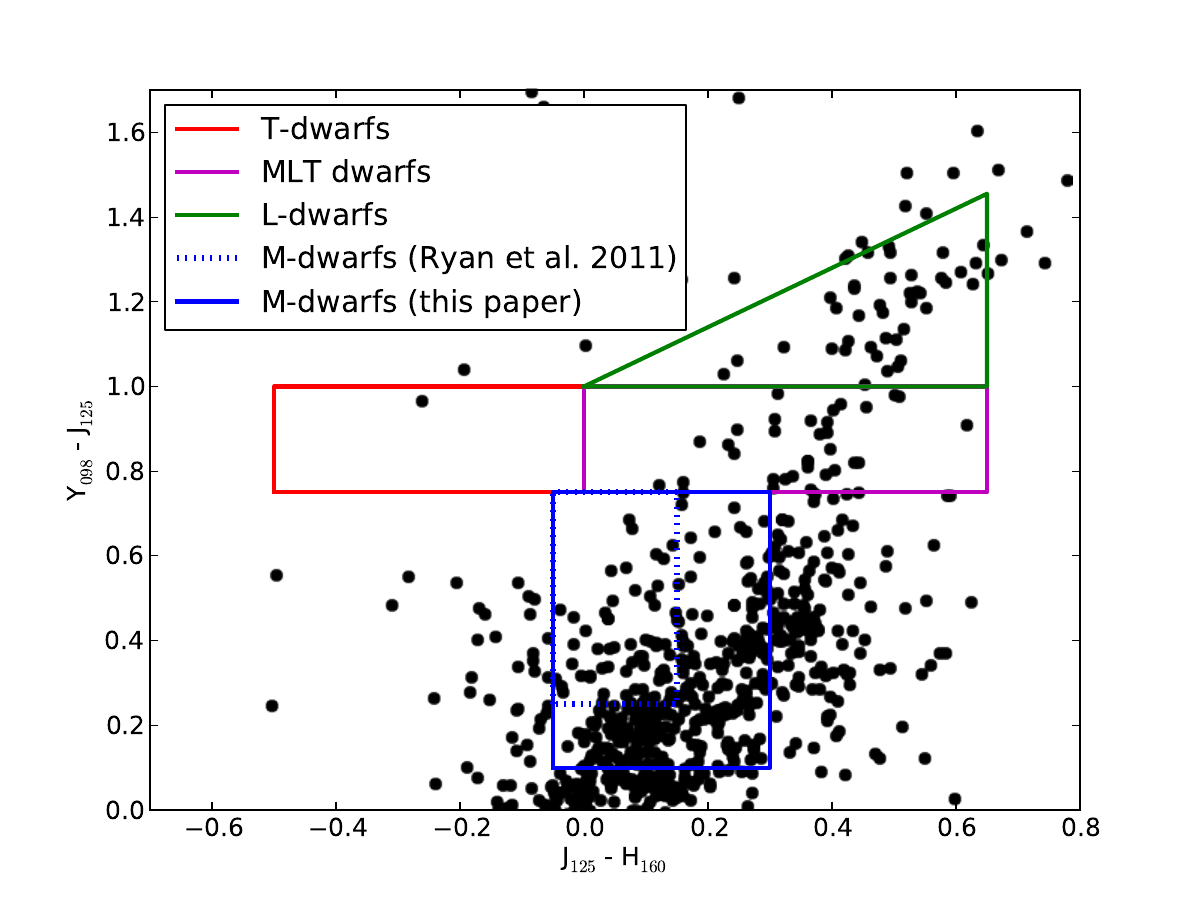}
\caption{$J_{F125W}-H_{F160W}$ versus $G_{F098M}-J_{F125W}$ color diagram, similar to Figure 4 from \cite{Ryan11}, to distinguish between L, T, and M-dwarfs. }
% \label{f:jh:gj:borg}
\label{f:jh:gj:borg}
\end{center}
\end{figure}

\section{Stars in the BoRG WFC3 Pure-parallel Fields}
\label{s:MWstars}

With the morphological and near-infrared color-color selection in hand, we now search the BoRG fields for M-dwarfs as well as other substellar types.
% morphological selection
Figure \ref{f:borg:r50} shows the same parameter space as Figure \ref{f:r50} but with all the BoRG objects. Based on the $r\_{50}$ selection criterion, we find \borgstars\  stars to 24 mag in {\em F125W} in the 59 WFC3 parallel fields. For comparison, Figure \ref{f:borg:morph} shows the other morphology criteria. 
The $r\_{50}$ selection criterion seems the most appropriate for the BoRG fields as the locus of stellar points is within the criterion lines. $S\_2$ appears equally applicable but the {\sc mu\_max/mag\_auto} criterion shows a discrepancy compared to Figure \ref{f:murat}. We suspect this is because the CANDELS data was originally at a different pixel size and this criterion is sensitive to the exact pixel scale of the image.
Figure \ref{f:borg:r50} validates therefore our adoption of the $r\_{50}$ criterion for the morphological selection of stars. 

% color-color M-dwarf selection
To identify the various subtypes of stars, we construct a NIR color-color diagram similar to Figure 3 in \cite{Ryan11}, as discussed above. Figure \ref{f:jh:gj:borg} shows this plot with the limits for our selection criteria for approximate types. In the BoRG {\sc sextractor} catalog, we identify \Ldwarfs\ L-dwarfs, \Tdwarfs\ T-dwarfs, \Mdwarfs\ M-dwarfs and \MLTdwarfs\ M, L or T-dwarfs based on the color cuts. The M-dwarf sample is the one of interest here.
Tables \ref{t:borg:Mdwarfs}, \ref{t:borg:Tdwarfs} and \ref{t:borg:Ldwarfs} list the properties of the identified stars which conform to the M, T and L-dwarf color cuts (Figure \ref{f:jh:gj:borg}).

\begin{figure}[htbp]
\begin{center}
	\includegraphics[width=0.5\textwidth]{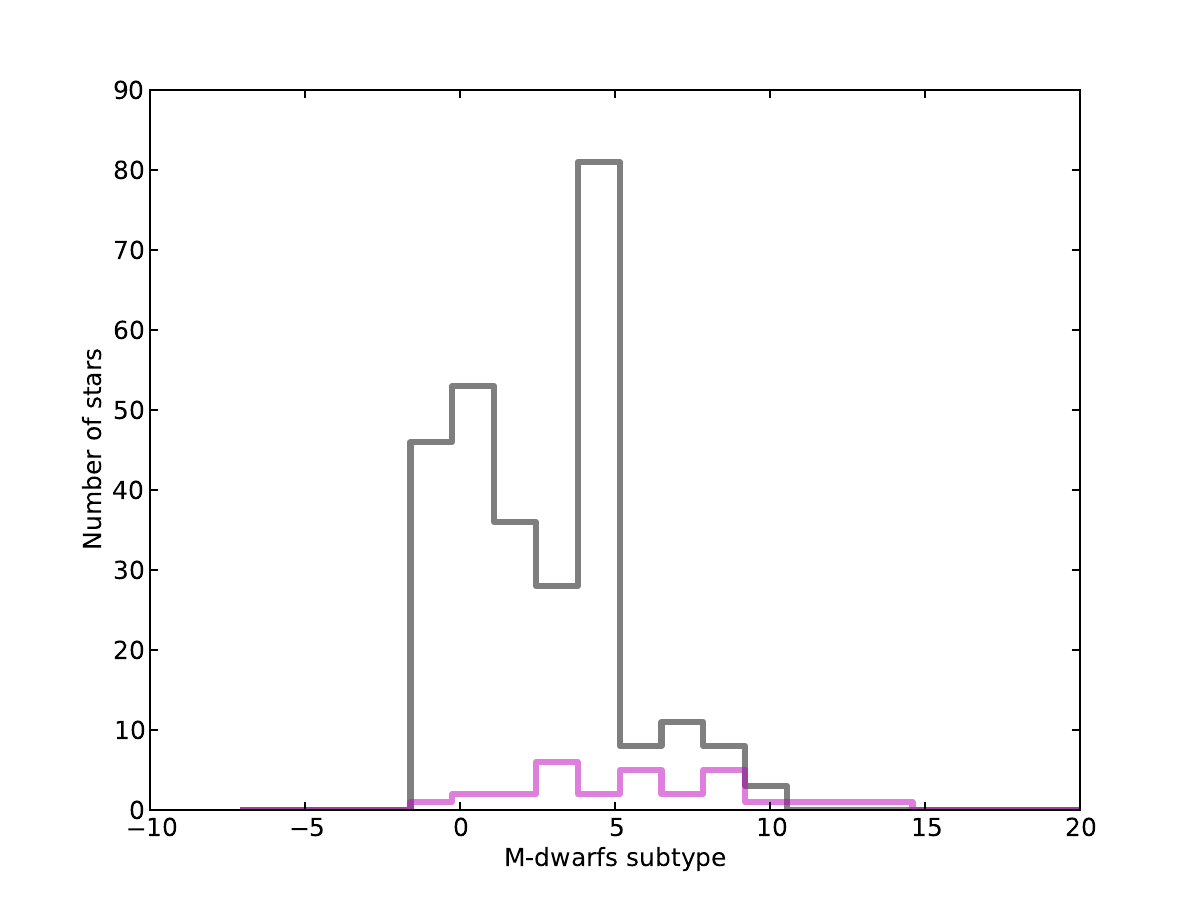}
\caption{The histogram of M-dwarf types in our BoRG, inferred from equation \ref{eq:type}. M0 and M1 are the most prevalent in the sample. As can be seen in the histogram, some of the stellar objects are not M-dwarfs but either other sub-stellar objects or possibly very distant (intergalactic?) M-giants. }
\label{f:borg:type}
\end{center}
\end{figure}

% subtyping
Subsequently, we sub-type the M-dwarfs according to equation \ref{eq:type}. Figure \ref{f:borg:type} shows the histogram of M-dwarf subtypes in the BorG fields as well as those objects in the MLT box in Figure \ref{f:jh:gj:borg} for comparison. In effect this is an additional optical-near-infrared color criterion for M-dwarf selection as we will discard any star that could not be sub-typed.
The dominant type in our M-dwarf sample is M0, specifically -1 to 1 type, not unexpectedly as these are the brightest and most numerous type of M-dwarf. There are almost as many later-type M-dwarfs (M4 and above) in the MLT color selection as in our exclusively M-dwarf box. However, as we note above, the later type classification is the most uncertain and in the MLT box, susceptible to L and T dwarf contamination.

% contamination
There is still the possibility of contamination by Giants, sub-dwarfs and AGNs but we are confident that the morphological selection, luminosity limit and the color-color restriction select a very clean sample of Milky Way M-dwarfs.

\subsection{Contamination by M-giants or M-type Sub-dwarfs}

\cite{Bochanski13} present a catalog of 404 M-giants from the UKIDSS survey DR8 (2400 deg$^2$, $K < 17$), with these M-giants among the most distant stars belonging to the Milky Way. In principle, M-giants could be selected by our color selection. However, the their on-sky density ($4.3 \times 10^{-5} $ M-giants/arcmin$^2$) makes it unlikely --not impossible-- that even a single one is included in our selection. One could make a similar plausibility argument against very nearby M-type subdwarfs: the volume probed at close distances is comparatively small for a pencil beam survey.\\

{\bf Please note that beyond this point there are corrected versions of many of the figures available in the ERRATUM appended below. }

\section{Discussion}
\label{s:disc}

Based on our bona-fide, morphologically and color selected M-dwarfs and the subsequent color sub-typing, we explore 
the North/South discrepancy, 
the Galactic thin disk scale-height as a function of subtype 
and whether or not we detect any additional Galactic component (Halo or Thick disk).

\subsection{The North-South Divide}
\label{s:ns}

\cite{Pirzkal09} found a North-South discrepancy between the number of M-dwarfs in the PEARS fields. They find 51 and 63 M0-9 dwarfs in the GOODS North and South fields respectively. Accounting for the different field sizes (41.61 arcmin$^2$ and 59.50 arcmin$^2$ respectively) this is equivalent to 1.23 and 1.06 M-dwarfs/arcmin$^2$, an excess of 11\% in the Northern field. Both GOODS fields are at similar Galactic Latitude and Pirzkal et. al. note that because the Sun is North of the Galactic plane, the Northern excess appears to be a contradiction. \cite{Widrow12} and \cite{Yanny13} find similar North-South discrepancies in the SDSS catalog of (dwarf) stars in the solar neighborhood. 

One of our science goals with this study was to establish if this a real discrepancy of due to natural field variance. 
The BoRG fields are over many more lines of sight and dividing these in Galactic Northern and Southern Fields: \borgnorth\ and \borgsouth\ square arcminute respectively.
We count the M-dwarfs in those fields above $20^\circ$ Galactic Longitude to avoid two Southern fields that sample more along the Galactic Plane.

% M-dwarfs are stellar objects in the J-H, Y-J color box (Figure \ref{f:jh:gj:borg}) and a limiting magnitude of $ m\_{125W} < 24.$. 
We find on average 1.157 $\pm$ 0.088  M-dwarfs of all types (M0-9) per arcmin$^2$ in the Northern fields and 1.293 $\pm$ 0.127 per arcmin$^2$ in the Southern Galactic fields (Table \ref{t:ns}), which agrees with the previously observed discrepancy. The Northern over density is the most pronounced in the early (M0-4) type M-dwarfs.
Taking into consideration that for these early-types the statistics are more secure, and early-types dominate the grism sample of \cite{Pirzkal09} (or the SDSS catalog), our counts corroborate the previously found North-South discrepancy. 
The Northern overdensity is most pronounced in the earliest types, gradually diminishing and reversing for late-type M-dwarfs (Table \ref{t:ns}).

\begin{table}[htdp]
\caption{\label{t:ns}Average number of M-dwarfs per square arcminute in Southern and Northern Galactic fields ($|l| ~ > ~ 20^\circ$). M-dwarf numbers refer to those objects in the blue box in Figure \ref{f:jh:gj:borg}. The last two columns are the ratio of M-dwarfs in Northern and Southern fields and the signal-to-noise of the noise.}
\begin{center}
\begin{tabular}{l l l l l}
				& M-dwarfs	&			&  			& \\
	 		 	& North		& South		& North 		& S/N\\
	 		 	& 			& 			& /South 		& \\
\hline
M-dwarfs & 1.157 $\pm$ 0.088 & 1.293 $\pm$ 0.127 & 0.89 & 8.02 \\ 
M0-4 & 0.859 $\pm$ 0.076 & 0.779 $\pm$ 0.099 & 1.10 & 6.45 \\ 
M5-10 & 0.210 $\pm$ 0.038 & 0.465 $\pm$ 0.076 & 0.45 & 4.11 \\ 
M0 & 0.304 $\pm$ 0.045 & 0.176 $\pm$ 0.047 & 1.73 & 3.27 \\ 
M1 & 0.156 $\pm$ 0.032 & 0.151 $\pm$ 0.043 & 1.03 & 2.81 \\ 
M2 & 0.115 $\pm$ 0.028 & 0.088 $\pm$ 0.033 & 1.31 & 2.23 \\ 
M3 & 0.088 $\pm$ 0.024 & 0.075 $\pm$ 0.031 & 1.17 & 2.03 \\ 
M4 & 0.196 $\pm$ 0.036 & 0.289 $\pm$ 0.060 & 0.68 & 3.58 \\ 
M5 & 0.095 $\pm$ 0.025 & 0.364 $\pm$ 0.068 & 0.26 & 3.07 \\ 
M6 & 0.007 $\pm$ 0.007 & 0.025 $\pm$ 0.018 & 0.27 & 0.82 \\ 
M7 & 0.041 $\pm$ 0.017 & 0.025 $\pm$ 0.018 & 1.62 & 1.22 \\ 
M8 & 0.041 $\pm$ 0.017 & 0.013 $\pm$ 0.013 & 3.23 & 0.93 \\ 
M9 & 0.020 $\pm$ 0.012 & 0.038 $\pm$ 0.022 & 0.54 & 1.22 \\ 
M10 & 0.007 $\pm$ 0.007 & 0.000 $\pm$ 0.000 & 0.00 & 0.00 \\

\hline
\end{tabular}
\end{center}
\end{table}%

\begin{figure}[htbp]
\begin{center}
	\includegraphics[width=0.5\textwidth]{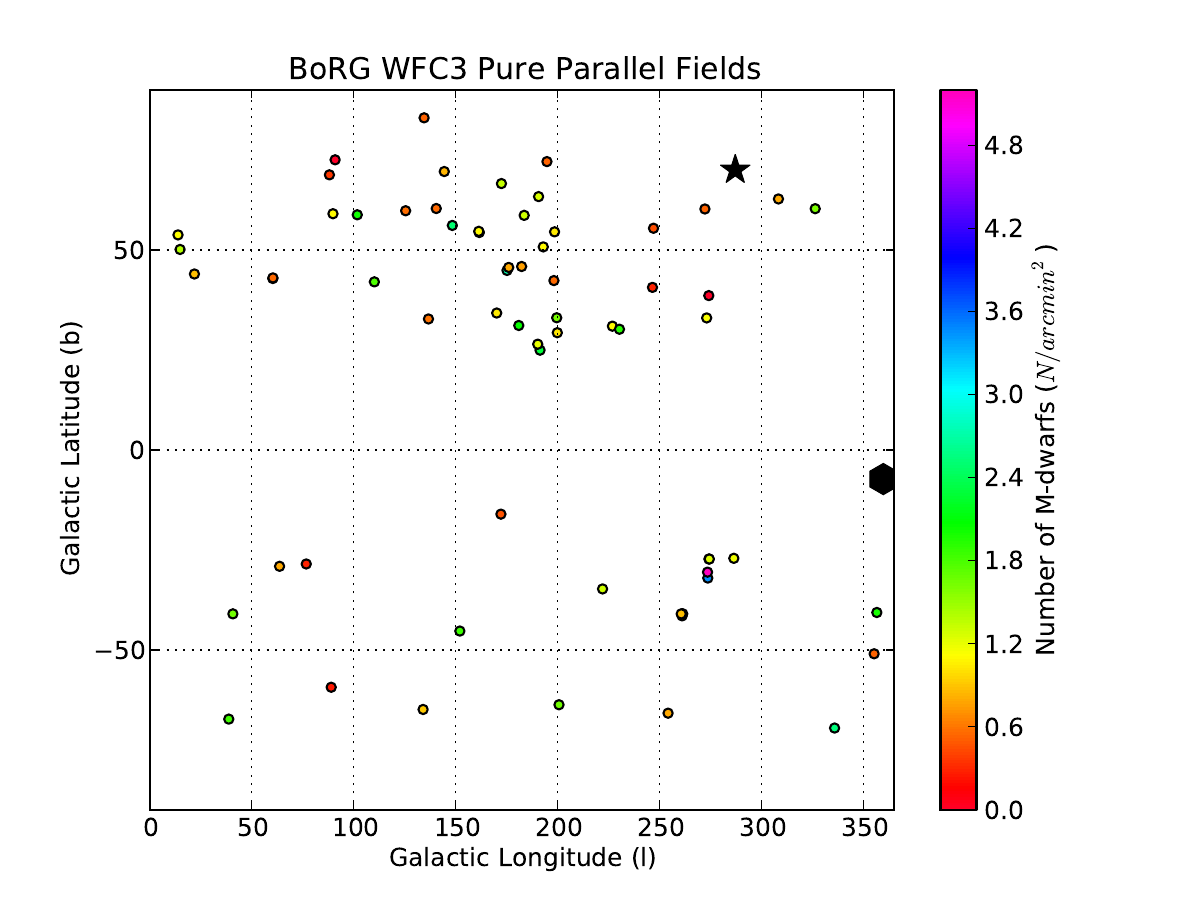}
\caption{The distribution of BoRG fields with the number of M-dwarfs indicated. One field, borg\_1230+0750 (star) stands out with 22 M-dwarfs ($\sim20$ stars/arcmin$^2$). We discard borg\_1815-3244 (black circle) for its low latitudes and line-of-sight through the plane of the disk and close of the center of the bulge.}
\label{f:MdwarfsFields}
\end{center}
\end{figure}

\begin{figure}[htbp]
\begin{center}
	\includegraphics[width=0.5\textwidth]{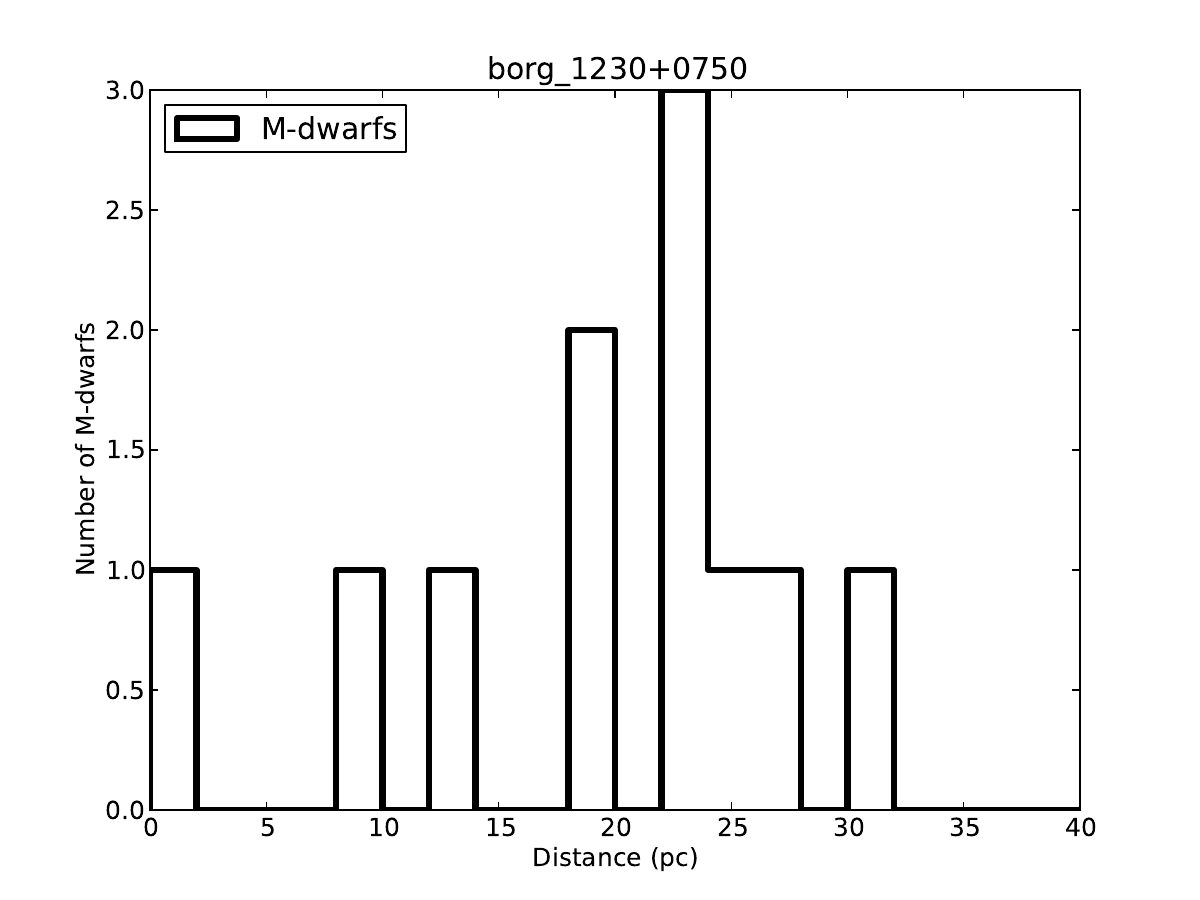}
\caption{The distribution of M-dwarf distances in borg\_1230+0750, with a concentration at 20-25 kpc. }
\label{f:sag:hist:d}
\end{center}
\end{figure}

\subsection{A Local Over-density of M-dwarfs}
\label{s:stream}

A single field, borg\_1230+0750 stands out from the other BoRG fields with a local density of 21.4 M-dwarfs/arcmin$^2$. 
Such over-densities of faint Galactic stars are a possible contaminant for identifications of over-densities of high-redshift clusters \citep[e.g.,][]{Trenti12}. 

The Galactic position of this over-density ($l=287.119670593^\circ, b=70.0322203957^\circ$) argues against a spiral arm or disk substructure to be the origin of this over-density. However, its position is exactly on the Sagittarius stellar stream \citep{Majewski03, Belokurov06}.
Figure \ref{f:sag:hist:d} shows the histogram of photometric inferred distances for the M-dwarfs in borg\_1230+0750. The distance of the over-density of M-dwarfs at 20-25 kpc. is also in line with these M-dwarfs belonging to the Sagittarius stream \citep{Newberg02, Belokurov14} at this Galactic longitude.

The relative ease with which this stellar stream was identified in the BoRG data, shows that the future EUCLID mission will identify virtually {\em all} streams in the Milky Way halo from the dwarf stars alone. % The factor four over-density in M-dwarf numbers may pose an issue for the correction of stellar contamination in the EUCLID data. 

\begin{figure}[htbp]
\begin{center}
	\includegraphics[width=0.5\textwidth]{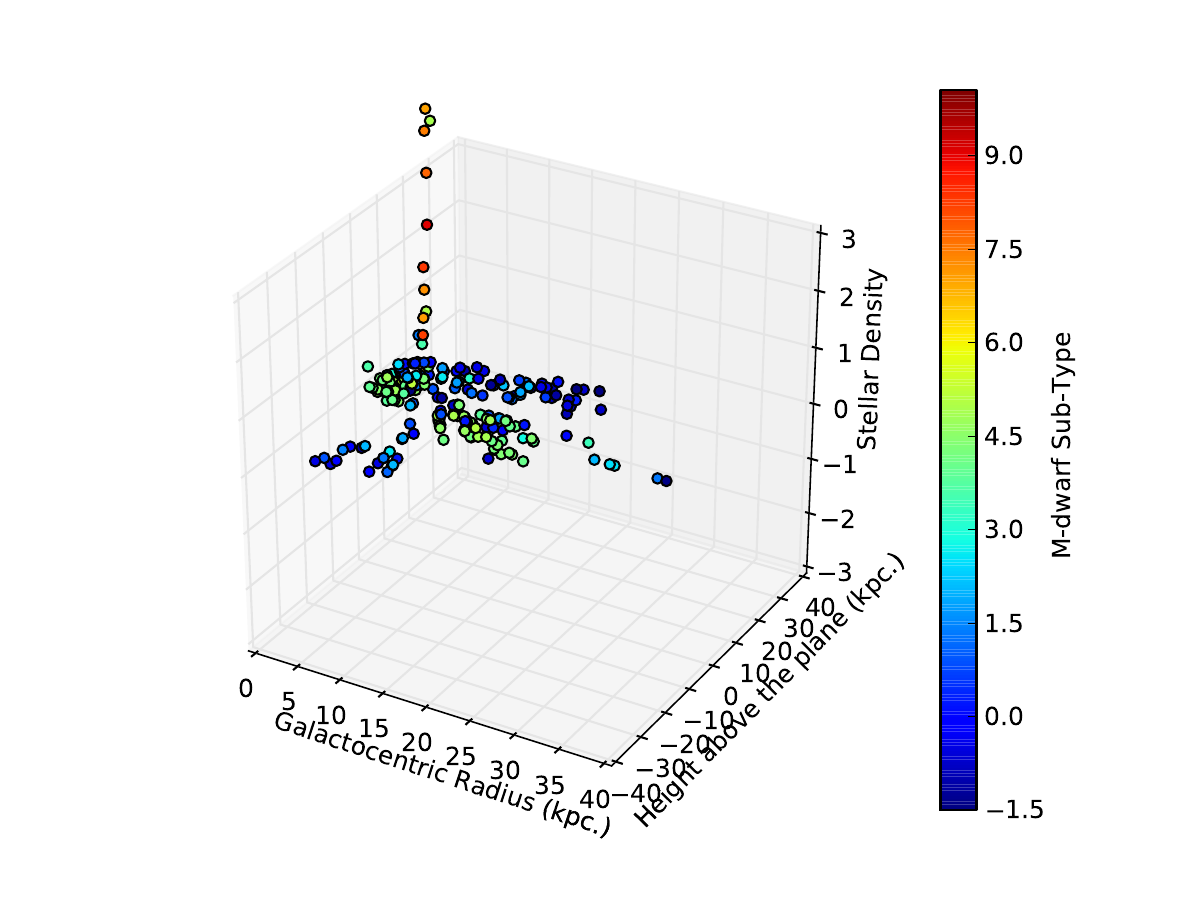}
\caption{The computed volume for each M-dwarf as a function of Galactocentric radius, height above the plane. M-dwarf types as inferred from their optical-nir color.}
\label{f:Mdwarf:3D}
\end{center}
\end{figure}

\begin{table}[htdp]
\caption{\label{t:N0} Local dwarf density from Figure 8 in \cite{Reid08} and \cite{Cruz07}.}
\begin{center}
\begin{tabular}{l l l}
Subtype & Number within 20 pc. &\\
\hline
		& Reid et al. 2008 & Cruz et al. 2007 \\
0		& 35	 	& 	\\
1		& 50 		& \\
2		& 140 	& \\
3		& 90 		& \\
4		& 350 	& \\
5		& 400 	& \\
6		& 190 	& \\
7		& 170	& 21\\
8		& 45		& 21\\
9		& 60		& 11\\
10		& 42		& 3\\
\hline
\end{tabular}
\end{center}
\end{table}%

\subsection{}

In the original catalog of M-dwarfs presented, there is an issue with the conversion from celestial coordinates to Galactic ones, done with {\sc pyephem} a wrapper around a trusted and vetted library {\em ephermis}. In the this updated astro-ph entry, we present the corrected coordinates (using {\sc AstroPy}) and distances based on AB magnitudes. We have amended the tables and Figures beyond this point accordingly.

\subsection{Thin Disk Scale Height.}
\label{s:z0}

Previous studies based on small numbers of fields have already given us several estimates of the scale-height of the thin disk of the Milky Way in substellar objects (Table \ref{t:z0}). The consensus appears to be a scale-height between 300 and 400 pc for the thin disk. For M-dwarfs, \cite{Juric08} find a thin disk of $z_0 = 300$ pc. and a thick disk with $z_0 = 900$ pc. from SDSS data. 

To estimate the distances to each M-dwarf, we compute the distance modulus from the inferred sub-type (and hence absolute magnitude) and the apparent magnitude in F125W. 
We compute the Galactic radius and height above the plane for all M-dwarfs, based on their Galactic longitude, latitude and the above distance, assuming the position of the Sun is 27 pc. above the plane, and at a Galactocentric radius of 8.5 kpc from the Galactic center. To compute the density at each M-dwarf's position, we compute the physical area at the inferred distance of the BoRG survey field in which the dwarf was found and a binwidth of 1pc.

We discard one field (borg\_1815-3244) for two reasons: it is at low Galactic Latitude with a line-of-sight straight through the Galactic bulge (Figure \ref{f:MdwarfsFields}), and secondly, there is only F600LP information available which makes subtyping of M-dwarfs more uncertain. This field contains 67 M-dwarf candidates but these are not used for the following analysis.

Figure \ref{f:Mdwarf:3D} shows the distribution of radius and height above the disk for the \Mdwarfs\ stars in the BoRG survey with their corresponding number density and type. The fainter late-sub-type M-dwarfs are in a localized thin disk and the bright early-sub-typ (e.g., M0) correspond closer to the thick disk component. 

We assume that the Galactic disk has the following parametric shape:
\begin{equation}
\rho(R,z) = \rho_0 ~ exp(-R/h) sech^2\left({z \over z_0}\right),
\end{equation}
\noindent where $\rho(R,z)$ is the dwarf number density in a point in the disk, $\rho_0$ is the central number density, $R$ is galactocentric radius, $h$ is the scale-length, $z$ is height above the plane, and $z_0$ is the scale-height of the disk.

To collapse Figure \ref{f:Mdwarf:3D}, we weight the volume densities with the exponential part of this equation since we aim to infer the vertical structure first, i.e., we assume a single scale-length and calculate $exp(-R/h)$ as a weight for each dwarf volume density.
In Figure \ref{f:Mdwarf:z}, the volume densities have all been scaled this way, assuming the scale-length ($h$) of the thin disk from \cite{Juric08} of 2.6 kpc. They found a scale-length for the thick disk of $h$=2.9 kpc. 
Alternatively, \cite{Bensby11} found a thin-disk scale-length of $h$=3.8 kpc and $h$=2.0 for the thick disk, corroborated by \cite{Cheng12a} and the trend of stellar population and scale-length in \cite{Bovy12a}. We use the Bensby et al. thin disk scale-length for dwarfs found in below 500 pc and the thick disk one for those more than 500 pc out of the plane.
The implicit assumption in this renormalization is that the scale-height does not change with radius, which is observed in external galaxies seen edge-on \citep[][Streich et al {\em in preparation}]{Comeron11a,Comeron11b,Comeron11c}.

%				h\_thin (kpc)	h\_thick (kpc)
%	Juric08		2.6			2.9
%	Bensby11		3.8			2.0
%	Cheng12a				1.8
%	Bovy12a		4.5-2.0

\begin{figure}[htbp]
\begin{center}
	\includegraphics[width=0.5\textwidth]{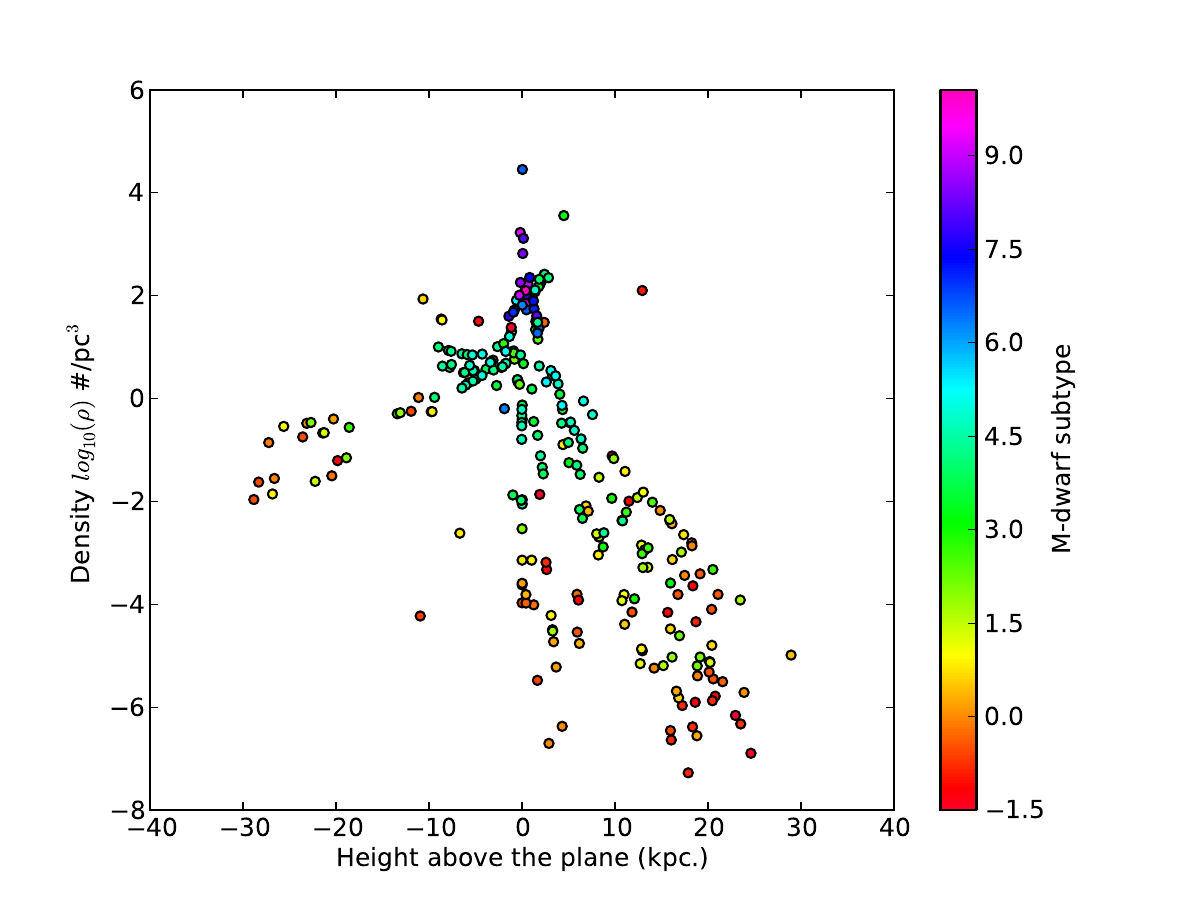}
\caption{The volume density of M-dwarfs as a function as height above the plane of the Milky Way disk. 
The volume densities were normalized with a scale-length of 2.6 kpc. Points are color-coded by the M-dwarf sub-type.}
\label{f:Mdwarf:z}
\end{center}
\end{figure}

Figure \ref{f:Mdwarf:z} shows more clearly how the later-type M-dwarfs (M4-8) in the BoRG fields are concentrated in the thin disk and the earlier types (M0-3) probe both thin and thick disks. It also shows that there may be several disk components in early types with different scale-heights above and below the plane of the disk. This would explain the North-South differential noted earlier by \cite{Pirzkal09}.

\begin{figure}[htbp]
\begin{center}
	\includegraphics[width=0.5\textwidth]{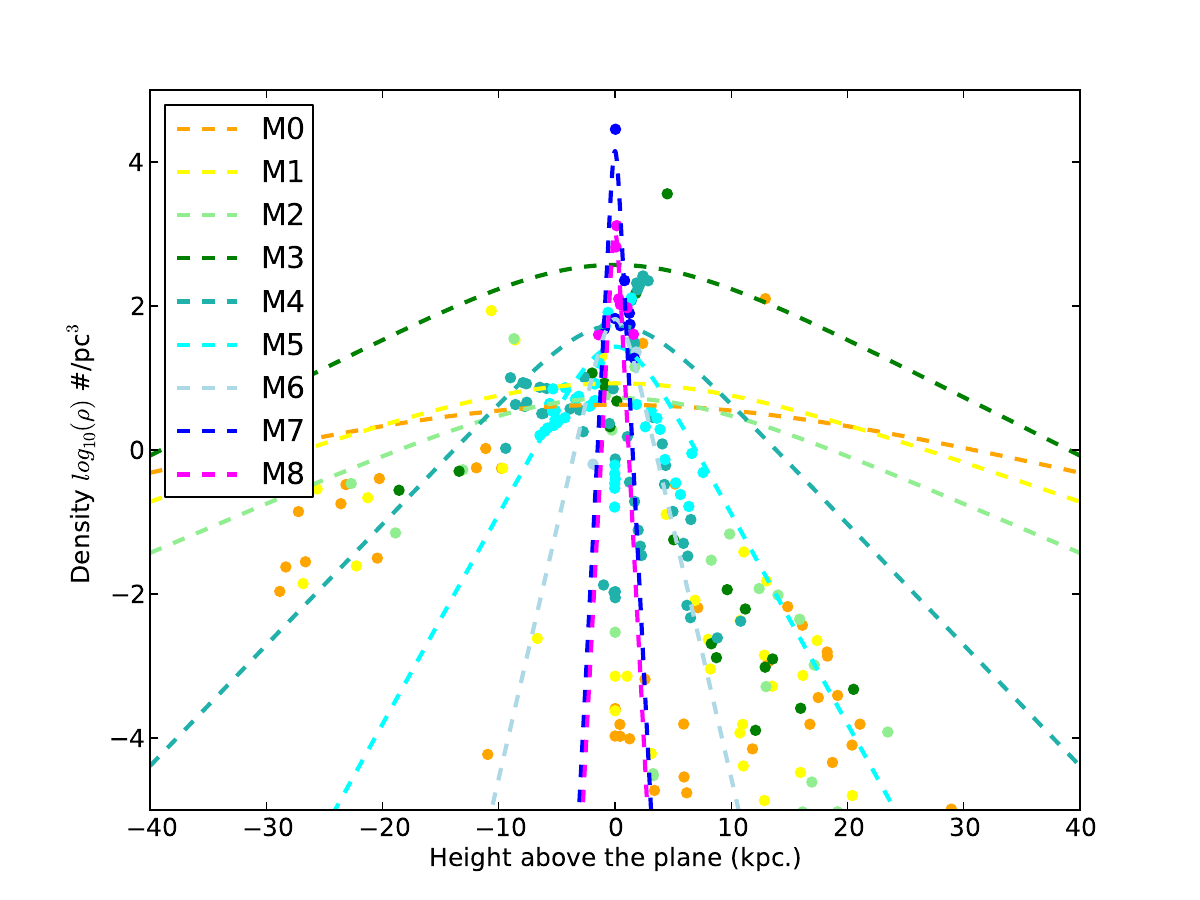}
\caption{The volume density of M-dwarfs as a function as height above the plane of the Milky Way disk. Points are grayscale-coded by the M-dwarf sub-type with best fits shown for each M-dwarf type.}
\label{f:Mdwarf:z-Mtype}
\end{center}
\end{figure}

% Naive fit
First we perform a naive fit on the vertical distribution of M-dwarfs as a function of sub-type: we fit the $sech^2(z/z_0)$ to their densities for each sub-type disregarding the different sampling or the existence more than a single disk component. Fit parameters, $\rho_0$ and $z_0$ are listed for each M-dwarf type in Table \ref{t:zfits}. The fit to the M4 dwarfs distribution is not a physical solution.
Figure \ref{f:Mdwarf:z-Mtype} shows the best fits to the densities and Figures \ref{f:Mtype-rho0} and \ref{f:Mtype-z0}, the central density and scale-height as a function of type. 

The central densities in Figure \ref{f:Mtype-rho0} are not scaled by the central volume density of dwarfs. For comparison, we plot the volume densities computed from the local number of M-dwarfs compiled in \cite{Reid08}, their Figure 7, which is based on the numbers reported in \cite{Reid04,Reid07} and \cite{Cruz07} for the latest M-dwarf types (M7-10). Table \ref{t:N0} lists the local densities as a function of M-dwarf type. We only note that both stay relatively constant with M-dwarf subtype. 

Figure \ref{f:Mtype-z0} shows the dependence of scale-height, $z_0$ with M-dwarf subtype. The progressive decline in scale-hight with subtype is in part due to the different heights sampled. Early types are a mix of thin and thick disk while later types were only found in the thin disk.

\cite{Bovy12a} argue that the different metallicity (and hence age) populations of the Milky Way all have their own distinct and unique scale-length and -height. They find that older populations are more concentrated (shorter scale-length) but the most vertically distributed (greatest scale-height). Similar results were found by  \cite{Bensby11} (using a thin/thick disk terminology) and \cite{Cheng12a}. 
% How does this compare to what we find for M-dwarfs?
The trend between scale-height and M-dwarf subtype in Figure \ref{f:Mtype-z0} fits with this general picture that there are not distinct thick and thin disks but a gradual transition in scales with stellar population. Because the majority of our fields are high Galactic latitude, we lack the statistics to perform a double fit (vertical and radial). We note however, that our use of the  \cite{Bensby11} thick- and thin-disk scale-lengths improved the vertical fits substantially over a single scale-length solution. 

% halo?

% Alternatively, \cite{Bensby11} found a thin-disk scale-length of $h$=3.8 kpc and $h$=2.0 for the thick disk, corroborated by \cite{Cheng12a} and the trend of stellar population and scale-length in \cite{Bovy12a}. We use the Bensby et al. thin disk scale-length for dwarfs found in below 500 pc and the thick disk one for those more than 500 pc out of the plane.

\begin{table}[htdp]
\caption{\label{t:zfits} The vertical profile fits to the full sample of identified M-dwarfs.}
\begin{center}
\begin{tabular}{l l l}
M-type & $\rho_0$ & $z_0$ \\
($\circ$) & ($\#/pc^3$) & (kpc) \\
\hline
\hline
0        & 4.25          & 22.85 $\pm$ 32.92\\ 
1        & 8.44          & 15.50 $\pm$ 14.33\\ 
2        & 5.29          & 12.62 $\pm$ 9.22\\ 
3        & 373.12        & 10.69 $\pm$ 14.41\\ 
4        & 53.93         & -5.17 $\pm$ 2.74\\ 
5        & 27.27         & 2.97 $\pm$ 1.39\\ 
6        & 65.78         & 1.24 $\pm$ 0.42\\ 
7        & 14185.51      & 0.28 $\pm$ 0.41\\ 
8        & 1066.36       & 0.28 $\pm$ 0.12\\ 

\hline
\end{tabular}
\end{center}
\end{table}%

\begin{figure}[htbp]
\begin{center}
	\includegraphics[width=0.5\textwidth]{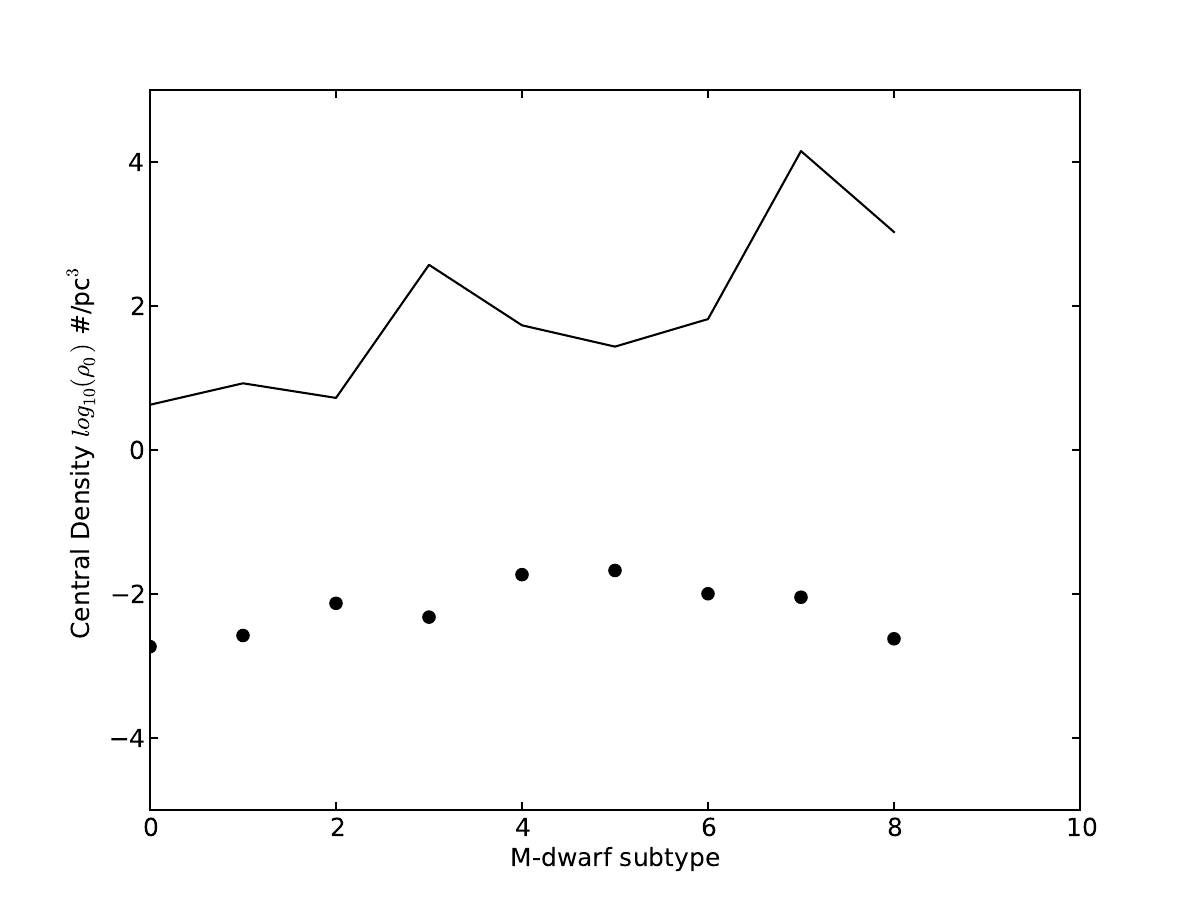}
\caption{The central density of the best fit as a function of the M-dwarf type. The black circles are the values from \protect\cite{Reid08}.  }
\label{f:Mtype-rho0}
\end{center}
\end{figure}

\begin{figure}[htbp]
\begin{center}
	\includegraphics[width=0.5\textwidth]{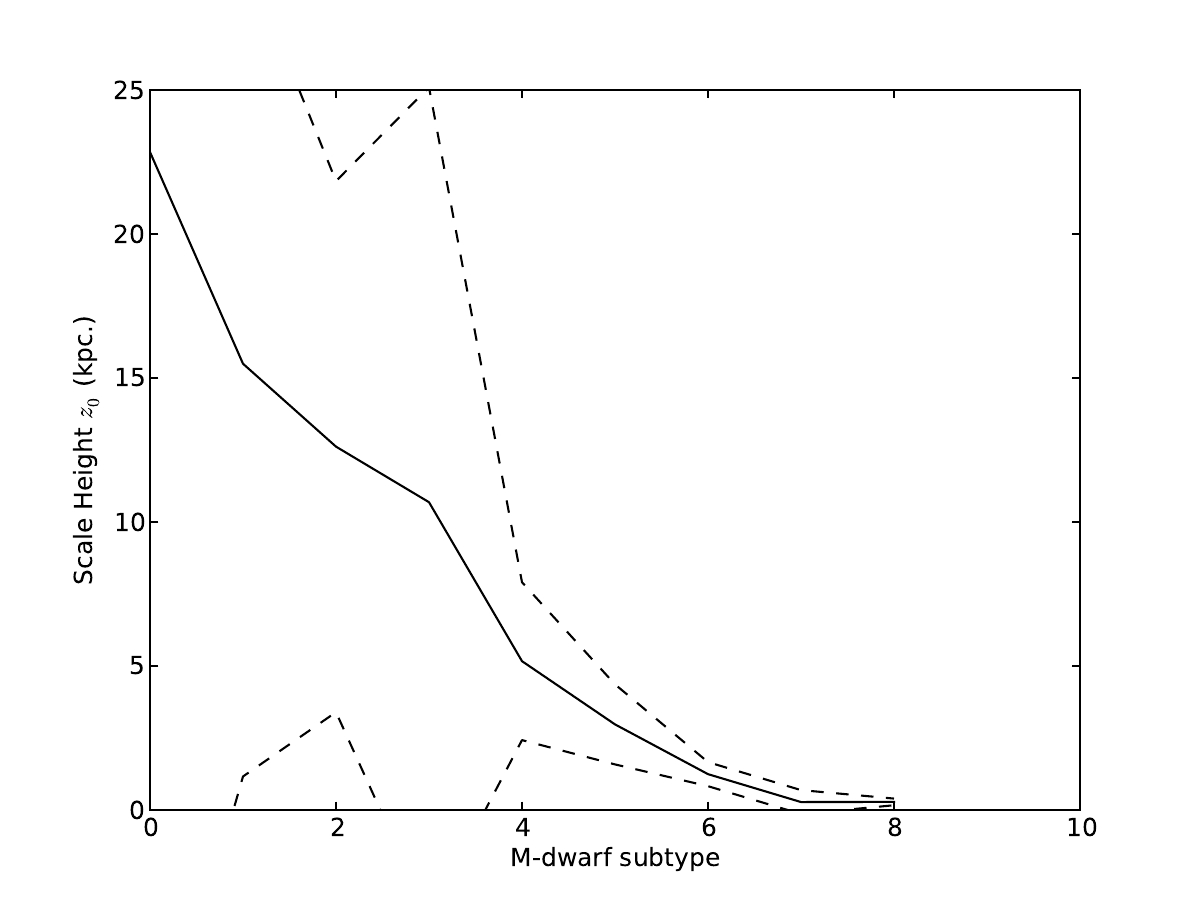}
\caption{The scale-height of the best fit as a function of the M-dwarf type. Dashed lines are the uncertainty in the fit.}
\label{f:Mtype-z0}
\end{center}
\end{figure}

\begin{figure*}[htbp]
\begin{center}
	\includegraphics[width=0.49\textwidth]{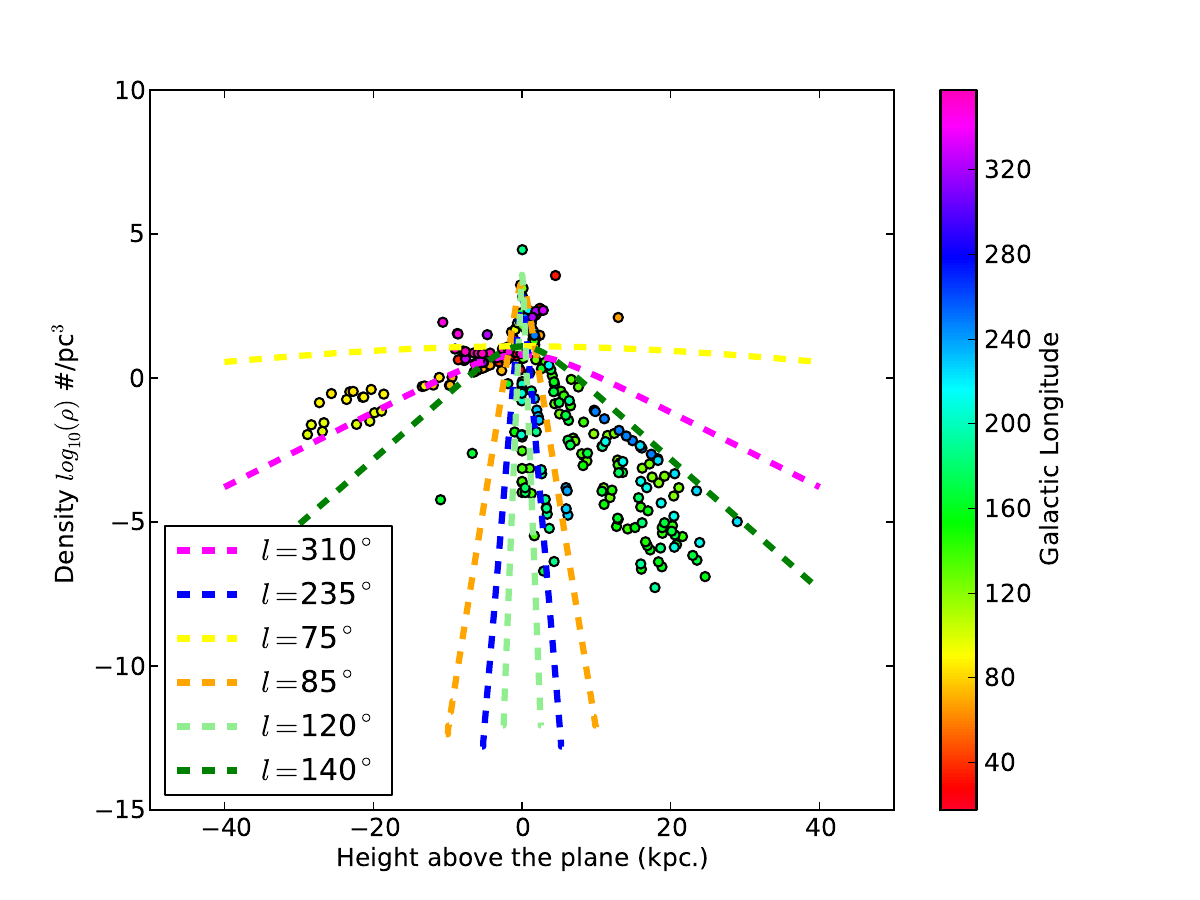}
	\includegraphics[width=0.49\textwidth]{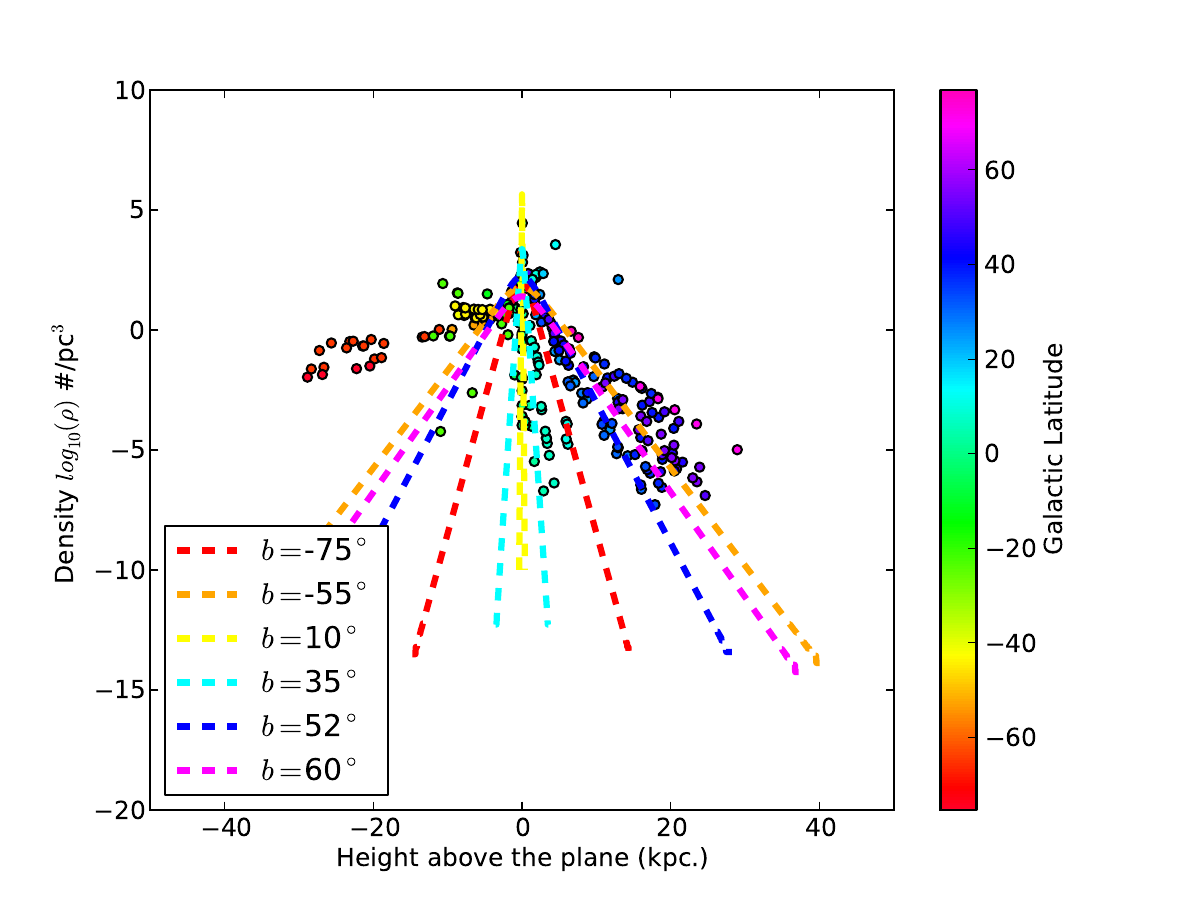}
\caption{The volume density of M-dwarfs as a function as height above the plane of the Milky Way disk. Points are color-coded by the dwarf's Galactic Longitude (left) and Latitude (right panel).}
\label{f:Mdwarf:z-MWcor}
\end{center}
\end{figure*}

There are several ``spurs'' of stars in the height-density plot of Figure \ref{f:Mdwarf:z-Mtype}. To investigate if these correspond to specific areas in the Milky Way, we replot the height-density plot for all M-dwarfs, coded with the Galactic position in Figure \ref{f:Mdwarf:z-MWcor}. Substructures become visible as a function of Galactic Longitude. For example, the upturn in M-dwarfs at 0 to -10 kpc above the plane is due to a few fields near $l \geq310^\circ$ and $b=-40^\circ$. The Longitude position strongly suggests the Milky Way Bulge as the origin of the additional M-dwarfs but the Latitude is quite low for much of the effect of the Galactic bulge.

To characterize the different scale-heights implied by the spurs in Figure \ref{f:Mdwarf:z-MWcor}, we selected Galactic longitude or latitude sections for a $sech^2$ fit, summarized in Tables \ref{t:Glatfits} and \ref{t:Glongfits}. 
%
% HALO
The $sech^2$ fits to these cuts show an expected variety of $z_0$ values. Given the vertical extent that the observations probe, contributions from stellar halo components are 
likely included (similar to the Sagittarius stream in \S \ref{s:stream}). For example, the spur of stars at 20 kpc at $l=-75^\circ$ and  $b=-75^\circ$ (yellow and red points in Figure \ref{f:Mdwarf:z-MWcor} left and right panel respectively) is at a higher density than one would expect from the different Latitude observations at $l=120^\circ$ (green point in Figure \ref{f:Mdwarf:z-MWcor}, left panel). 
However, that spur is the result of a single field and this may be simple cosmic variance in our counts. To reliably detect halo substructure or even typical scales, using M-dwarfs, one needs to image a much greater area of sky continuously (i.e., with EUCLID).

\begin{table}[htdp]
\caption{\label{t:Glatfits}
The fits to selected Galactic Latitude cuts.}
\begin{center}
\begin{tabular}{l l l}
$b$ & $\rho_0$ & $z_0$ \\
($\circ$) & ($\#/pc^3$) & (kpc) \\
\hline
\hline
%\input{../Glat_fit.tex}
%-75.00 	 & 390.84 	 & 0.77 \\ 
%-55.00 	 & 175.37 	 & 2.26 \\ 
%10.00 	 & 1256790.87 	 & 0.02 \\ 
%35.00 	 & 6675.49 	 & 0.18 \\ 
%52.00 	 & 518.97 	 & 0.56 \\ 
%60.00 	 & 10.04 	 & 2.26 \\ 
-75.00   & 140.03        & 0.77 $\pm$ 0.00 \\ 
-55.00   & 60.60         & 2.14 $\pm$ 0.07 \\ 
10.00    & 451455.50     & 0.02 $\pm$ 0.01 \\ 
35.00    & 2394.64       & 0.19 $\pm$ 0.00 \\ 
52.00    & 172.83        & 1.48 $\pm$ 0.27 \\ 
60.00    & 25.13         & 1.98 $\pm$ 0.01 \\ 
\hline
\end{tabular}
\end{center}
\end{table}%

\begin{table}[htdp]
\caption{The fits to selected Galactic Longitude cuts.}
\begin{center}
\begin{tabular}{l l l}
$l$ & $\rho_0$ & $z_0$ \\
($\circ$) & ($\#/pc^3$) & (kpc) \\
\hline
\hline
% \input{../Glong_fit.tex}
%310.00 	 & 22.61 	 & 13.56 \\ 
%235.00 	 & 2050.63 	 & 0.23 \\ 
%75.00 	 & 281.83 	 & 1.01 \\ 
%85.00 	 & 5398.05 	 & 0.54 \\ 
%120.00 	 & 10976.64 	 & 0.13 \\ 
%140.00 	 & 22.71 	 & 1.26 \\ 
310.00   & 6.45          & 6.68 $\pm$ 0.38 \\ 
235.00   & 716.35        & 0.28 $\pm$ 0.00 \\ 
75.00    & 12.64         & 32.24 $\pm$ 77.73 \\ 
85.00    & 1921.33       & 0.54 $\pm$ 0.00 \\ 
120.00   & 3875.14       & 0.13 $\pm$ 0.01 \\ 
140.00   & 12.58         & 3.84 $\pm$ 3.12 \\ 
\hline
\end{tabular}
\end{center}
\label{t:Glongfits}
\end{table}%

\begin{figure}[htbp]
\begin{center}
	\includegraphics[width=0.49\textwidth]{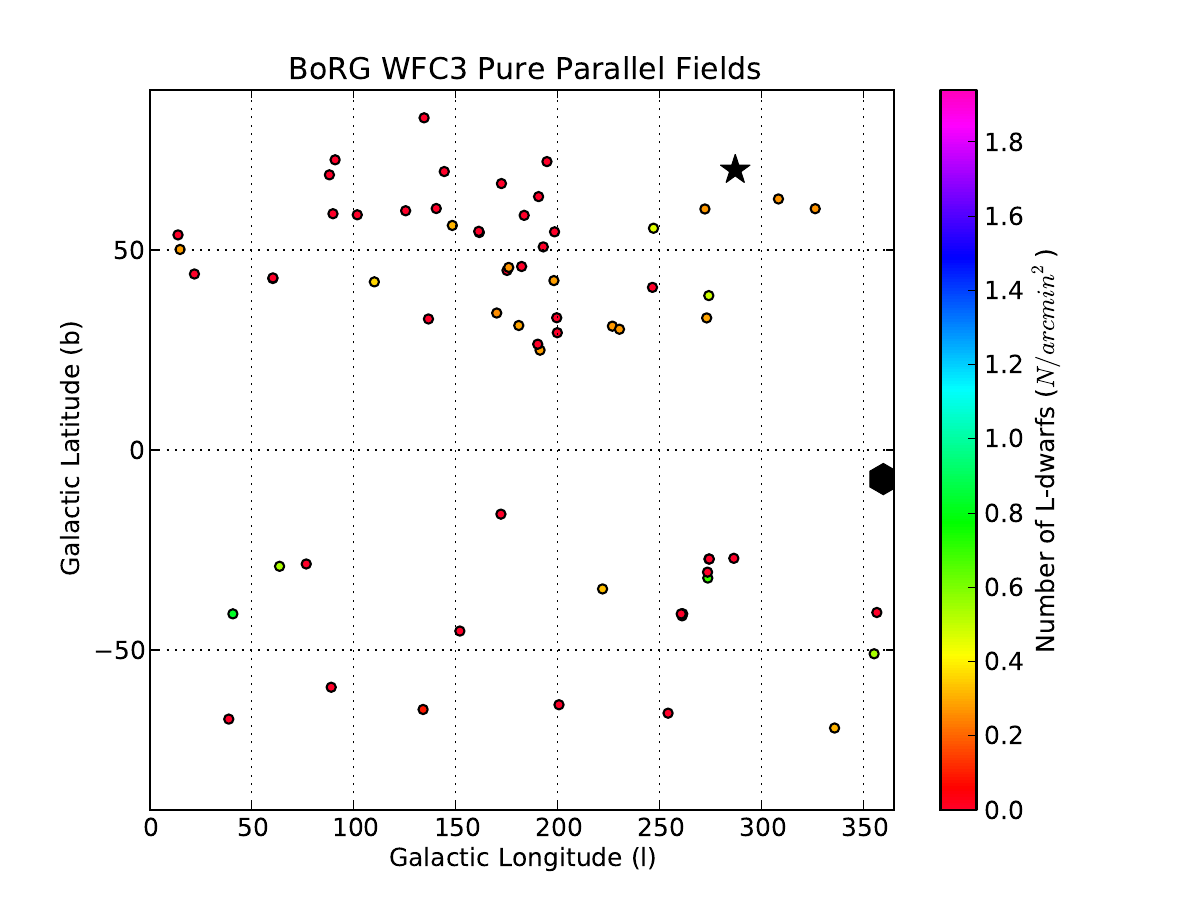}
\caption{The distribution of BoRG fields with the number of L-dwarfs indicated. Other symbols identical to Figure \ref{f:MdwarfsFields}.}
\label{f:LdwarfsFields}
\end{center}
\end{figure}

\subsection{L and T Dwarfs}
\label{s:LT}

We do find a number of stellar objects in the L and T color selection criteria from \cite{Ryan11}. 
%\subsubsection{T-dwarfs}
There is one object, brighter than 24 mag in {\em F125W} in these fields that fall into the T-dwarf box. This is not the sole T-dwarf identified 
by \cite{Ryan11} which is a 25.08 mag object. Table \ref{t:borg:Tdwarfs} lists the properties of the single $m<24$T-dwarf stars in BoRG.
%\subsubsection{L-dwarfs}
The color-color selection also identifies \nL\ L-dwarfs in the BoRG survey. Table \ref{t:borg:Ldwarfs} lists their properties.
Neither T nor L dwarfs have a reasonably well-calibrated relation between optical-near infrared color and sub-type \citep[see e.g.,][]{Burningham13} to allow for a distance estimate similar to the M-dwarfs, nor do we have the statistics in hand to infer a scale-height for these objects.

\cite{Wilkins14} reiterate that L and T-dwarfs pose a risk of contaminating high redshift ($z\sim7$) color-selected samples and show how some dwarf spectra may even masquerade as objects with emission lines. They generated their color selections similar to \cite{Ryan11} and the boxes in Figure \ref{f:jh:gj:ers} do generally agree with the derived colors. 
\cite{Wilkins14} claim HST morphological information is insufficient to cull L- and T-dwarfs from a high-redshift galaxy sample.
We have shown that dwarfs can be unequivocally identified by their stellar morphology using simple tools such as {\sc sextractor} down to 24. AB mag in {\em undithered} WFC3 data and a reliably star/galaxy separation is not only crucial for high-redshift studies but very likely possible down to much lower luminosities. 
The second issue is the surface density of these later dwarf types. For the first time, we can give an indication for random high-latitude, space-based observations based on the number of L-dwarfs found (Table \ref{t:borg:Ldwarfs}). Figure \ref{f:LdwarfsFields} shows the surface density of L-dwarfs selected by our morphology and color criteria. It shows a WFC3 exposure has $<1$ L-dwarf that can be selected against morphologically. Extrapolating to lower luminosities (including improved star/galaxy separation), the contamination problem may not be as dire as predicted.
For example, \cite{Wilkins14} assume that the local brown dwarf density (0.015 stars/pc$^3$) holds throughout all the Milky Way components (thin ad thick disks and halo). However, the spatial density drops exponentially with height above the place. If L- and T-dwarfs follow a similar scale-height as the M-dwarfs examined here ($z_0 \sim 300$ pc), then contaminating dwarf stars at 1-2 kpc distance will be very rare, even before morphological selection.

% I see you reference the new Wilkins paper. ÊI think that paper has a number of problems and I'd recommend you point them out. ÊMost importantly, Wilkins et al. explicitly assume the local number density of brown dwarfs (0.015 stars/ pc3) is true throughout the galaxy, including the thick disk and halo. ÊObviously this cannot be correct and even they admit to this. ÊBut at no point do they attempt to account for this gross error. ÊIn my 2011 paper, we argued that the number counts of BDs declines for J>25ish mag due to the exponential galactic distribution. ÊSimply put, for a T-dwarf (of the ilk they worry corrupts the hi-z counts) to have the apparent mag of the hi-z galaxies, it would have to be at 1-2 kpc (something they admit), however the number of T-dwarfs that can be at those distance is *VERY* small since their scale height seems to be 300-400 pc. ÊThat said, I don't think Wilkins et al. add anything to discussion, instead they focus on information that was known in 2003 (colors of LTdwarfs are similar to hi-z galaxies) and omit (or unaware of) half a dozen papers attempting to characterize this exact issue. ÊI would recommend you point out these failures of their paper, particularly in the places where you discuss their concern that hi-z surveys are/will be contaminated by these stars. ÊI think their conclusion is incorrect, these stars will not be a problem because they simply are not common enough --- of course there maybe a thick disk or halo population, but that will bring into question issues regarding the cooling times of these stars.

\section{Conclusions}
\label{s:concl}

% Is there still a North/South difference? What is the scale-height in M-dwarfs? Does it depend on the subtypes? 

\begin{itemize}
\item Stars can be selected best using {\sc sextractor}'s half-light radius ($r\_{50}$), corroborating previous identifications (Figures \ref{f:r50} and \ref{f:borg:r50}).
\item M-dwarfs can be reliably sub-typed using a near-infrared and optical color (Figure \ref{f:typing}).
\item Relatively more early-type M-dwarfs are identified in Northern fields as opposed to Southern Galactic ones. This confirms the dichotomy \cite{Pirzkal09} noted (Table \ref{t:ns}).
\item We report an over-density in one of our fields, borg\_1230+0750, consistent with the M-dwarf members of the Sagittarius stream at a distance of 20-25 kpc. (Figure \ref{f:sag:hist:d}).
\item A naive, single-component fit of the vertical distribution of M-dwarfs shows a steady decline of scale-height, $z_0$, with M-dwarf sub-type (Figure \ref{f:Mtype-z0}).
\item The grouping of spurs of stars at different Galactic coordinates strongly hints at substructure in the Milky Way disk/halo of M-dwarfs (Figure \ref{f:Mdwarf:z-MWcor}).
\end{itemize}

\section{Future Work}

Future work with WFC3 (pure-parallel) programs on Milky Way Dwarfs will use the ever-increasing number of sight-lines available to 
build a comprehensive catalog of M,L and T dwarfs belonging to the thin and thick disks and stellar halo of our Milky Way. 
New pure-parallel observational programs with HST/WFC3 are needed to expand the number of sight-lines and improve statistics.
At lower Galactic latitudes, this could easily be achieved with lower integration times than the BoRG or HIPPIES programs.
Once such a proposed program is completed fully, the total tally of M-dwarfs should give a first idea of the size and shape of the Milky Way disks, thin 
and thick, and depending on the total statistics, the Galactic Halo. 

In parallel, spectroscopic identification if Milky Way dwarf stars by the HST-3D near-infrared grism spectra \citep[e.g.,][]{Brammer12}.
The ongoing imaging and grism campaigns on the CANDELS fields \citep{Koekemoer11,Grogin11} will improve the calibration of the photometric 
identifications of Galactic Dwarfs in the pure-parallel surveys as well as add additional statistics to the star counts.

%\subsection{EUCLID}

The general technique of morphologically identifying stars in space-based imaging, identifying dwarfs from near-infrared colors and  subsequently sub-type these using an optical-near-infrared color will be perfectly suited for the EUCLID mission \citep{Laureijs11}. The wide-band optical filter can be used to morphologically identify the stars and the three near-infrared filters to hone in on the Galactic dwarfs. Finally, the optical-nir color and complementary grism spectroscopy can then be used to subtype these dwarfs. 
With the wide survey area to faint point source sensitivity (all-sky 24 mag, 26 mag in deep fields) the EUCLID mission will yield an accurate measure of the shape and size of our Milky Way in sub-solar dwarfs as well as an accurate census of Halo substructure similar to the Sagittarius stream. Limiting factors may be the 
PSF in the near-infrared, leading to confusion issues and the width of the optical filter complicating subtyping. However, the two 
independent-angle grism spectra of all the objects in the field should unequivocally identify the Galactic dwarf stars.

%\subsection{JWST}

Milky Way dwarfs are unlikely to be the focus of James Webb Space Telescope observations but they will certainly feature in them. The ubiquity of M-dwarfs in the BoRG fields points to
their possible use as fine guidance for JWST image and NIRspec imaging/MOS registration.

\acknowledgments

The authors would like tot thank N. Reid and K. Cruz for their help with the local density of M-dwarfs and R. Benjamin for useful discussions on Galactic substructures.
The lead author thanks the European Space Agency for the Fellowship program and its support.
We acknowledge the support of HST Archival grant number AR-12134, General Observer Grants GO-11700, GO-12572, and G0-12905
and the European Space Agency for support of this work.
MT  was partially supported by the European Commission through the Marie
Curie Career Integration Fellowship PCIG12-GA-2012-333749"
This work is based in part on observations taken by the CANDELS Multi-Cycle Treasury Program with the NASA/ESA HST, which is operated by the Association of Universities for Research in Astronomy, Inc., under NASA contract NAS5-26555.
This research has made use of the NASA/IPAC Extragalactic Database (NED) which is operated by the Jet Propulsion Laboratory, California Institute of Technology, under contract with the National Aeronautics and Space Administration. 
This research has made use of NASA's Astrophysics Data System.

%
%\bibliographystyle{apj} 
%\bibliography{/Users/bholwerd/Desktop/Science/Bib/Bibliography}
\newpage

{\Large {\bf ERRATUM: }``Milky Way Red Dwarfs in the BoRG Survey;\\
Galactic scale-height and the distribution of dwarfs stars in WFC3 imaging'' (2014, ApJ, 788, 77)}

\section*{Abstract}
In the catalog of M-dwarfs presented in \cite{Holwerda14}, there is an issue with the conversion from celestial coordinates to Galactic ones, done with {\sc pyephem} a wrapper around a trusted and vetted library {\em ephermis}. Here we present the corrected coordinates (using {\sc AstroPy}) and distances based on AB magnitudes. We have amended the tables and Figures accordingly. 
The relation between vertical scale-height ($z_0$) and M- dwarf subtype found in H14 is no longer present. We find a scale-height of 600 pc for all types, in part due to the presence of a second Galactic structural component.

\section{Mapping the M-dwarf Distribution}

We re-computed the Galactic coordinates for the BoRG fields using the {\sc AstroPy} package \citep{Astropy-Collaboration13a} and the number of M-dwarfs in each, shown in the new Figure \ref{f:MdwarfsFields}. To estimate the distances to each M-dwarf, we compute the distance modulus from the inferred sub-type (and hence absolute magnitude) and the apparent magnitude in $J_{F125W}$ from \cite{Hawley02}, converted from Vega magnitudes to AB. 
We compute the Galactic radius and height above the plane for all M-dwarfs, based on their Galactic longitude, latitude and the inferred photometric distances, assuming the position of the Sun 27 pc. above the plane and 8.5 kpc from the Galactic center. 

The net difference with H14 is that the vertical distribution observed for these M-dwarfs is much closer to the exponential drop-off one would expect for fields out of the plane of the Galaxy.
%\clearpage
%
\setcounter{figure}{21}
\begin{figure*}[b]
\begin{center}
	\includegraphics[width=\textwidth]{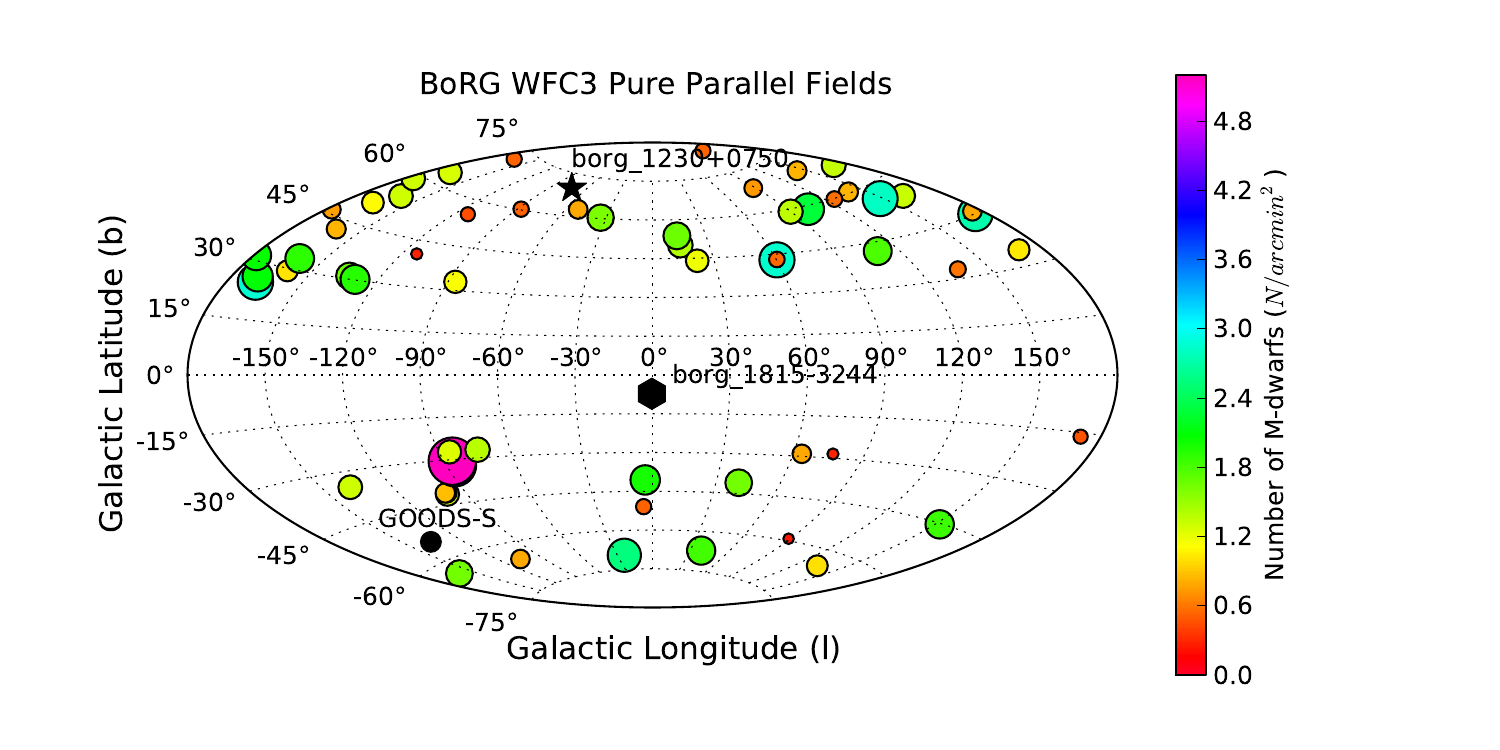}
\caption{The Aitoff projection of the distribution of BoRG fields with the surface density of M-dwarfs of all types indicated. One field, borg\_1230+0750 (star) stands out with 22 M-dwarfs ($\sim20$ stars/arcmin$^2$). We discard borg\_1815-3244 (black circle) for its low latitudes and line-of-sight through the plane of the disk and close of the center of the bulge.}
\label{f:MdwarfsFields}
\end{center}
\end{figure*}
\clearpage

\setcounter{figure}{23}

\begin{figure}[htbp]
\begin{center}
	\includegraphics[width=0.5\textwidth]{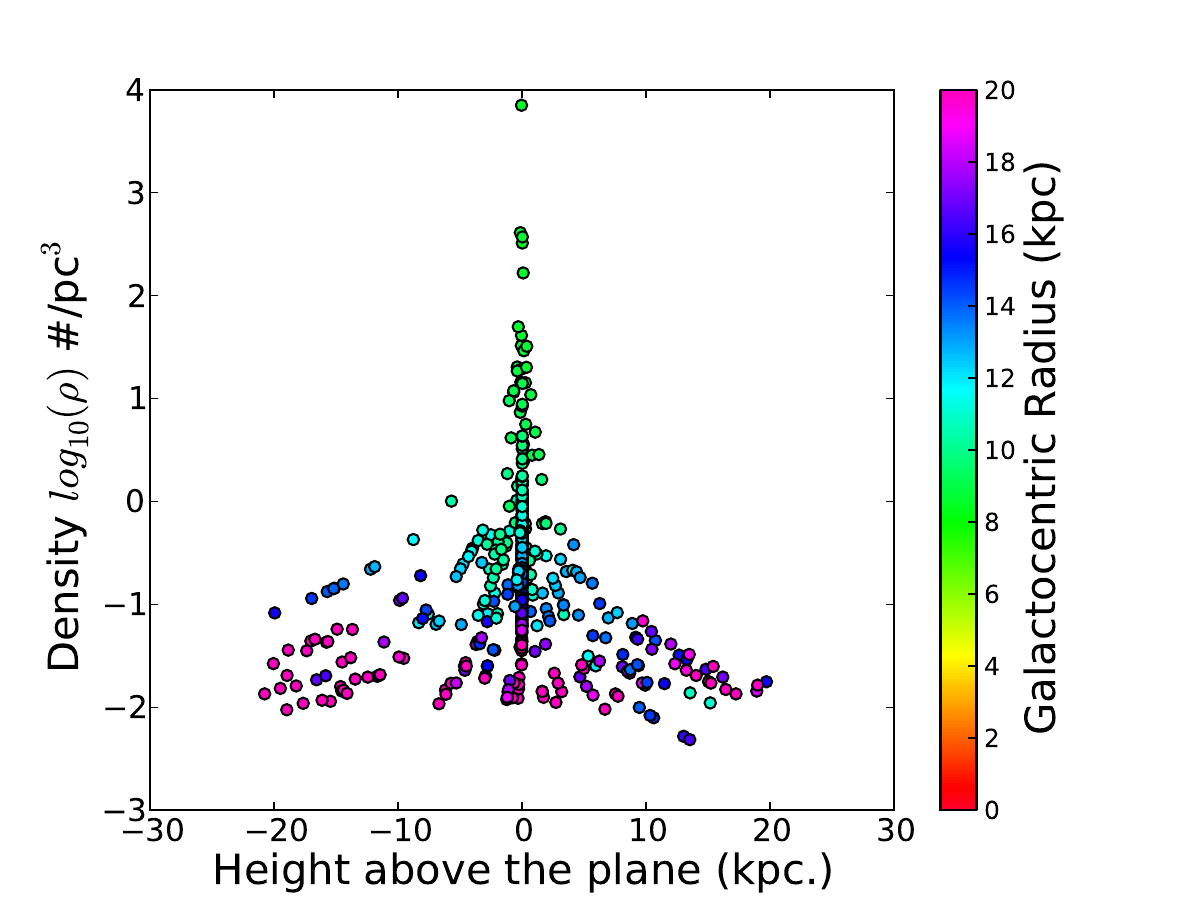}
\caption{The volume density of M-dwarfs as a function as height above the plane of the Milky Way disk. The volume densities were normalized with a scale-length of 2.6 kpc and a depth of 1pc at each star's position. Points are color-coded by the inferred Galactic radius. The majority of M-dwarfs are found close to the plane of the disk.  }
\label{f:Mdwarf:Rz}
\end{center}
\end{figure}

Figure \ref{f:Mdwarf:Rz} shows the vertical distribution of M-dwarfs color-coded by their Galactocentric radius. There is a clear disk component with all the M-dwarfs are $\sim6-10$ kpc from the center of the Milky Way and another component which shows a gradient with radius and is well above the plane of the disk. We now know this component to be the Halo part of the vertical distribution (van Vledder et al {\em accepted}) but it is sometimes treated as the thick disk by other authors \cite[e.g.,][]{Ryan15}. This second component is much clearer in the corrected Galactic positions compared to H14.
 The majority of stars close to the plane of the disk are $\sim 8 $kpc from the center of the Galaxy. This is because the majority of BoRG fields are pointing out of the disk at an angle away from the Galactic Center (to avoid confusion with Galactic Objects) and thus probe the vertical extent of the disk close to the position of the Sun. 

We assume there is no need to correct the densities in this part of the disk for the exponential decline with radius. Figure \ref{f:Mdwarf:Rz} does show a radial dependence of the outer parts of the vertical distribution on the radius. 
Our assumption is that the scale-height does not change significantly with radius, which is observed in external galaxies seen edge-on \citep[][Streich et al {\em in preparation}]{Comeron11a,Comeron11b,Comeron11c}.

\begin{figure}[htbp]
\begin{center}
	\includegraphics[width=0.5\textwidth]{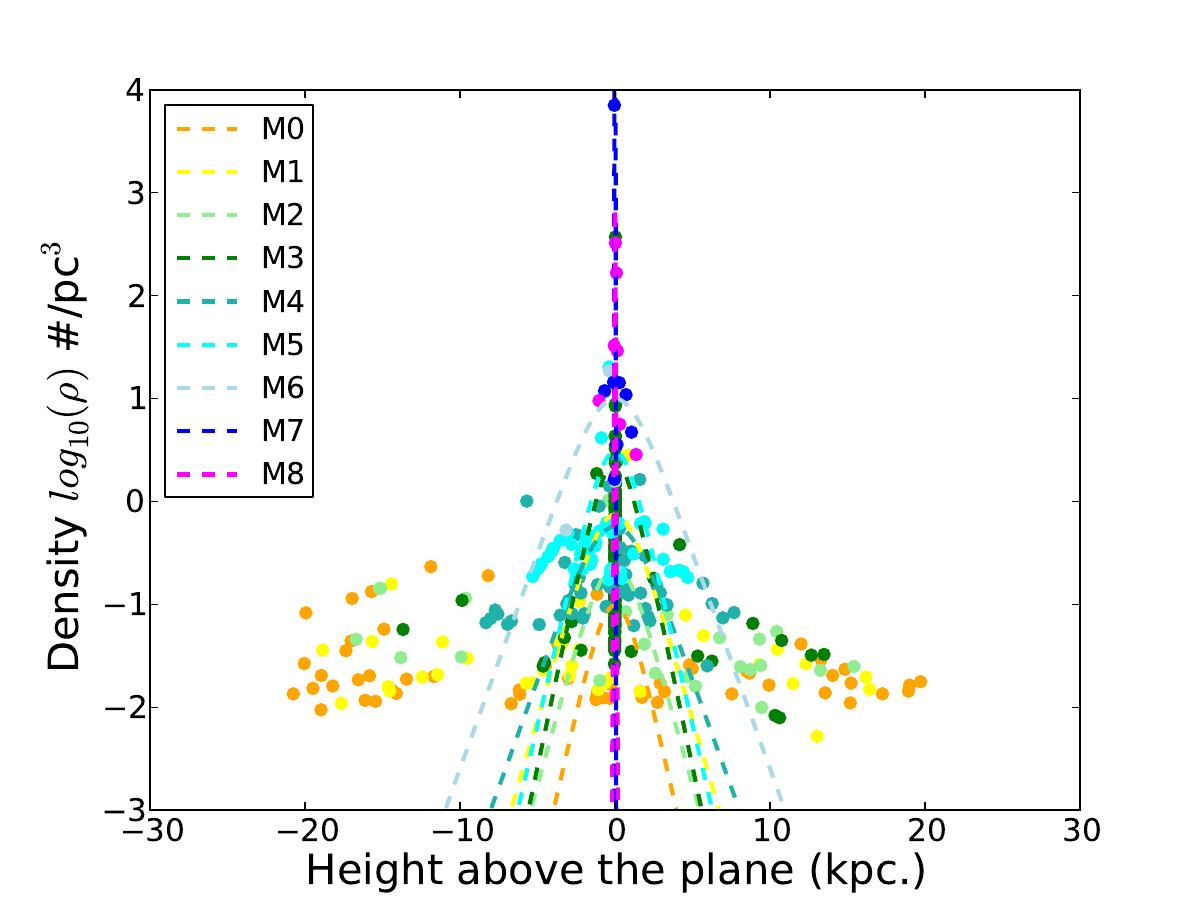}
\caption{The volume density of M-dwarfs as a function as height above the plane of the Milky Way disk. Points and fits are color-coded by the M-dwarf sub-type.}
\label{f:Mdwarf:z}
\end{center}
\end{figure}

We assume that the Galactic disk has the following parametric shape:
\begin{equation}
\rho(R,z) = \rho_0 ~ exp(-R/h) sech^2\left({z \over 2 \times z_0}\right),
\end{equation}
\noindent where $\rho(R,z)$ is the dwarf number density in a point in the disk, $\rho_0$ is the central number density, $R$ is Galactocentric radius, $h$ is the scale-length, $z$ is height above the plane, and $z_0$ is the exponential scale-height of the disk. 
Figure \ref{f:Mdwarf:z} shows how the later-type M-dwarfs (M4-8) in the BoRG fields are concentrated in the thin disk and the earlier types (M0-3) probe both thin disk and the other component. We fit the $sech^2$ distribution to each photometric subtype, as well as early-, late- and all-types of M-dwarfs, summarized in the new Table \ref{t:zfits} and the new Figures \ref{f:Mtype-rho0} and \ref{f:Mtype-z0}.

% DIFFERENCES WITH H14
Figures \ref{f:Mtype-rho0} and \ref{f:Mtype-z0} are significantly different from the versions in H14 in two ways:
the relation between vertical scale-height ($z_0$) and M- dwarf subtype found in H14 is no longer present in Figure \ref{f:Mtype-z0}.
Secondly, the densities inferred are much closer to the local ($< 20$ pc of the Sun) \cite{Reid04} and \cite{Cruz07} values (Figure \ref{f:Mtype-rho0}).
% The central densities in Figure \ref{f:Mtype-rho0} are scaled to the central volume density of dwarfs in the center of the Milky Way; assuming a typical radius of $\sim12$ kpc for our measurement and a scale-length of 2.6 kpc. For comparison, we plot the volume densities computed from the local number of M-dwarfs compiled in \cite{Reid08}, their Figure 7, which is based on the numbers reported in \cite{Reid04,Reid07} and \cite{Cruz07} for the latest M-dwarf types (M7-10) locally ($< 20$ pc of the Sun). Table \ref{t:N0} lists the local densities as a function of M-dwarf type. We only note that both stay relatively constant with M-dwarf subtype. We have scaled these densities to a central density assuming a radius of 8.5 kpc for the position of the Sun and a scale-length of 2.6. The resulting central densities for all M-dwarf types is very similar for all subtypes. 
We note there is an offset of $\sim 0.1$ dex between our central density and the one implied by Reid et al. The likely culprit is the additional numbers of M-dwarfs in the secondary component well above the plane of the disk, i.e., we are including the density of Halo stars in our disk fit. A 0.1 dex offset suggest of order 10\% of the stars in are, in fact, Halo stars (see also van Vledder et al. {\em accepted}).

%Figure \ref{f:Mtype-z0} shows the dependence of scale-height, $z_0$ with M-dwarf subtype. There is no clear relation between scale-height and M-dwarf subtype in Figure \ref{f:Mtype-z0}. This may be because the photometric sub-typing we do for these objects is not accurate enough to subtype M-dwarfs accurately or because the completeness changes dramatically with M-dwarf subtype. We note that, using this same data, van Vledder et al. {\em submitted} find that improved disk+halo models do not show a dependence on M-dwarf photometric subtype population.

The values in Figure \ref{f:Mtype-z0} and Table \ref{t:zfits} are still too high to be consistent with previous work on the vertical scale of stars in the thin disk. The vertical $sech^2$ profile does not describe the density distribution well above $\sim500$pc and this new distributions highlights the need to include another component well above and in addition to the disk. 

\setcounter{table}{9}

\begin{table}[htdp]
\caption{\label{t:zfits} The vertical profile fits to the full sample of identified M-dwarfs. Values are scaled so to be compatible with exponential fits to the Galactic profiles ($\rho_0/4$ and $z_0/2$).}
\begin{center}
\begin{tabular}{l l l}
M-type & $\rho_0$ & $z_0$ \\
($\circ$) & ($\#/pc^3$) & (kpc) \\
\hline
\hline
0        & 0.02          & 0.67 $\pm$ 0.73\\ 
1        & 0.19          & 0.84 $\pm$ 0.89\\ 
2        & 0.06          & 0.80 $\pm$ 1.31\\ 
3        & 0.60          & 0.61 $\pm$ 12.36\\ 
4        & 0.14          & 1.04 $\pm$ 0.67\\ 
5        & 0.77          & 0.67 $\pm$ 0.74\\ 
6        & 2.49          & 1.03 $\pm$ 5.42\\ 
7        & \dots      & \dots\\
8        & 181.04        & -0.01 $\pm$ 0.02\\ 
\hline
\end{tabular}
\end{center}
\end{table}%

\begin{figure}[htbp]
\begin{center}
	\includegraphics[width=0.5\textwidth]{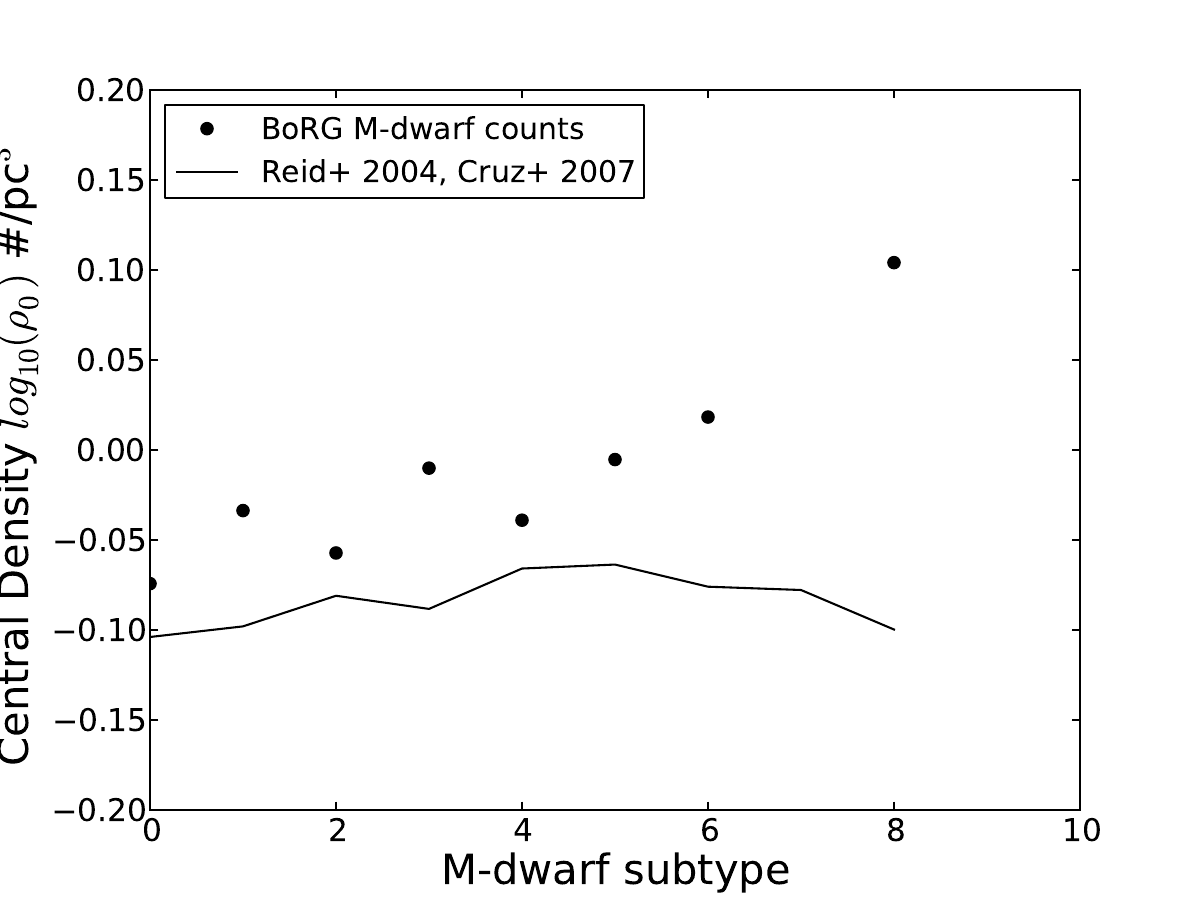}
\caption{The in-plane central density -- the number of M-dwarfs belonging to the disk at the center of the Milky Way-- for the best fit to the vertical distribution of M-dwarfs as a function of the M-dwarf subtype. The black circles are the values from \protect\cite{Reid08} for the immediate Solar neighborhood ($<20$ pc), renormalized to the center of the Milky Way. }
\label{f:Mtype-rho0}
\end{center}
\end{figure}

\begin{figure}[htbp]
\begin{center}
	\includegraphics[width=0.5\textwidth]{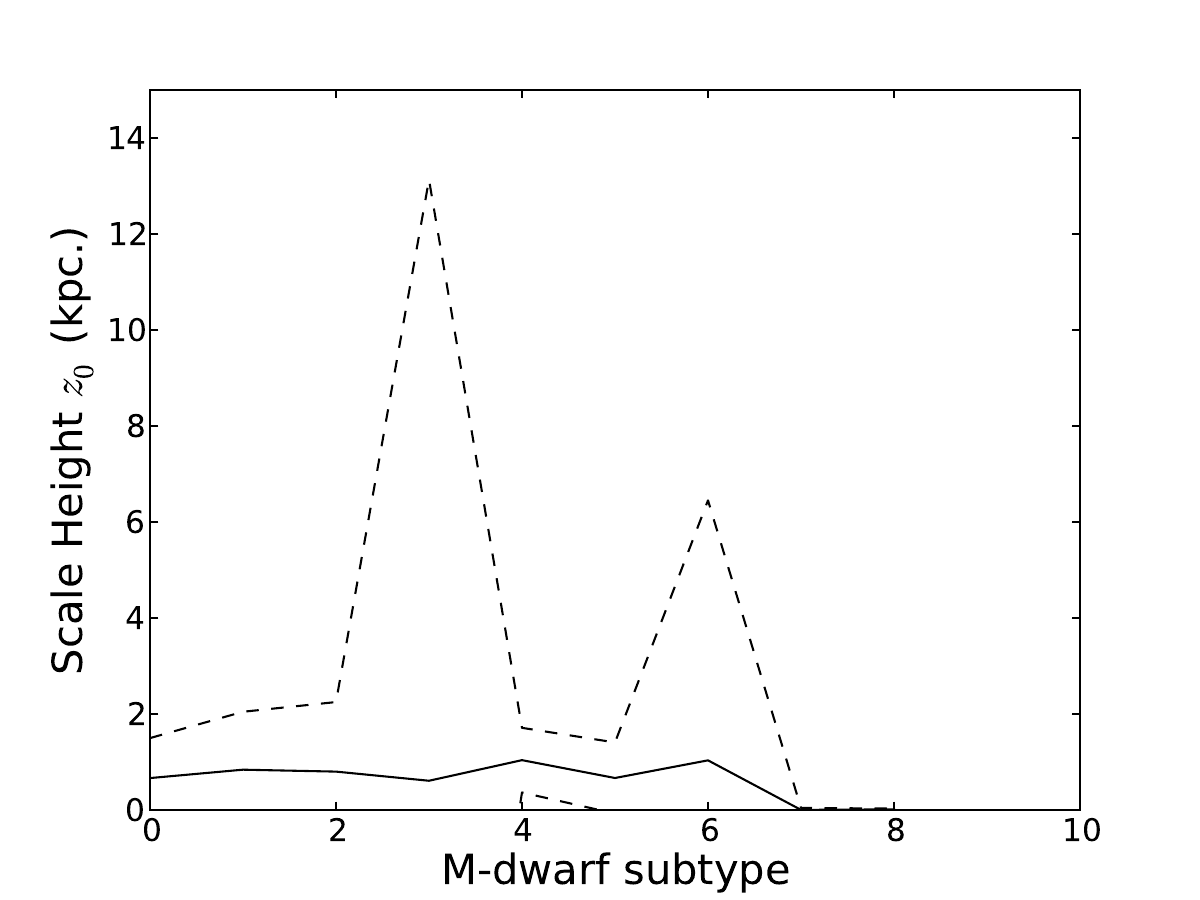}
\caption{The scale-height ($z_0/2$) of the best vertical fit as a function of the M-dwarf subtype. Dashed lines are the uncertainty in the fit. A constant value of 600 pc would be consistent with most of these fits. This is $\sim10\times$ greater than expected from previous work.}
\label{f:Mtype-z0}
\end{center}
\end{figure}

\begin{figure*}
\begin{center}
	\includegraphics[width=\textwidth]{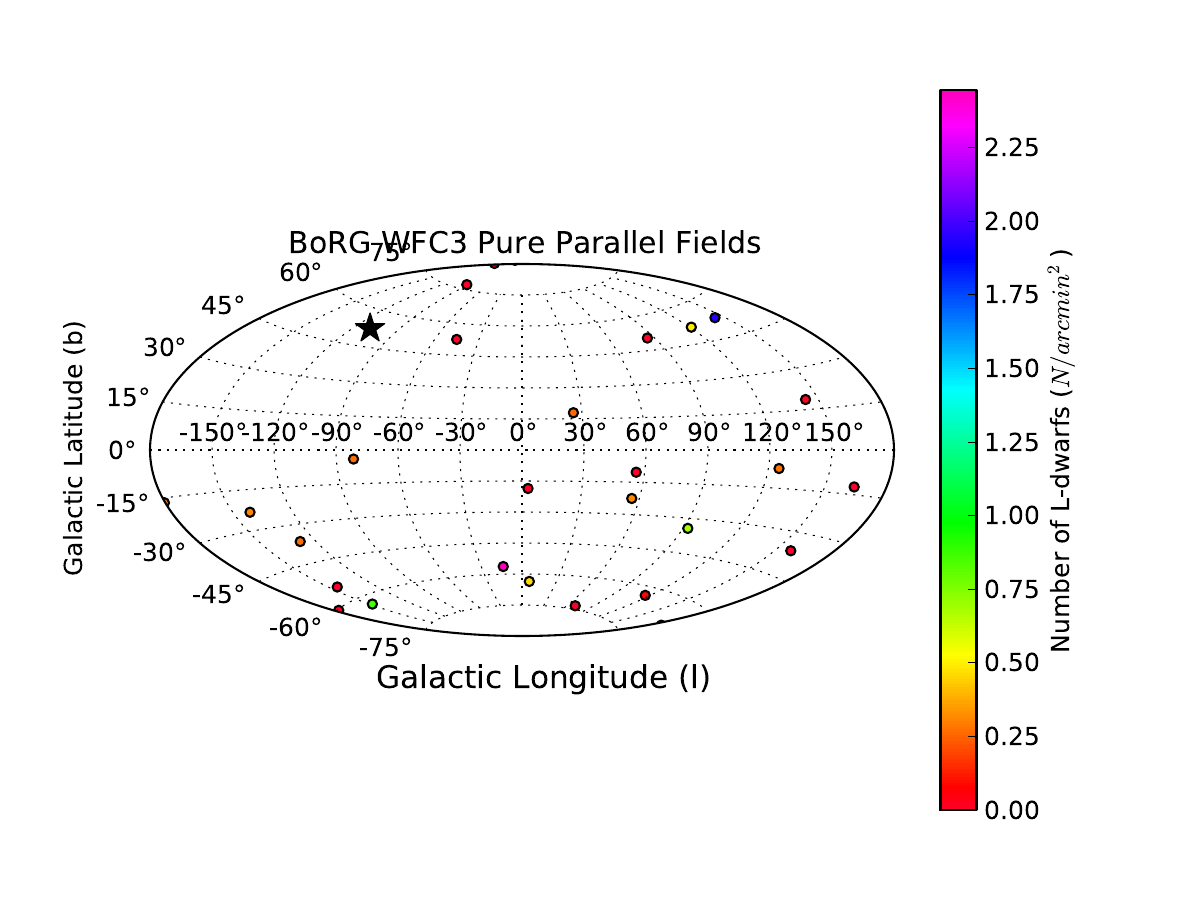}
\caption{The distribution of BoRG fields with the number of L-dwarfs indicated. Other symbols identical to Figure \ref{f:MdwarfsFields}.}
\label{f:LdwarfsFields}
\end{center}
\end{figure*}

For completeness, we include Figure \ref{f:LdwarfsFields} which shows the surface density of L-dwarfs selected by our morphology and color criteria.

\section{Conclusions}
\label{s:concl}

% Is there still a North/South difference? What is the scale-height in M-dwarfs? Does it depend on the subtypes? 
The majority of conclusions for \cite{Holwerda14} remain the same, only the last two change:

\begin{itemize}
\item The secondary component visible in the vertical distribution of M-dwarfs is likely the Halo and not the thick disk of the Milky Way (Figures \ref{f:Mdwarf:Rz} and \ref{f:Mdwarf:z}).
\item A naive, single-component fit of the vertical distribution of M-dwarfs shows no dependence on M-dwarf subtype and a scale-height for all M-dwarf subtypes that is still too high to be consistent with previous measures of the scale-height: $z_0\sim600$ pc, (Figure \ref{f:Mtype-z0}), a result of the second structural component.
\end{itemize}

\acknowledgments

The authors would like to thank the referee, K. Cruz for her help with the erratum.
This research made use of Astropy, a community-developed core Python package for Astronomy \citep{Astropy-Collaboration13a}. This research made use of matplotlib, a Python library for publication quality graphics \citep{Hunter07}. PyRAF is a product of the Space Telescope Science Institute, which is operated by AURA for NASA. This research made use of SciPy \citep{scipy}.

\clearpage
\newpage

 %\begin{landscape}
% BORG13
\addtocounter{table}{-1} 
\begin{table*}
\caption{\label{t:Mdwarfs}The M-dwarfs found in the BoRG13 data with photometry, corrected Galactic coordinates and positions.}
\rotatebox{90}{
\tiny{
\begin{tabular}{l l l l l l l l l l l l l l l l l l l l l l}
Field  	& RA  	& DEC 	& $l$ 	& $b$ 	& $m_{F098W}$ 	& $m_{F125W}$   	& $m_{F125W}$ 	&  $m_{F606W}$  	&  Mtype 	& M-m   	& Dist 	&  Height &  Radius  	& Volume &  Area \\ 
		& J2000 	& J2000 	& $\circ$ 	& $\circ$ 	& mag 			& mag 			& mag			& mag			& 		& mag 	& kpc.	& kpc	& kpc.	& kpc$^2$	& arcmin$^2$			 \\
\hline
\hline
borg\_0110-0224.551 & 17.501759 & -2.418265 & 133.918256 & -64.893095  & 23.82 $\pm$ 0.04 & 23.46 $\pm$ 0.02 & 23.28 $\pm$ 0.03 & 25.17 $\pm$ 0.09 & 2.04 $\pm$ 0.32 & 15.59 & 13.10 $\pm$ 1.22 & 9.47 & 14.06 & 100.63 & 10.88 \\ 
borg\_0110-0224.719 & 17.503586 & -2.415663 & 133.921454 & -64.890220  & 24.33 $\pm$ 0.05 & 23.97 $\pm$ 0.03 & 23.87 $\pm$ 0.04 & 25.83 $\pm$ 0.14 & 2.56 $\pm$ 0.50 & 15.84 & 14.71 $\pm$ 1.81 & 10.62 & 14.74 & 126.76 & 10.88 \\ 
borg\_0110-0224.820 & 17.528020 & -2.411389 & 133.976421 & -64.881854  & 19.68 $\pm$ 0.00 & 19.51 $\pm$ 0.00 & 19.45 $\pm$ 0.00 & 22.77 $\pm$ 0.01 & 7.30 $\pm$ 0.04 & 7.70 & 0.35 $\pm$ 0.01 & 0.28 & 8.65 & 0.07 & 10.88 \\ 
borg\_0110-0224.1016 & 17.553100 & -2.408888 & 134.033565 & -64.875102  & 24.54 $\pm$ 0.09 & 23.90 $\pm$ 0.04 & 23.72 $\pm$ 0.05 & 25.39 $\pm$ 0.13 & 1.30 $\pm$ 0.46 & 16.29 & 18.10 $\pm$ 2.55 & 13.04 & 16.18 & 191.93 & 10.88 \\ 
borg\_0110-0224.1414 & 17.559120 & -2.400783 & 134.044244 & -64.866085  & 24.08 $\pm$ 0.06 & 23.71 $\pm$ 0.06 & 23.44 $\pm$ 0.04 & 24.40 $\pm$ 0.05 & -1.43 $\pm$ 0.27 & 16.37 & 18.79 $\pm$ 456.20 & 13.53 & 16.48 & 206.91 & 10.88 \\ 
borg\_0110-0224.1529 & 17.564067 & -2.398369 & 134.054731 & -64.862858  & 24.22 $\pm$ 0.13 & 23.91 $\pm$ 0.04 & 23.78 $\pm$ 0.05 & 25.76 $\pm$ 0.17 & 2.51 $\pm$ 0.60 & 15.78 & 14.33 $\pm$ 2.02 & 10.33 & 14.59 & 120.39 & 10.88 \\
\hline
\end{tabular}
}
}
\end{table*}

\addtocounter{table}{-1} 
\begin{table*}
\caption{Table \ref{t:Mdwarfs} -- {\em continued}}
\rotatebox{90}{
\tiny{
\begin{tabular}{l l l l l l l l l l l l l l l l l l l l l l}
Field  	& RA  	& DEC 	& $l$ 	& $b$ 	& $m_{F098W}$ 	& $m_{F125W}$   	& $m_{F125W}$ 	&  $m_{F606W}$  	&  Mtype 	& M-m   	& Dist 	&  Height &  Radius  	& Volume &  Area \\ 
		& J2000 	& J2000 	& $\circ$ 	& $\circ$ 	& mag 			& mag 			& mag			& mag			& 		& mag 	& kpc.	& kpc	& kpc.	& kpc$^2$	& arcmin$^2$			 \\
\hline
\hline
borg\_0751+2917.46 & 117.702830 & 29.265557 & 191.372718 & 24.948643  & 20.53 $\pm$ 0.00 & 20.06 $\pm$ 0.00 & 19.91 $\pm$ 0.00 & 23.47 $\pm$ 0.04 & 7.79 $\pm$ 0.15 & 8.03 & 0.40 $\pm$ 0.01 & -0.05 & 8.87 & 0.03 & 3.47 \\ 
borg\_0751+2917.153 & 117.715495 & 29.272066 & 191.369754 & 24.961181  & 20.45 $\pm$ 0.00 & 20.10 $\pm$ 0.00 & 20.12 $\pm$ 0.00 & 23.77 $\pm$ 0.06 & 8.68 $\pm$ 0.19 & 7.78 & 0.36 $\pm$ 0.03 & -0.04 & 8.83 & 0.02 & 3.47 \\ 
borg\_0751+2917.271 & 117.695308 & 29.275232 & 191.360309 & 24.945472  & 20.92 $\pm$ 0.00 & 20.68 $\pm$ 0.00 & 20.43 $\pm$ 0.00 & 22.93 $\pm$ 0.03 & 3.87 $\pm$ 0.09 & 11.45 & 1.95 $\pm$ 0.08 & -0.36 & 10.27 & 0.71 & 3.47 \\ 
borg\_0751+2917.354 & 117.717431 & 29.277303 & 191.364861 & 24.964436  & 21.34 $\pm$ 0.00 & 20.93 $\pm$ 0.00 & 20.77 $\pm$ 0.00 & 24.78 $\pm$ 0.13 & 9.30 $\pm$ 0.45 & 8.61 & 0.53 $\pm$ 0.02 & -0.08 & 8.98 & 0.05 & 3.47 \\ 
borg\_0751+2917.464 & 117.698929 & 29.278551 & 191.357935 & 24.949517  & 23.40 $\pm$ 0.02 & 23.08 $\pm$ 0.01 & 23.01 $\pm$ 0.02 & 25.27 $\pm$ 0.14 & 3.67 $\pm$ 0.48 & 13.85 & 5.88 $\pm$ 1.56 & -1.13 & 13.83 & 6.46 & 3.47 \\ 
borg\_0751+2917.509 & 117.709001 & 29.280669 & 191.358778 & 24.958522  & 21.71 $\pm$ 0.01 & 21.56 $\pm$ 0.01 & 21.38 $\pm$ 0.01 & 26.96 $\pm$ 27.23 & 14.56 $\pm$ 92.40 & 119.67 & \dots  & \dots & \dots & \dots  & 3.47 \\
\hline
\end{tabular}
}
}
\end{table*}

\addtocounter{table}{-1} 
\begin{table*}
\caption{Table \ref{t:Mdwarfs} -- {\em continued}}
\rotatebox{90}{
\tiny{
\begin{tabular}{l l l l l l l l l l l l l l l l l l l l l l}
Field  	& RA  	& DEC 	& $l$ 	& $b$ 	& $m_{F098W}$ 	& $m_{F125W}$   	& $m_{F125W}$ 	&  $m_{F606W}$  	&  Mtype 	& M-m   	& Dist 	&  Height &  Radius  	& Volume &  Area \\ 
		& J2000 	& J2000 	& $\circ$ 	& $\circ$ 	& mag 			& mag 			& mag			& mag			& 		& mag 	& kpc.	& kpc	& kpc.	& kpc$^2$	& arcmin$^2$			 \\
\hline
\hline
borg\_0922+4505.861 & 140.412665 & 45.099140 & 175.122919 & 44.901473  & 24.25 $\pm$ 0.05 & 24.00 $\pm$ 0.04 & 23.73 $\pm$ 0.04 & 25.49 $\pm$ 0.16 & 1.31 $\pm$ 0.57 & 16.39 & 18.93 $\pm$ 2.76 & 1.64 & 21.91 & 70.52 & 3.66 \\ 
borg\_0926+4000.93 & 141.367553 & 39.998027 & 182.329464 & 45.855741  & 24.01 $\pm$ 0.16 & 23.79 $\pm$ 0.07 & 23.64 $\pm$ 0.08 & 24.90 $\pm$ 0.12 & -0.01 $\pm$ 0.49 & 16.45 & 19.46 $\pm$ 3.44 & -0.76 & 22.05 & 81.12 & 3.98 \\ 
borg\_0926+4000.147 & 141.379236 & 40.002023 & 182.323599 & 45.864650  & 23.46 $\pm$ 0.05 & 22.94 $\pm$ 0.04 & 22.65 $\pm$ 0.04 & 25.98 $\pm$ 0.36 & 6.52 $\pm$ 1.23 & 11.13 & 1.69 $\pm$ 0.69 & -0.04 & 9.67 & 0.61 & 3.98 \\ 
borg\_0926+4000.399 & 141.403410 & 40.012853 & 182.307772 & 45.883054  & 23.81 $\pm$ 0.06 & 23.69 $\pm$ 0.06 & 23.62 $\pm$ 0.07 & 24.99 $\pm$ 0.14 & 0.63 $\pm$ 0.52 & 16.08 & 16.45 $\pm$ 2.88 & -0.64 & 19.95 & 57.97 & 3.98 \\ 
borg\_0926+4426.163 & 141.388766 & 44.417316 & 176.000758 & 45.652281  & 23.46 $\pm$ 0.02 & 23.18 $\pm$ 0.02 & 22.92 $\pm$ 0.02 & 25.48 $\pm$ 0.90 & 4.02 $\pm$ 3.06 & 13.95 & 6.16 $\pm$ 5.99 & 0.46 & 12.81 & 7.58 & 3.71 \\ 
borg\_0926+4426.166 & 141.368669 & 44.415785 & 176.004767 & 45.638119  & 23.95 $\pm$ 0.02 & 23.79 $\pm$ 0.03 & 23.74 $\pm$ 0.04 & 26.30 $\pm$ 0.92 & 4.72 $\pm$ 3.11 & 13.46 & 4.93 $\pm$ 4.76 & 0.37 & 11.95 & 4.85 & 3.71 \\
\hline
\end{tabular}
}
}
\end{table*}

\addtocounter{table}{-1} 
\begin{table*}
\caption{Table \ref{t:Mdwarfs} -- {\em continued}}
\rotatebox{90}{
\tiny{
\begin{tabular}{l l l l l l l l l l l l l l l l l l l l l l}
Field  	& RA  	& DEC 	& $l$ 	& $b$ 	& $m_{F098W}$ 	& $m_{F125W}$   	& $m_{F125W}$ 	&  $m_{F606W}$  	&  Mtype 	& M-m   	& Dist 	&  Height &  Radius  	& Volume &  Area \\ 
		& J2000 	& J2000 	& $\circ$ 	& $\circ$ 	& mag 			& mag 			& mag			& mag			& 		& mag 	& kpc.	& kpc	& kpc.	& kpc$^2$	& arcmin$^2$			 \\
\hline
\hline
borg\_1230+0750.2486 & 187.471959 & 7.820086 & 287.128967 & 70.027773  & 22.49 $\pm$ 0.02 & 22.38 $\pm$ 0.02 & 22.28 $\pm$ 0.03 & 22.99 $\pm$ 0.02 & -1.71 $\pm$ 0.11 & 120.49 & \dots  & \dots & \dots & \dots  & 0.52 \\ 
borg\_1230+0750.3901 & 187.447223 & 7.825443 & 287.055567 & 70.026957  & 23.52 $\pm$ 0.05 & 23.11 $\pm$ 0.10 & 22.83 $\pm$ 0.07 & 25.29 $\pm$ 0.14 & 3.62 $\pm$ 0.56 & 13.88 & 5.98 $\pm$ 1.93 & -5.69 & 10.54 & 0.99 & 0.52 \\ 
borg\_1230+0750.4408 & 187.461291 & 7.827180 & 287.093877 & 70.032062  & 24.15 $\pm$ 2.07 & 23.88 $\pm$ 0.05 & 23.81 $\pm$ 0.08 & 25.51 $\pm$ 0.15 & 1.75 $\pm$ 0.55 & 16.01 & 15.92 $\pm$ 2.61 & -15.19 & 13.94 & 7.04 & 0.52 \\ 
borg\_1230+0750.4516 & 187.450540 & 7.827297 & 287.063559 & 70.029562  & 22.98 $\pm$ 0.03 & 22.82 $\pm$ 0.02 & 22.71 $\pm$ 0.03 & 23.71 $\pm$ 0.04 & -0.73 $\pm$ 0.15 & 15.48 & 12.45 $\pm$ 1.26 & -11.88 & 12.75 & 4.31 & 0.52 \\ 
borg\_1230+0750.5771 & 187.478632 & 7.831401 & 287.139635 & 70.040361  & 24.14 $\pm$ 0.08 & 23.94 $\pm$ 0.06 & 23.89 $\pm$ 0.08 & 25.06 $\pm$ 0.16 & 0.02 $\pm$ 0.57 & 16.60 & 20.90 $\pm$ 3.55 & -19.94 & 15.63 & 12.13 & 0.52 \\ 
borg\_1230+0750.7878 & 187.484283 & 7.834039 & 287.153652 & 70.044287  & 24.13 $\pm$ 0.07 & 23.51 $\pm$ 0.04 & 23.36 $\pm$ 0.06 & 24.89 $\pm$ 0.09 & 0.93 $\pm$ 0.34 & 15.90 & 15.12 $\pm$ 2.05 & -14.42 & 13.66 & 6.34 & 0.52 \\
\hline
\end{tabular}
}
}
\end{table*}

\addtocounter{table}{-1} 
\begin{table*}
\caption{Table \ref{t:Mdwarfs} -- {\em continued}}
\rotatebox{90}{
\tiny{
\begin{tabular}{l l l l l l l l l l l l l l l l l l l l l l}
Field  	& RA  	& DEC 	& $l$ 	& $b$ 	& $m_{F098W}$ 	& $m_{F125W}$   	& $m_{F125W}$ 	&  $m_{F606W}$  	&  Mtype 	& M-m   	& Dist 	&  Height &  Radius  	& Volume &  Area \\ 
		& J2000 	& J2000 	& $\circ$ 	& $\circ$ 	& mag 			& mag 			& mag			& mag			& 		& mag 	& kpc.	& kpc	& kpc.	& kpc$^2$	& arcmin$^2$			 \\
\hline
\hline
borg\_1815-3244.336 & 273.630284 & -32.751479 & 359.743625 & -7.275362  & 20.91 $\pm$ 0.03 & 20.81 $\pm$ 0.02 & 20.76 $\pm$ 0.03 & 22.81 $\pm$ 0.91 & 3.01 $\pm$ 3.09 & 12.68 & 3.43 $\pm$ 2.97 & 0.01 & 11.90 & 2.07 & 3.27 \\ 
borg\_1815-3244.355 & 273.628645 & -32.751258 & 359.743177 & -7.274039  & 21.66 $\pm$ 0.07 & 21.55 $\pm$ 0.05 & 21.54 $\pm$ 0.07 & 23.58 $\pm$ 0.95 & 3.12 $\pm$ 3.21 & 13.42 & 4.83 $\pm$ 4.21 & 0.01 & 13.30 & 4.12 & 3.27 \\ 
borg\_1815-3244.361 & 273.623142 & -32.751044 & 359.741201 & -7.269841  & 22.64 $\pm$ 0.14 & 22.23 $\pm$ 0.12 & 22.26 $\pm$ 0.30 & 24.30 $\pm$ 1.18 & 3.24 $\pm$ 4.04 & 14.10 & 6.62 $\pm$ 6.98 & -0.00 & 15.07 & 7.73 & 3.27 \\ 
borg\_1815-3244.372 & 273.627252 & -32.750901 & 359.742948 & -7.272835  & 23.26 $\pm$ 0.22 & 22.98 $\pm$ 0.14 & 23.02 $\pm$ 0.19 & 25.07 $\pm$ 1.07 & 3.30 $\pm$ 3.65 & 14.85 & 9.34 $\pm$ 9.88 & -0.01 & 17.77 & 15.38 & 3.27 \\ 
borg\_1815-3244.375 & 273.617257 & -32.750999 & 359.738923 & -7.265437  & 21.01 $\pm$ 0.04 & 20.80 $\pm$ 0.03 & 20.68 $\pm$ 0.03 & 22.73 $\pm$ 0.91 & 2.77 $\pm$ 3.10 & 12.67 & 3.42 $\pm$ 2.96 & 0.01 & 11.89 & 2.06 & 3.27 \\ 
borg\_1815-3244.396 & 273.617470 & -32.750665 & 359.739306 & -7.265440  & 21.70 $\pm$ 0.07 & 21.39 $\pm$ 0.04 & 21.37 $\pm$ 0.05 & 23.42 $\pm$ 0.93 & 3.13 $\pm$ 3.16 & 13.26 & 4.48 $\pm$ 3.89 & 0.01 & 12.94 & 3.53 & 3.27 \\
\hline
\end{tabular}
}
}
\end{table*}

\addtocounter{table}{-1} 
\begin{table*}
\caption{Table \ref{t:Mdwarfs} -- {\em continued}}
\rotatebox{90}{
\tiny{
\begin{tabular}{l l l l l l l l l l l l l l l l l l l l l l}
Field  	& RA  	& DEC 	& $l$ 	& $b$ 	& $m_{F098W}$ 	& $m_{F125W}$   	& $m_{F125W}$ 	&  $m_{F606W}$  	&  Mtype 	& M-m   	& Dist 	&  Height &  Radius  	& Volume &  Area \\ 
		& J2000 	& J2000 	& $\circ$ 	& $\circ$ 	& mag 			& mag 			& mag			& mag			& 		& mag 	& kpc.	& kpc	& kpc.	& kpc$^2$	& arcmin$^2$			 \\
\hline
\hline
borg\_1815-3244.2391 & 273.648867 & -32.738169 & 359.762826 & -7.283023  & 21.31 $\pm$ 0.05 & 21.18 $\pm$ 0.04 & 21.15 $\pm$ 0.05 & 23.20 $\pm$ 0.92 & 3.07 $\pm$ 3.14 & 13.05 & 4.08 $\pm$ 3.54 & 0.01 & 12.54 & 2.93 & 3.27 \\ 
borg\_1815-3244.2411 & 273.649471 & -32.737717 & 359.763467 & -7.283263  & 22.05 $\pm$ 0.09 & 21.84 $\pm$ 0.06 & 21.70 $\pm$ 0.07 & 23.75 $\pm$ 0.95 & 2.70 $\pm$ 3.21 & 13.71 & 5.51 $\pm$ 4.81 & 0.00 & 13.97 & 5.35 & 3.27 \\ 
borg\_1815-3244.2414 & 273.647984 & -32.737928 & 359.762694 & -7.282254  & 21.28 $\pm$ 0.05 & 21.14 $\pm$ 0.03 & 21.03 $\pm$ 0.04 & 23.07 $\pm$ 0.92 & 2.80 $\pm$ 3.12 & 13.01 & 3.99 $\pm$ 3.46 & 0.01 & 12.46 & 2.81 & 3.27 \\ 
borg\_1815-3244.2423 & 273.635752 & -32.738100 & 359.757723 & -7.273221  & 19.97 $\pm$ 0.02 & 19.84 $\pm$ 0.01 & 19.85 $\pm$ 0.01 & 21.89 $\pm$ 0.89 & 3.20 $\pm$ 3.03 & 11.71 & 2.20 $\pm$ 1.89 & 0.02 & 10.68 & 0.85 & 3.27 \\ 
borg\_1815-3244.2510 & 273.612674 & -32.737735 & 359.748959 & -7.255861  & 21.78 $\pm$ 0.11 & 21.57 $\pm$ 0.07 & 21.59 $\pm$ 0.10 & 23.63 $\pm$ 0.98 & 3.21 $\pm$ 3.35 & 13.44 & 4.88 $\pm$ 5.10 & 0.01 & 13.35 & 4.20 & 3.27 \\ 
borg\_1815-3244.2514 & 273.629024 & -32.737249 & 359.755832 & -7.267814  & 22.67 $\pm$ 0.14 & 22.54 $\pm$ 0.10 & 22.35 $\pm$ 0.11 & 24.39 $\pm$ 0.99 & 2.53 $\pm$ 3.38 & 14.41 & 7.61 $\pm$ 6.72 & -0.01 & 16.05 & 10.20 & 3.27 \\
\hline
\end{tabular}
}
}
\end{table*}

\addtocounter{table}{-1} 
\begin{table*}
\caption{Table \ref{t:Mdwarfs} -- {\em continued}}
\rotatebox{90}{
\tiny{
\begin{tabular}{l l l l l l l l l l l l l l l l l l l l l l}
Field  	& RA  	& DEC 	& $l$ 	& $b$ 	& $m_{F098W}$ 	& $m_{F125W}$   	& $m_{F125W}$ 	&  $m_{F606W}$  	&  Mtype 	& M-m   	& Dist 	&  Height &  Radius  	& Volume &  Area \\ 
		& J2000 	& J2000 	& $\circ$ 	& $\circ$ 	& mag 			& mag 			& mag			& mag			& 		& mag 	& kpc.	& kpc	& kpc.	& kpc$^2$	& arcmin$^2$			 \\
\hline
\hline
borg\_1815-3244.4016 & 273.630301 & -32.731098 & 359.761828 & -7.265908  & 21.31 $\pm$ 0.07 & 21.17 $\pm$ 0.05 & 21.04 $\pm$ 0.05 & 23.09 $\pm$ 0.93 & 2.74 $\pm$ 3.17 & 13.04 & 4.05 $\pm$ 3.53 & 0.01 & 12.52 & 2.90 & 3.27 \\ 
borg\_1815-3244.4042 & 273.615880 & -32.731286 & 359.755979 & -7.255253  & 20.27 $\pm$ 0.03 & 20.05 $\pm$ 0.02 & 19.96 $\pm$ 0.03 & 22.01 $\pm$ 0.90 & 2.87 $\pm$ 3.07 & 11.92 & 2.42 $\pm$ 2.09 & 0.02 & 10.90 & 1.03 & 3.27 \\ 
borg\_1815-3244.4083 & 273.617450 & -32.731099 & 359.756764 & -7.256336  & 21.50 $\pm$ 0.06 & 21.38 $\pm$ 0.05 & 21.32 $\pm$ 0.06 & 23.37 $\pm$ 0.94 & 2.97 $\pm$ 3.19 & 13.25 & 4.47 $\pm$ 3.89 & 0.01 & 12.93 & 3.52 & 3.27 \\ 
borg\_1815-3244.4172 & 273.637713 & -32.723422 & 359.771600 & -7.267865  & 21.79 $\pm$ 0.07 & 21.61 $\pm$ 0.05 & 21.57 $\pm$ 0.06 & 23.62 $\pm$ 0.94 & 3.05 $\pm$ 3.20 & 13.48 & 4.96 $\pm$ 4.32 & 0.01 & 13.42 & 4.34 & 3.27 \\ 
borg\_1815-3244.4178 & 273.611336 & -32.724976 & 359.759821 & -7.248936  & 21.43 $\pm$ 0.05 & 21.23 $\pm$ 0.04 & 21.12 $\pm$ 0.04 & 23.17 $\pm$ 0.92 & 2.81 $\pm$ 3.13 & 13.10 & 4.16 $\pm$ 3.61 & 0.01 & 12.63 & 3.05 & 3.27 \\ 
borg\_1815-3244.4196 & 273.627084 & -32.726965 & 359.764249 & -7.261592  & 17.27 $\pm$ 0.00 & 17.15 $\pm$ 0.00 & 16.95 $\pm$ 0.00 & 19.00 $\pm$ 0.88 & 2.51 $\pm$ 2.99 & 9.02 & 0.64 $\pm$ 0.55 & 0.02 & 9.13 & 0.07 & 3.27 \\
\hline
\end{tabular}
}
}
\end{table*}

\addtocounter{table}{-1} 
\begin{table*}
\caption{Table \ref{t:Mdwarfs} -- {\em continued}}
\rotatebox{90}{
\tiny{
\begin{tabular}{l l l l l l l l l l l l l l l l l l l l l l}
Field  	& RA  	& DEC 	& $l$ 	& $b$ 	& $m_{F098W}$ 	& $m_{F125W}$   	& $m_{F125W}$ 	&  $m_{F606W}$  	&  Mtype 	& M-m   	& Dist 	&  Height &  Radius  	& Volume &  Area \\ 
		& J2000 	& J2000 	& $\circ$ 	& $\circ$ 	& mag 			& mag 			& mag			& mag			& 		& mag 	& kpc.	& kpc	& kpc.	& kpc$^2$	& arcmin$^2$			 \\
\hline
\hline
borg\_1815-3244.6368 & 273.632076 & -32.716968 & 359.775141 & -7.260667  & 22.63 $\pm$ 0.13 & 22.41 $\pm$ 0.09 & 22.37 $\pm$ 0.11 & 24.42 $\pm$ 0.99 & 3.04 $\pm$ 3.37 & 14.28 & 7.18 $\pm$ 6.32 & -0.00 & 15.63 & 9.09 & 3.27 \\ 
borg\_1815-3244.6457 & 273.626287 & -32.718826 & 359.771201 & -7.257217  & 17.82 $\pm$ 0.00 & 17.65 $\pm$ 0.00 & 17.59 $\pm$ 0.00 & 19.64 $\pm$ 0.88 & 2.96 $\pm$ 3.00 & 9.52 & 0.80 $\pm$ 0.69 & 0.02 & 9.30 & 0.11 & 3.27 \\ 
borg\_1815-3244.6462 & 273.626119 & -32.717768 & 359.772080 & -7.256600  & 20.63 $\pm$ 0.03 & 20.19 $\pm$ 0.01 & 20.07 $\pm$ 0.02 & 22.11 $\pm$ 0.90 & 2.74 $\pm$ 3.04 & 12.06 & 2.58 $\pm$ 2.23 & 0.02 & 11.06 & 1.18 & 3.27 \\ 
borg\_1815-3244.6499 & 273.631052 & -32.719064 & 359.772866 & -7.260878  & 19.87 $\pm$ 0.02 & 19.74 $\pm$ 0.01 & 19.75 $\pm$ 0.02 & 21.79 $\pm$ 0.89 & 3.20 $\pm$ 3.04 & 11.61 & 2.10 $\pm$ 1.81 & 0.02 & 10.58 & 0.78 & 3.27 \\ 
borg\_1815-3244.6502 & 273.627646 & -32.716892 & 359.773463 & -7.257332  & 20.65 $\pm$ 0.04 & 20.35 $\pm$ 0.02 & 20.32 $\pm$ 0.03 & 22.37 $\pm$ 0.91 & 3.06 $\pm$ 3.09 & 12.22 & 2.78 $\pm$ 2.41 & 0.02 & 11.26 & 1.36 & 3.27 \\ 
borg\_1815-3244.6520 & 273.621825 & -32.719025 & 359.769266 & -7.253985  & 22.08 $\pm$ 0.10 & 21.98 $\pm$ 0.09 & 21.88 $\pm$ 0.10 & 23.93 $\pm$ 0.98 & 2.84 $\pm$ 3.33 & 13.85 & 5.88 $\pm$ 5.17 & 0.00 & 14.33 & 6.09 & 3.27 \\
\hline
\end{tabular}
}
}
\end{table*}

\clearpage

\newpage

\end{document}